\documentclass[
 aip, cha, amsmath, amssymb, reprint
]{revtex4-2}

\usepackage{graphicx}
\usepackage{dcolumn}
\usepackage{bm}
\usepackage[utf8]{inputenc}
\usepackage[T1]{fontenc}
\usepackage{mathptmx}
\usepackage{etoolbox}
\usepackage[colorlinks=true,linkcolor=blue,urlcolor=blue,citecolor=blue]{hyperref}
\usepackage{physics}
\usepackage{tikz,pgfplots}
\pgfplotsset{compat=1.18}
\usepackage[noend]{algpseudocode} 

\makeatletter
\def\@email#1#2{%
 \endgroup
 \patchcmd{\titleblock@produce}
  {\frontmatter@RRAPformat}
  {\frontmatter@RRAPformat{\produce@RRAP{*#1\href{mailto:#2}{#2}}}\frontmatter@RRAPformat}
  {}{}
}%
\makeatother

\AtBeginDocument{%
  \setlength{\abovedisplayskip}{10pt}
  \setlength{\belowdisplayskip}{10pt}
} 

\allowdisplaybreaks

\begin{document}

\preprint{AIP/123-QED}

\title{Revisiting the Dynamical Properties of Pedlosky's Two-Layer Model for Finite Amplitude Baroclinic Waves}

\author{Nicolas De Ro}
\email{nicolas.dero@unige.ch}
\affiliation{Department of Theoretical Physics, University of Geneva, 24 quai Ernest-Ansermet, 1211 Genève 4, Switzerland}
 
\author{Jonathan Demaeyer}
\affiliation{Royal Meteorological Institute of Belgium, Meteorological and Climatological Research, Avenue Circulaire, 3, 1180 Brussels,
Belgium}

\author{Stéphane Vannitsem}
\affiliation{School of Physical and Mathematical Sciences \& The Asian School of the Environment, Nanyang Technological University, Singapore}

\date{\today}

\begin{abstract}
\hfill \\
Baroclinic instability is a fundamental mechanism driving atmospheric dynamics. In this work, we revisit Pedlosky’s two-layer model for finite amplitude baroclinic waves -- a seminal framework for studying the unstable growth of finite perturbations -- leveraging modern nonlinear techniques and computational resources. We show that the geophysical state of the baroclinic wave exhibits a rich diversity of dynamical regimes governed by the level of dissipation induced by Ekman boundary layers. In the inviscid limit, we demonstrate that the model is integrable. Upon increasing dissipation, the system undergoes a complex sequence of bifurcations. On one hand, deterministic chaos, identified by means of the Lyapunov exponents, provides a genuine mechanism for destabilization of the wave. On the other hand, in regimes where the wave equilibrates, dependence on the initial condition is crucial, eventually leading to the coexistence of multiple attractors. We study the governing equations of the model and their truncation to a finite-dimensional system of ordinary differential equations, together with the minimal low-order truncated system which is structurally equivalent to the Lorenz model. Its bifurcation diagram allows for elucidating the transition of the wave amplitude from stable equilibration to periodic oscillations -- terminating in homoclinic orbits -- and, ultimately, deterministic chaos through a period-doubling route. We finally comment on the robustness of these features for higher-dimensional models.
\end{abstract}

\maketitle

\begin{quotation}
The dynamics of Earth’s atmosphere and oceans are governed by equations of motion derived from fundamental physical principles. Observations indicate that atmospheric and oceanic flows are generally time-dependent and erratic. The emergence of such solutions through instabilities is an essential element of the understanding of large-scale motions. Baroclinic instability originates from the presence of a horizontal temperature gradient in a rotating fluid that is subject to vertical wind shear. The growth of these unstable waves generates storms and influences global atmospheric and ocean circulations. While studying the growth of these waves is a difficult task in the real Earth system, modeling the atmosphere or the ocean as two layers of stratified, rotating fluids offers an insightful framework for investigation, capturing the essential dynamics. Here, we use such a representation, originally proposed by Joseph Pedlosky, to explore the nonlinear growth of finite amplitude baroclinic waves, which exhibit deterministic chaos and other complex features as dissipation is introduced. This study revisits the two-layer model through the lens of modern analytical and numerical techniques allowing for a detailed description of the nature of the solutions generated by the system. Among the key findings, the presence of multiple coexisting attractors with intricate basin boundaries reveals an additional complexity of the system which renders the problem of forecasting much harder.
\end{quotation}

\section{\label{section1}Introduction}

The atmosphere in the extratropics displays a variability whose origin lies in the presence of differential heating between the Tropics and the Poles. This imposes gradients of temperature that in turn are inducing exchanges of air masses between the Tropics and the Poles. At large scales, this transport is responsible for the development of the high and low pressure zones moving eastward in the extratropics \cite{Holton, VallisBook, Pedlosky_book}. The dynamics of high and low pressure zones display an erratic evolution whose description is at the center of more than a hundred years of research. This erratic behavior was already pointed out by \cite{Poincare} who suggested that this dynamics is associated with the natural instability of atmospheric flows. Nowadays, this remarkable intuition is embedded in the theories of dynamical systems and chaos which allows to understand the complexity of the atmospheric dynamics, see for example \onlinecite{Nicolisbook,Vannitsem0,Vannitsem1}.

Baroclinic instability is nowadays considered as the major element inducing the development of high and low pressure zones over the extratropics. This instability is the result of a vertical velocity shear, which in turn is associated with a horizontal temperature gradient \cite{Holton}. This feature of the atmosphere has been considerably studied over the past decades, in particular to understand the emergence of new solutions beyond the instability. Such developments were in particular led by Joseph Pedlosky who developed an important model describing these new solutions \cite{Pedlosky1, Pedlosky2, Pedlosky3}. The current paper is revisiting the model proposed by Pedlosky through the lens of modern dynamical systems theory, and evaluating how chaos is emerging in this system.

New insights on the dynamics are gained thanks to a powerful solutions continuation software, known as AUTO~\cite{auto07p}. Combined with the current computational power \footnote{That was not available at the time of the first attempt to characterize the bifurcation diagram of the model~\cite{Pedlosky3}.} and the computation of the Lyapunov exponents~\cite{Benettin_1, Benettin_2, Benettin_3, Kuptsov2012}, it is shown that the model possesses interesting similarities with the Lorenz model~\cite{Lorenz1963}, but also notable differences, raising questions about the universality of Lorenz' model and its link with more general geophysical models~\cite{Curry1978, Roy2007, Saiki2017}. In particular, the present work details the route to chaos in Pedlosky's model, which proves to be of a different kind from the Lorenz model.

Section~\ref{sec:PM} introduces the Pedlosky model of baroclinic instability and its ordinary differential equations (ODEs) representation. In Section~\ref{sec:stability}, the linear stability of the model's equilibria is assessed, unraveling the particular structure of its Jacobian matrix. This sets the stage for the investigation in~\ref{sec:toy_model} of \emph{a toy model} -- a highly-truncated 3-dimensional representation of Pedlosky model's dynamics -- whose bifurcation diagram structure and route to chaos are elucidated. It helps gain understanding of the dynamical behavior of higher-dimensional, less severely truncated representation of the Pedlosky model, as shown in Section~\ref{sec:full_model}. Finally, we present the conclusion and outlook of this analysis in Section~\ref{sec:conclusion_outlooks}.

\section{\label{sec:PM}Pedlosky's model}

We present Pedlosky's two-layer model and recall the primary assumptions and ideas underlying its derivation. Once the main object of this work -- the system of ODEs -- is defined, we will highlight some of its basic features and demonstrate that the system is integrable for specific parameter values.

\subsection{\label{sec:origin}Origin of the model}

A key process at play in the atmosphere is baroclinic instability, known to be the primary source of large-scale variability in the extratropical atmosphere. This instability originates in vertical wind shear, which is related to the horizontal temperature gradients present in the atmosphere. These gradients result from the differential heating of the atmosphere between the Tropics and the Poles \cite{Pedlosky_book, VallisBook}. Once the vertical shear flow destabilizes, new types of solutions should emerge through the development of finite perturbations beyond the bifurcation point. Pedlosky's model is precisely an attempt to describe such a perturbation close to the bifurcation point. 

\begin{figure}[b!]
\includegraphics[scale=0.6]{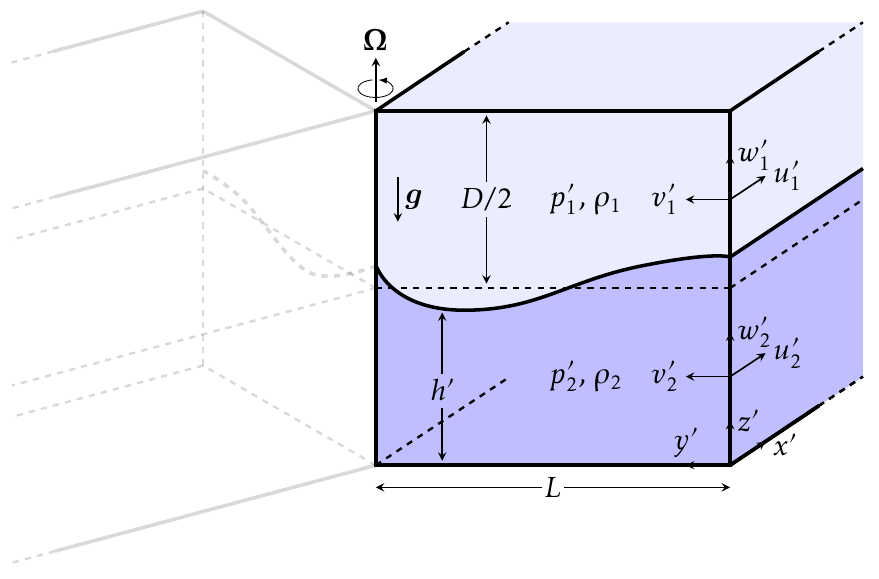}
\caption{Schematic representation of the two-layer atmospheric model. The layers are distinguished by color within a semi-infinite domain of width $L$ and height $D$, rotating about the indicated axis. At rest, the interface lies in the mid-plane $z'=D/2$, between the rigid boundaries at $z'=0$ and $z'=D$. Under relative motion, the interface shifts to height $z'=h'$. Adapted from Ref.~\onlinecite{Pedlosky1}.}
\label{fig:model} 
\end{figure}

The two-layer atmosphere serves as a minimal model for baroclinic instability, which was developed by Pedlosky in a series of papers \cite{Pedlosky1, Pedlosky2, Pedlosky3}.  We first concisely review the derivation of the model's governing equations (see Eqs.~\eqref{PM:MainEquations} below), focusing on the geophysical principles rather than exhaustive mathematical details. The system consists of two layers of homogeneous, immiscible fluids with distinct, constant densities -- where the denser fluid underlies the lighter one -- confined within a semi-infinite rectangular domain subject to gravity, see Fig.~\ref{fig:model}. The domain rotates at an angular velocity $\Omega$ about the vertical axis passing through $(0,L,0)$ and $(0,L,D)$. 

The equations of motion, expressed in terms of dimensional variables (denoted by a prime symbol), comprise five equations per layer. Three of these are the equations for each component of the velocity field in the rotating coordinate frame,
\begin{align}\label{PM:EqDim1}
     \dv{\bm{u}'_n}{t'}&\equiv \partial_t' \bm{u}'_n+(\bm{u}'_n\cdot \bm{\nabla}')\bm{u}'_n\nonumber \\
     &=-2\, \bm{\Omega}\times \bm{u}'_n+\bm{g}-\frac{1}{\rho_n}\bm{\nabla}'p'_n+\nu \bm{\nabla}'^2 \bm{u}_n'\, ,
\end{align}
and one is the continuity equation $\bm{\nabla}'\cdot \bm{u}_n'=0$, which reduces to the incompressibility condition since the density of each fluid layer is constant. In addition to these equations, there is an equation for the kinematic boundary condition at the interface of the two fluids, $\dd h'/\dd t=w_n'$.

 In Eq.~\eqref{PM:EqDim1}, it is implicitly assumed that the effective gravity coincides with the true gravity as the effects of the centrifugal force are neglected. The frictional term arises from the pressure tensor, which for a general velocity field $\bm{u}=\bm{u}(t,\bm{x})$ takes the form \cite{Landau}, 
 \begin{equation}
     \text{P}_{ij}=p \,\delta_{ij}-\mu \left(\partial_i u_j+\partial_j u_j-\frac{2}{3}\, \bm{\nabla}\cdot \bm{u} \,\delta_{ij}\right)-\zeta\, \bm{\nabla}\cdot \bm{u} \, \delta_{ij}\, ,
 \end{equation}  
 where $p$ is the pressure field, $\mu$ is the dynamic (shear) viscosity and $\zeta$ the bulk viscosity.

All the equations of motion can be brought to a dimensionless form by dividing the primed variable by an associated characteristic value (the resulting dimensionless variables are denoted without the prime symbol),
 \begin{subequations}
 \begin{align}
     &\epsilon\left(\partial_t u_n+ u_n\partial_x u_n+v_n \partial_y u_n+w_n\partial_z u_n\right)\nonumber \\
     &\qquad=  v_n - \partial_x p_n+ \frac{E}{2}\left( \delta^2(\partial_x^2+\partial_y^2)+\partial_z^2\right) u_n\, , \label{PM:Eq1} \\
    &\epsilon\left(\partial_t v_n+ u_n\partial_x v_n+v_n \partial_y v_n+w_n\partial_z v_n\right)&\nonumber \\
    & \qquad= -u_n - \partial_y p_n+ \frac{E}{2}\left( \delta^2(\partial_x^2+\partial_y^2)+\partial_z^2 \right) v_n\, , \label{PM:Eq2} \\
    & \delta^2 \epsilon \left(\partial_t w_n+ u_n\partial_x w_n+v_n \partial_y w_n+w_n\partial_z w_n \right)\nonumber \\
    &\qquad= -\partial_z p_n+\delta^2\frac{E}{2}\left( \delta^2(\partial_x^2+\partial_y^2)+\partial_z^2 \right) w_n\, , \label{PM:Eq3} \\
    & \partial_x u_n+\partial_y v_n+\partial_z w_n=0\, , \label{PM:Eq4} \\
    & \frac{\epsilon F}{2}\left(\partial_th+u_n\partial_x h+v_n\partial_y h\right)=w_n \label{PM:Eq5}\, .
 \end{align}
 \end{subequations}
 
 Eqs.~\eqref{PM:Eq1}-\eqref{PM:Eq5} are at the core of Pedlosky's work in the set of papers cited above. They describe the dynamics of the fluid in the two-layer model under the assumptions introduced previously. Several dimensionless parameters of physical relevance appear in these equations. $\delta\equiv D/L$ is the aspect ratio between the vertical and horizontal scales of the problem. In the limit where $\delta \rightarrow 0$, the equations describe a horizontal motion. $\epsilon\equiv U/2 \Omega L$ is the Rossby number, representing the ratio of inertial forces in the rotating frame to the Coriolis acceleration. $E\equiv \nu / \Omega D^2$ (with $\nu=\mu/\rho$, the kinematic viscosity) is the Ekman number, the ratio of frictional forces per unit of mass to the Coriolis acceleration. The frictional terms introduce second-order derivatives in the velocity equations. In the limit $E \rightarrow 0$, the equations describe a geostrophic flow. For the dynamics of interest, $\epsilon\gg E$, so the frictional terms are neglected initially but are reintroduced later through a separate mechanism. Finally, $F\equiv 8\rho_2 \Omega^2 L^2/ D g (\rho_2-\rho_1)$ is the Froude number, representing the ratio of inertial forces to the external gravity field. It appears only in Eq.~\eqref{PM:Eq5} for the kinematic boundary condition at the interface.

Together with appropriate (unspecified) boundary and initial conditions, Eqs.~\eqref{PM:Eq1}-\eqref{PM:Eq5} form a closed system. Although the system determines the five unknown fields $f_i$ ($i=1,\cdots,5$), it is important to note that only the velocities and interface height are governed by prognostic equations. The resulting solutions take the form $f_i=f_i(t, \bm{x}\, ;\, \{\delta, \epsilon, E, F \})$.

Exploiting the smallness of the Rossby number, a formal asymptotic expansion of the equations of motion in $\epsilon$ can be performed,
 \begin{equation}\label{PM:FieldExpansion}
     f_i(t, \bm{x}\, ;\, \{\delta, \epsilon, E, F \}) = \sum_{k=0}^\infty \epsilon^k f_i^{(k)}(t, \bm{x}\, ;\, \{\delta, E, F \})\, .
 \end{equation}
The validity of such a procedure is discussed in Ref.~\onlinecite{RossbyExpansion}. At zeroth-order in $\epsilon$, the set of equations \eqref{PM:Eq1}-\eqref{PM:Eq5} describes a hydrostatic and geostrophic flow, with $\psi_n\equiv p_n^{(0)}$, where the zeroth-order pressure field serves as the geostrophic stream function. Eqs.~\eqref{PM:Eq1} and \eqref{PM:Eq2} can be cast into the vorticity equation 
\begin{equation}\label{PM:VorticityEquation}
    \dv{_{0,n}}{t} \xi_n^{(0)}=\partial_z w_n^{(1)}\, ,
\end{equation}
where $\dd_{0,n}/\dd t\equiv \partial_t + u_n^{(0)}\partial_x + v_n^{(0)}\partial_y=\partial_t - \partial_y \psi_n \partial_x + \partial_x \psi_n\partial_y$. In this expression, $\xi_n^{(0)}$ represents the vertical component of the zeroth-order geostrophic vorticity $\bm{\xi}_n^{(0)}=\bm{\nabla}\times \bm{u}_n^{(0)}$. Noting that the left-hand side of the above equation is independent of $z$, and assuming that the density jump between the two layers is small (so that $h^{(0)}\simeq \psi_2-\psi_1$), it can be integrated with respect to $z$ within each layer,
\begin{subequations}\label{PM:FinalTransit}
\begin{align}
    &\dv{_{0,1}}{t}\left( \bm{\nabla}^2 \psi_1+F \left(\psi_2-\psi_1\right)\right) = 2 w_1^{(1)}(t,x,y,1)\, ,\\
    &\dv{_{0,2}}{t}\left( \bm{\nabla}^2 \psi_2+F \left(\psi_1-\psi_2\right)\right) = -2 w_1^{(1)}(t,x,y,1)\,.
\end{align}
\end{subequations}
In Ref.~\onlinecite{ViscosityResult}, the authors show (see Eq.~(31) therein) that
\begin{subequations}\label{PM:EkmanLayerVertical}
\begin{align} 
    & w_1^{(1)}(t,x,y,1) = -r\, \xi_{1}^{(0)}(t,x,y,1)\, , \\
    & w_2^{(1)}(t,x,y,0) =  r\, \xi_{2}^{(0)}(t,x,y,0)\, ,
\end{align}
\end{subequations}
with $r = \sqrt{E}/(2\epsilon)$. See also \cite{VallisBook} for a pedagogical derivation of this result. This is a generic result for the vertical velocity showing that, due to frictional forces, geostrophic vorticity induces an ascending motion. With this result in hand, Eqs.~\eqref{PM:FinalTransit} can be rewritten as
\begin{subequations}\label{PM:Final}
\begin{align}
    &\dv{_{0,1}}{t}\left( \bm{\nabla}^2 \psi_1+F \left(\psi_2-\psi_1\right)\right) = -r \bm{\nabla}^2 \psi_1\, ,\\
    &\dv{_{0,2}}{t}\left( \bm{\nabla}^2 \psi_2+F \left(\psi_1-\psi_2\right)\right) = r \bm{\nabla}^2 \psi_2\,,
\end{align}
\end{subequations}
and describe the dynamics of the fluid within each layer of the model up to first-order correction in the Rossby number. The right-hand-side contains a viscosity term since $r$ is proportional to $\sqrt{\nu}$. In the absence of viscosity, the quantities in parenthesis are conserved along fluid elements.

Eqs.~\eqref{PM:Final} admit a trivial solution given by $u_1^{(0)} = U_1$, $u_2^{(0)} = U_2$, with $U_1 \neq U_2$ two constants and $v_1^{(0)} = v_2^{(0)}=0$, \textit{i.e.} $\psi_n = - U_n y$. This corresponds to a uniform and stationary (basic) zonal flow. The fundamental question is: how does a small perturbation superimposed to this trivial solution evolve under Eqs.~\eqref{PM:Final}? Will it grow, equilibrate or decay over time? Of particular interest is the regime in which the perturbation, which is associated with the baroclinic instability, exhibits unstable growth. To analyze this, we write
\begin{equation}\label{PM:PerturbationExpression}
    \psi_n(t,x,y) = -U_n y + \varphi_n(t,x,y)\, ,
\end{equation}
where $\varphi(t,x,y)$ is an arbitrarily small perturbation of the trivial solution. The linear stability problem is obtained by substituting the geostrophic stream function into Eqs.~\eqref{PM:Final}, retaining only terms linear in $\varphi_n$, and making the ansatz of wave solutions ($m\in \mathbb{Z}$)
\begin{subequations}\label{PM:WaveAnsatz}
 \begin{align}\label{wave_solutions}
     & \varphi_1=\Re A \, e^{i\, k(x-c t)}\sin m \pi y\, , \\
     & \varphi_2=\Re \alpha A \, e^{i\, k(x-c t)}\sin m \pi y \, ,
 \end{align}
\end{subequations}
for the baroclinic waves. These represent two-dimensional waves with spatial oscillations in the $y$-direction and of wavelength $2/m$, propagating in the $x$-direction with wave speed $c$ and frequency $c k / 2 \pi$. Here, $A$ is the amplitude of the wave in the first layer and $\alpha A$ is the amplitude of the wave in the second one. The corresponding dispersion relation $c=c(k)$ is given by
\begin{align}\label{PM:DispersionRelation}
    c_{\pm}
    &=\frac{U_1+U_2}{2}\pm\frac{\sqrt{(a^4-4F^2) (U_1-U_2)^2-4 \frac{F^2 r^2}{k^2}}}{2(a^2+2F)}\nonumber \\
    &\quad -i\,\frac{r}{k}\frac{(a^2+F)}{(a^2+2F)}\, ,
\end{align}
where $a^2 = k^2 + \pi^2 m^2$. The wave becomes unstable whenever the imaginary part $c_i$ of the wave speed $c$ is strictly positive. The marginal scenario corresponds to the case where $c_i = 0$, which yields a condition for the critical Froude number associated with the wave of speed $c_+$,
\begin{equation}
    F_c = \frac{a^2}{2}+\frac{2 r^2 a^2}{k^2(U_1-U_2)^2}\, .
\end{equation}
We note that this is different from the critical Froude number associated with the wave of speed $c_-$. In particular, for $F>F_c$, the imaginary part of $c_+$ is positive, whereas that of $c_-$ is negative. On the other hand, when $F<F_c$, the imaginary parts of both $c_\pm$ are always negative. Hence, in the former regime one wave experiences instability while the other is damped, whereas in the latter, both waves are damped.

Having identified the instability criterion for one of the waves, the Froude number is fixed near the marginal case, $F = F_c + \Delta$, with $|\Delta|\ll F_c$ and $\Delta>0$. Since the critical Froude number is a function of $r$, requiring $|\Delta|\ll F_c$ imposes a relation between $r$ and $\Delta$. Two extreme regimes can be identified: $r=0$, corresponding to an inviscid fluid, and large $r$, corresponding to a highly viscous fluid. We work in the intermediate regime in which $r$ scales as $\mathcal{O}(\sqrt{\Delta})$. 

One also has to investigate over which time scale the wave grows and how it depends on $\Delta$. For $F = F_c + \Delta$ and $r \leq C\sqrt{\Delta}$ for some constant $C$, the imaginary part of the dispersion relation \eqref{PM:DispersionRelation} scales as $\mathcal{O}(\sqrt{\Delta})$. Consequently, the growth of the wave occurs over a long time scale. Indeed, since $e^{i\, k(x-c t)} \propto e^{k c_i t}$, and since $c_i$ scales as $\mathcal{O}(\sqrt{\Delta})$, the development of the wave takes place on time scales of order $\mathcal{O}(1/\sqrt{\Delta})$. This suggests the introduction of a slow time variable $T$ defined such that the fast time $t$ is given by $t=T/\sqrt{\Delta}$, and $c_i t=\mathcal{O}(1)$. The variable $T$ will therefore play the role of the physical time scale used throughout this work. The perturbation $\varphi$, earlier introduced in Eq.~\eqref{PM:PerturbationExpression}, is now a function of both slow and fast times, \textit{i.e.} $\varphi_n=\varphi_n(t, T, x,y)$. Differential operators are also subject to this time separation, $\dd_{0,n}/ \dd t=\partial_t+ \sqrt{\Delta}\, \partial_T + u_n^{(0)}\partial_x + v_n^{(0)}\partial_y$. 

Exploiting the smallness of $\Delta$, the perturbation $\varphi$ to the basic zonal flow can be asymptotically expanded as $\varphi_n=\Delta^{1/2}\,\varphi^{(1)}_n+\Delta\,\varphi^{(2)}_n+\Delta^{3/2}\,\varphi^{(3)}_n+\mathcal{O}(\Delta^2)$. Substituting this into Eqs.~\eqref{PM:Final} yields a hierarchy of equations for the perturbations $\varphi^{(1)}_n$, $\varphi^{(2)}_n$ and $\varphi^{(3)}_n$, obtained by collecting terms of equal order in $\Delta$. The first two problems admit wave solutions of the form given in Eqs.~\eqref{wave_solutions}, with a slowly varying amplitude. At these orders, the governing equations have the generic form $\mathcal{L}\varphi^{(i)}_n=g(T,y)$ ($i=1,2$), where $\mathcal{L}$ is a linear differential operator involving derivatives with respect to $t$ and $x$, and $g$ depends only on $T$ and $y$. Consequently, any function $\Phi(T,y)$ independent of $t$ and $x$ solves the associated homogeneous problem and may be interpreted as a correction to the basic zonal flow.

Such zonal-flow corrections, however, are not dynamically forced until nonlinear wave-wave interactions appear in the third-order problem, and therefore only the zonal-flow correction arising at second order in $\Delta$ needs to be retained, as earlier additions merely introduce unnecessary degrees of freedom. For clarity, the perturbations in each layer at first and second order in $\Delta$ are given by Eqs.~\eqref{wave_ansatz}.

At third order, nonlinearities generate secular contributions whose elimination is necessary to preserve the validity of the asymptotic expansion, thereby providing a dynamical constraint on the zonal-flow correction. In other words, the zonal-flow corrections are required to balance the growth of the secular forcing terms generated by nonlinear wave interactions. In addition, solvability of the third-order problem requires to satisfy the Fredholm alternative~\cite{nicolisbook2}, which yields an evolution equation for the wave amplitude. Taken together, these two constraints lead to the following system of coupled partial differential equations (PDEs) for the (real) wave amplitude $A(T)$, which is the amplitude of the baroclinic wave \footnote{Hereafter referred simply as the \textit{wave amplitude}.}, and the zonal-flow correction $\Phi(y,T)$,
\begin{subequations}\label{PM:MainEquations}
\begin{align}
    & \partial_T\left(\partial_y^2 \Phi -a^2\Phi\right)+\gamma \partial_y^2\Phi=\Big(\dv{A^2}{T}+2\gamma A^2\Big)\sin 2 m\pi y \label{PM:MainEquations1} \, ,\\
    & \dv[2]{A}{T}+\frac{3\gamma}{2}\dv{A}{T}+A\left(\int_0^1\sin\left(2 m\pi y\right)\partial_y^2\Phi\, \dd y-1\right)=0\, , \label{PM:MainEquations2}
\end{align}
\end{subequations}
where $\gamma=r/(\omega \sqrt{\Delta})$ and $\omega=k(U_1-U_2)/2a$. 

Beyond the mathematical derivation, it is crucial to emphasize the model's geophysical relevance. Although the two-layer model is a simplified representation of the atmosphere, it contains the minimal ingredients allowing to characterize baroclinic instability, known for giving rise to large-scale perturbations occurring at mid-latitudes. Moreover, its strength lies in its mathematical tractability. Specifically, Eqs.~\eqref{PM:MainEquations}
 \begin{widetext}
     \begin{subequations}\label{wave_ansatz}
 \begin{align}
     &\varphi_1(x,y,t,T)= \sqrt{\Delta} \Re A(T) \, e^{i\, k(x-c t)}\sin m \pi y+\Delta \,\Phi(y,T)\, ,\\
     &\varphi_2(x,y,t,T)=\sqrt{\Delta} \Re\left( A(T) -\frac{4i\,\sqrt{\Delta}}{k(U_1-U_2)} \,\left[ \frac{r}{\sqrt{\Delta}}A(T)+A'(T)\right] \right) \, e^{i\, k(x-c t)}\sin m \pi y-\Delta\,\Phi(y,T)\,.
 \end{align}
 \end{subequations}
 \end{widetext}
address a fundamental question: given a uniform zonal flow, through which mechanisms can a tiny perturbation of the flow evolve into a large-scale, measurable disturbance? Addressing this question is the main motivation of the present work. More precisely, the growth dynamics of the wave amplitude $A(T)$ are expected to depend on viscosity. Increased viscosity leads to stronger dissipation and should, \textit{a priori}, stabilize the wave. Conversely, one may ask how the wave amplitude evolves in the weak-viscosity regime. Can the wave reach a state of maximal instability -- namely, chaos? Answering these questions requires solving Eqs.~\eqref{PM:MainEquations}, which at first sight appears to be a challenging task. The strategy consists in transforming these equations, as we explain in the next subsection.

\subsection{\label{sec:model}The model as a system of ODEs}

The system of PDEs \eqref{PM:MainEquations} can be transformed into a system of ODEs involving the functions $A(T)$, $B(T)$ and $V_k(T)$ with $k\in \mathbb{N}=\{1, 2, 3,\cdots\}$,
\begin{subequations}\label{PM:model}
\begin{align}
    &\dv{A}{T}=B-\gamma A\, ,  \label{PM:model1} \\
    &\dv{B}{T}=-\frac{\gamma}{2}B+ \frac{\gamma^2}{2}A+A-A\sum_{k=1}^\infty f(k)\left( A^2+V_k\right) , \label{PM:model2}\\
    & \dv{V_k}{T}= \gamma\left(g(k) A^2-h(k) V_k\right) ,\label{PM:model3} 
\end{align}
\end{subequations}
where the functions $f(k)$, $g(k)$, and $h(k)$ are defined by
\begin{subequations}\label{PM:function}
\begin{align}
    & f(k) = \frac{2 m^2}{\pi^2}\frac{ h(k)}{\left[(k-1/2)^2-m^2\right]^2}\, , \label{PM:functionf}\\
    & g(k) = h(k) + \frac{a^2/2\pi^2}{(k-1/2)^2+a^2/4\pi^2}\, ,\label{PM:functiong}\\
    & h(k) = \frac{(k-1/2)^2}{(k-1/2)^2+a^2/4\pi^2}\, . \label{PM:functionh}
\end{align}
\end{subequations}
Appendix~\ref{appendixA} provides a step-by-step derivation of the transformation from the PDE system \eqref{PM:MainEquations} to the ODE system \eqref{PM:model} above. In particular, it is shown that Eq.~\eqref{PM:MainEquations2} gives rise to an equation for $\ddot{A}$ (or equivalently, for $\dot{A}$ and $\dot{B}$), while Eq.~\eqref{PM:MainEquations1} yields an evolution equation for the modes $V_k$. The initial conditions have the form
\begin{subequations}\label{PM:ic}
\begin{align}
        & A(T=0)=A_0\, ,\\
        & B(T=0)= \gamma A_0+\dv{A}{u}\Big|_{u=0}\, ,\\
        & V_k(T=0)=-A_0^2\, ,
\end{align}
\end{subequations}
and correspond to a vanishing zonal-flow correction at $T=0$. Given values for $(A_0,B_0)$ and parameters $(\gamma, a,m)$, the system \eqref{PM:model} can be readily integrated. In the following, we always assume $\gamma\geq 0$.

We further note that the zonal-flow correction can be expressed in terms of the wave amplitude $A$ and the modes $V_k$ as 
\begin{align}\label{PM:Phi_expression}
\Phi(T,y)
&= \frac{1}{2\pi^3}
\sum_{k=1}^{\infty}
\frac{m }{\left[(k-1/2)^2 - m^2\right]
\left[(k-1/2)^2 + a^2/4\pi^2\right]} \nonumber \\
&\quad \times \left(A^2(T) + V_k(T)\right)
\cos\!\left[(2k-1)\pi y\right] \, .
\end{align}
and is defined up to an arbitrary linear function of $y$, since $\Phi$ enters the PDE system \eqref{PM:MainEquations} only through the combination $\partial_y^2 \Phi$.

The model~\eqref{PM:model} depends on three parameters. As a result, the possible bifurcations undergone by the system can be localized in a three-dimensional parameter space. We choose to emphasize the dependence $\gamma$ due to its predominant physical importance, while $a$ and $m$ are absorbed into the definitions of the functions \eqref{PM:functionf}-\eqref{PM:functionh}. In particular, we will demonstrate that varying $\gamma$ leads to drastically different asymptotic dynamics.

For conciseness, the model~\eqref{PM:model} can be recast in the compact form $\dot{\bm{X}}= \bm{F}(\bm{X})$ where $\bm{X}\equiv \left(A, B, V_1, V_2,\cdots \right)$ and $\bm{F}$ is a vector-valued function defined componentwise by $F_1(\bm{X})=B-\gamma A$, $F_2(\bm{X})=-\gamma/2B+ \gamma^2 A /2+A-A\sum_{k=1}^\infty f(k)\left( A^2+V_k\right)$ and $F_i(\bm{X})=\gamma\left(g(i-2) A^2-h(i-2) V_{i-2}\right)$ for $i=3,4,\cdots$.

The model~\eqref{PM:MainEquations}, expressed in the form of the ODE system \eqref{PM:model}, is central to the analysis carried out in this work. In particular, it will be used to investigate the dynamics of the wave amplitude and to extract relevant physical features of Pedlosky's model. However, numerical integration of Eqs.~\eqref{PM:model} necessarily requires truncation to a finite number of modes $V_k$. A natural question therefore arises: how many of these modes are required to faithfully reproduce the dynamics of the PDE system?

\subsection{\label{sec:truncation}Effect of the truncation}

The transformation from the PDE system \eqref{PM:MainEquations} into the ODE system \eqref{PM:model} is an exact mathematical mapping. There is, however, a price to pay: the dynamics of the wave amplitude is now described by an infinite number of modes $V_k$. In practice, the numerical integration of the system, and the study of its solutions, require the introduction of a cut-off $k=k_c$ such that for a given $k_c$, the infinite-dimensional system \eqref{PM:model} becomes $(k_c+2)$-dimensional.

We observe that as $k\rightarrow \infty$, the function $f(k)\sim k^{-4}$, while $g(k)\sim 1+k^{-2}$ and $h(k) \sim 1$. Consequently, as the index $k$ increases in the sum in Eq. \eqref{PM:model2}, fewer and fewer $V_k$ terms contribute significantly to the time evolution of $B$, and the dynamics of $A$ and $B$ nearly decouple from the higher-order modes. Aside from its practical relevance, this observation also provides a theoretical justification for introducing a cut-off value $k_c$, above which the modes $V_k$ are neglected. This approximation yields a valid system of ODEs, since Eq.~\eqref{PM:model3} does not couple modes $V_k$ with distinct values of $k$, ensuring that the truncated system remains self-contained.

Choosing an appropriate value of $k_c$ requires balancing two competing considerations: a large value, which captures more accurately the dynamics of the PDE system, and a small value, which allows for efficient numerical integration of the truncated system. Here, we ask how rapidly solutions of the truncated model \eqref{PM:model} converge, as $k_c$ increases, to known solutions of the PDE system \eqref{PM:MainEquations}, corresponding to the limit $k_c\rightarrow \infty$. Understanding the rate of this convergence will guide our choice of the cut-off value.

\begin{figure}[b!]
\includegraphics[scale=0.28]{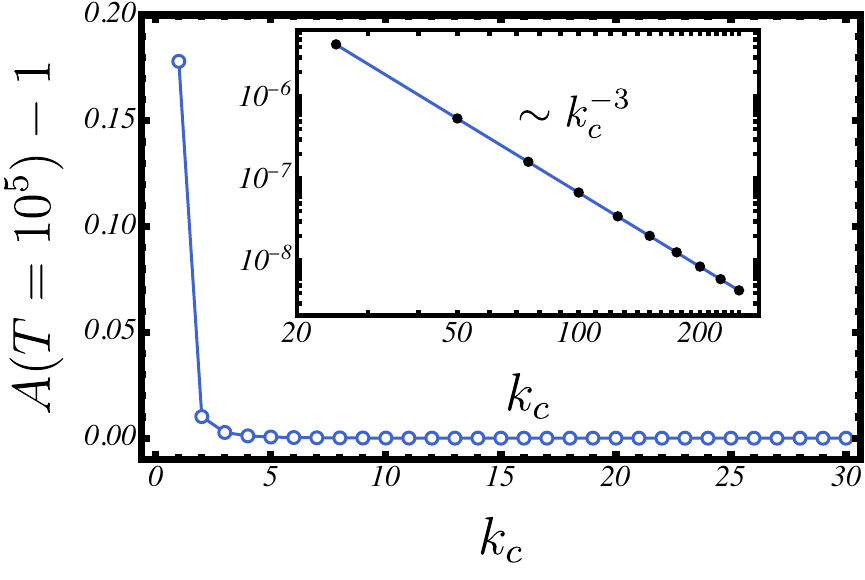}\quad
\includegraphics[scale=0.28]{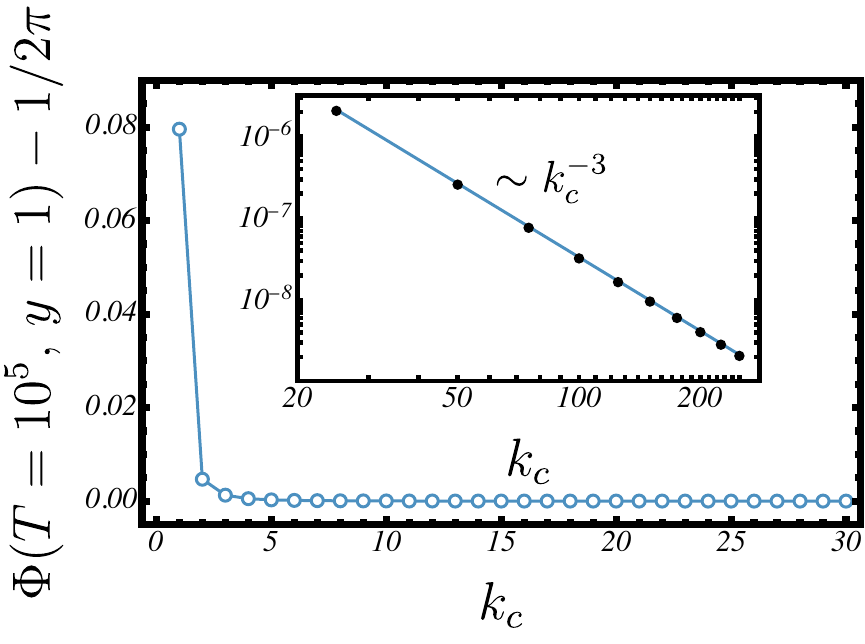}\vspace{1em}
\includegraphics[scale=0.28]{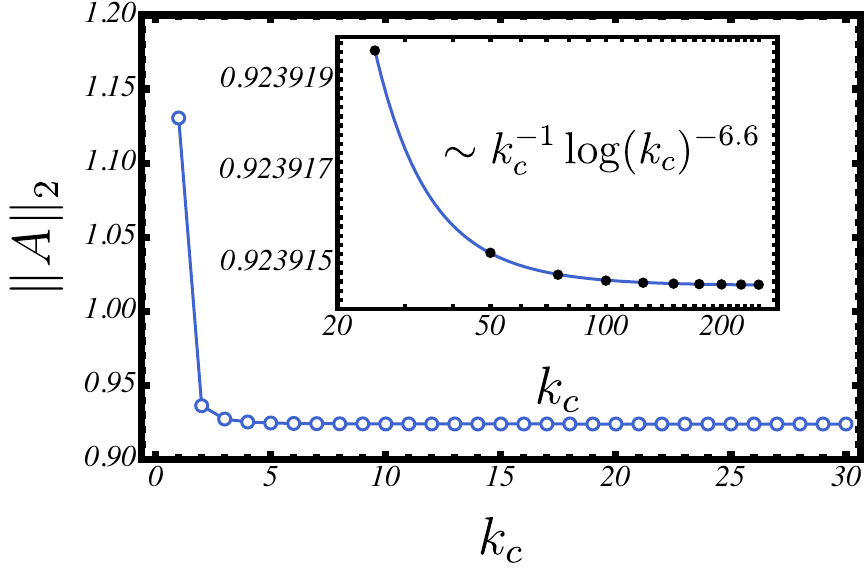}\quad
\includegraphics[scale=0.28]{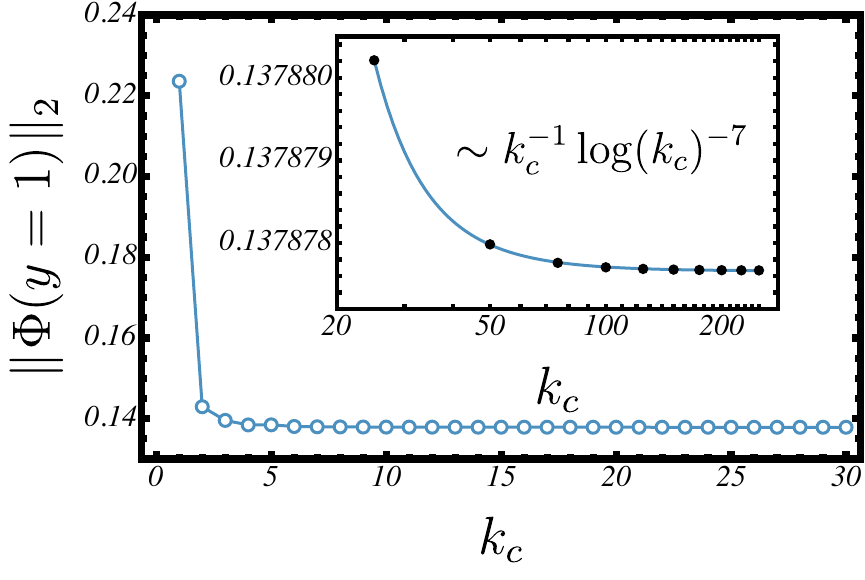}
\caption{\textbf{(Upper row)} Convergence of $A(T=10^5)$ and $\Phi(T=10^5, y=1)$, obtained by numerical integration of the truncated model \eqref{PM:model} (with $\gamma=1$), toward the respective fixed points of the PDE system, as a function of the cut-off $k_c$. A fit of the data in the inset indicates that the rate of convergence scales as $k_c^{-3}$. \textbf{(Lower row)} Convergence of the $L^2$-norm of periodic orbits, obtained by numerical integration of the truncated model \eqref{PM:model} (with $\gamma=0.05$), and by isolation of the stable periodic orbit. A fit of the inset data suggests a scaling of the form $k_c^{-1} \log(k_c)^{-\alpha}$. In both cases, the parameters used are $a=\pi \sqrt{2}$ and $m=1$, and the initial conditions are $(A_0,B_0)=(1,0)$. General conclusions cannot be drawn from the fitting functions alone, as these likely depend on the model parameters and the choice of initial conditions.}
\label{fig:solutions_kc} 
\end{figure}

The simplest steady-state solutions $A(T)$ and $\Phi(T,y)$ of the PDE system are its fixed points, whose expressions will be derived later in Sec.~\ref{sec:stability} (see Eqs.~\eqref{PM:formula_pi} and \eqref{formula_phi} below). To investigate convergence toward these fixed points, we select a value of $\gamma$ for which the solutions of the truncated model are known to converge to fixed points for all initial conditions, for example $\gamma=1$ (as will be justified in later sections). The truncated model is then numerically integrated, and the large-time values of $A(T)$ and $\Phi(T,y)$ are obtained for various values of the cut-off $k_c$. The results are shown in the upper row of Fig.~\ref{fig:solutions_kc}. The main observation is that both $A(T)$ and $\Phi(T,y)$ converge to the fixed points of the PDE system at a rate proportional to $k_c^{-3}$, at least for the parameter values considered here.

On the other hand, we also verify the convergence of nontrivial solutions of the truncated model to the corresponding solutions of the PDE system. In particular, it is known (see later sections of this work) that for small values of $\gamma$ (for example $\gamma=0.05$), a stable periodic orbit exists and is reached from suitably chosen initial conditions, which persists as the cut-off $k_c$ increases. Under the assumption that this stable orbit is also a solution of the PDE system, we investigate the rate of convergence of the $L^2$-norm with respect to $k_c$. We recall that, given a periodic time series $X(t)\in\mathbb{R}$ of period $T$, the $L^2$-norm $||X||_2$ is defined as $||X||_2 = \sqrt{\int_0^T X^2(t) \, \dd t/T}$. The convergence of the $L^2$-norm is shown in the lower row of Fig.~\ref{fig:solutions_kc}, where the norm is computed for each stable periodic orbit of the truncated model with cut-off $k_c$. In this case, the rate of convergence differs from that observed for fixed points, and follows the empirical form $k_c^{-1} \log(k_c)^{-\alpha}$, with $\alpha\simeq 7$ for the selected parameter values, indicating a slower rate of convergence. In particular, our numerical simulation indicates that the stable periodic orbit persists as a solution of the PDE system.

While the previous discussion is somewhat restrictive, as it only considers two different types of solutions of the PDE model, it is nonetheless useful to guide a choice of an appropriate value of $k_c$. For subsequent analysis of the model \eqref{PM:model}, see Sec.~\ref{sec:full_model}, we take $k_c=20$, which corresponds roughly to an error of around $10^{-5}$ for both the fixed point and the $L^2$-norm. We believe that the dynamics observed at this truncation value should adequately capture the dynamics of the PDE model. Further comments will be addressed in the Sec.~\ref{sec:Higher_dimension}.

\subsection{\label{sec:symmetry}Symmetries of the model}

A careful inspection of the model \eqref{PM:model} shows that it is invariant under the discrete symmetry mapping $(A,B,V_k)$ to $(-A,-B,V_k)$. This implies that the flow satisfies the relations
\begin{subequations}\label{PM:symmetry_1}
\begin{align}
    & X_1\left(T,A, B, V_1, V_2, \cdots\right)=-X_1\left(T,-A, -B, V_1, V_2, \cdots\right) ,\\
    & X_2\left(T,A, B, V_1, V_2, \cdots\right)=-X_2\left(T,-A, -B, V_1, V_2, \cdots\right) ,\\
    & X_i\left(T,A, B, V_1, V_2, \cdots\right)=X_i\left(T,-A, -B, V_1, V_2, \cdots\right) ,
\end{align}
\end{subequations}
where $X_i$ denotes the $i$th component of the flow ($i=3, 4, \cdots$ in the last equation above). In the $(A,B)$ plane, this symmetry implies that the image under central inversion about the origin of any solution with initial conditions $(A_0,B_0)$ is also a solution with initial conditions $(-A_0, -B_0)$. This symmetry has interesting consequences for the dynamics of the model. In particular, it is reflected in the basins of attraction as well as in the structural aspect of its bifurcation diagram (see Secs.~\ref{sec:toy_model} and \ref{sec:full_model} below).

Furthermore, any solution always rotates clockwise in the $(A,B)$ plane. Indeed, the evolution equation for $A$ is $\dot{A}=B-\gamma A$. If, at some time $T$, $A(T)=0$, then $\dot{A}(T)$ has the sign of $B(T)$. Similarly, if $B(T)=0$, then $\dot{A}(T)$ has the opposite sign of $A(T)$.

\subsection{Case where \texorpdfstring{$\gamma = 0$}{gamma = 0}}\label{sec:gamma_zero}

We now turn to the study of the inviscid equilibration of the wave amplitude $A(T)$, corresponding to the case $\gamma=0$ (recall that $\gamma$ is proportional to the square root of the kinematic viscosity). In this limit, the dynamics of the system \eqref{PM:model} reduces to that of a Hamiltonian system. Our results show that, in this regime, the system exhibits long-lasting, smooth, and periodic oscillations, as previously reported in Refs.~\onlinecite{Pedlosky1, Pedlosky2}. 

Taking $\gamma=0$, the system \eqref{PM:model} reduces to
\begin{subequations}
\begin{align}
    &\dv{A}{T}=B,\\
    &\dv{B}{T}=A-A \sum_{k=1}^\infty f(k)\left( A^2+V_k\right),\\
    &\dv{V_k}{T}=0\,.
\end{align}
\end{subequations}
Using the initial conditions in Eqs.~\eqref{PM:ic}, we have $V_k(T)=-A_0^2$ for all $k$. Denoting by $s(k_c,a,m)\equiv \sum_{k=1}^{k_c} f(k)$, the above equations reduce to a single second-order ODE for the wave amplitude $A(T)$,
\begin{equation}\label{equation_gamma_0}
    \dv[2]{A}{T}-(1+sA_0^2)A+s A^3=0\, ,
\end{equation}
which is a special case of an unforced, undamped Duffing oscillator. In this limit, the system is Hamiltonian, with Hamiltonian function
\begin{equation}\label{Hamiltonian}
    H(A,B)=\frac{1}{2}B^2-\frac{1}{2}(1+sA_0^2) A^2+\frac{1}{4} s A^4\, .
\end{equation}
The problem is equivalent to that of a mechanical point moving in a double-well potential centered at the origin, $V(A)=-(1+sA_0^2) A^2/2+ s A^4/4$, which features two stable equilibrium points (global minima) at $A_\pm=\pm \sqrt{(1+s A_0^2)/s}$ with $V(A_+)=V(A_-)<0$ and one unstable equilibrium (local maximum) at $A_0=0$, where $V(A_0)=0$. Consequently, under Hamiltonian dynamics, the motion of the wave amplitude $A(T)$ remains bounded for all energies.

It can be explicitly verified that Hamilton's equations reproduce Eq.~\eqref{equation_gamma_0}. Since the system is one-dimensional, it is inherently integrable. For a given initial condition $(A_0, B_0)$ such that $H(A_0,B_0)=E$, the dynamics takes place along the level set defined by $H(A,B)=E$, which, to our knowledge, does not belong to any standard family of quartic plane curves. Each point on this curve is mapped by the four-fold symmetry $(A,B)\rightarrow (\pm A, \pm B)$ (for energies $E\leq 0$, however, the choice of $A_0$ confines the trajectory to single well and thus breaks this symmetry).

\begin{figure}[b!]
\includegraphics[scale=0.28]{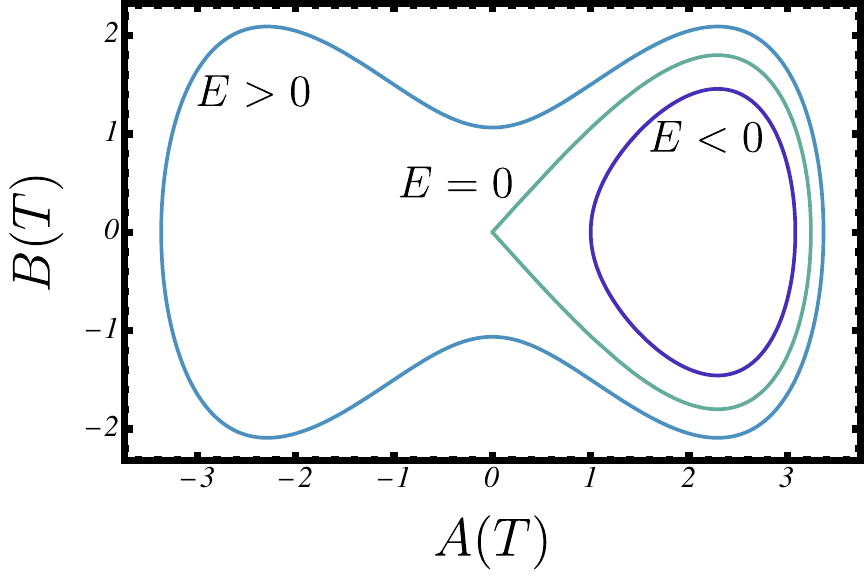}\quad
\includegraphics[scale=0.28]{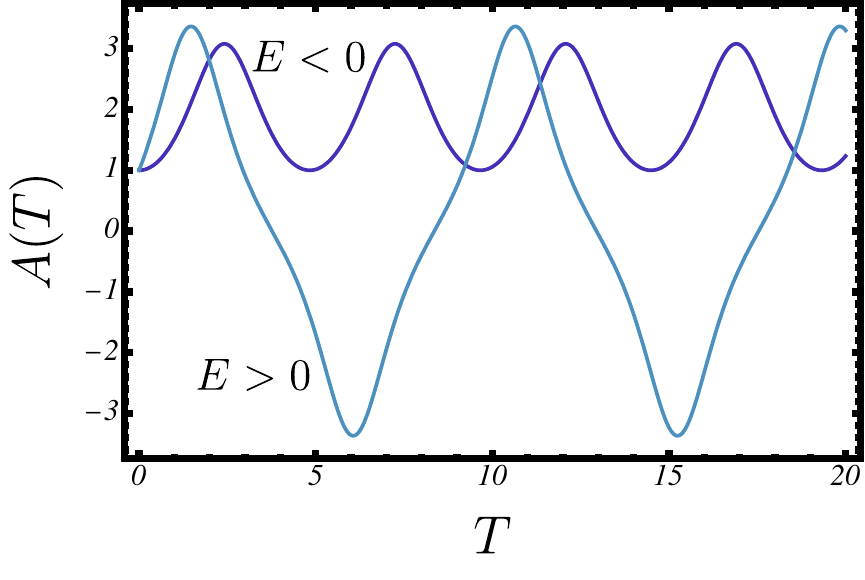}
\caption{Trajectories of the Hamiltonian system~\eqref{Hamiltonian} are shown for various initial conditions on the energy shells $E<0$ and $E>0$. The left panel displays the trajectories in the phase space $(A,B)$, together with the separatrix curve corresponding to $E=0$. We note that the trajectories are consistent with the intuition of a mechanical particle moving in the potential $V(A)$. The right panel shows the corresponding periodic evolution of the wave amplitude $A(T)$. The parameters used are $a=\pi \sqrt{2}$, $m=1$ and $k_c=10$. The initial conditions are $(A_0,B_0)=(1,0)$ yielding $E\simeq -0.559$, and $(A_0,B_0)=(1,3/2)$ yielding $E\simeq 0.566$.}
\label{fig:Integrable_solution} 
\end{figure}

We show in Appendix~\ref{appendixGamma0} that the solution for $A(T)$ depends on the sign of the energy. For $E<0$, the solution is given by Eq.~\eqref{exact_A_1} where $F$ denotes the incomplete elliptic integral of the first kind and $\text{sn}$ the Jacobi elliptic sine function. For $E>0$, the solution is instead given by Eq.~\eqref{exact_A_2}. In both expressions, the functions $\varphi$, $\phi$ and the parameters $u_\pm$ are defined by
\begin{subequations}
\begin{align}
&\varphi(x)=\arcsin\left(\sqrt{\left(\frac{u_+}{u_+-u_-} \right)\left(1-\frac{x^2}{u_+} \right)}\right) ,\\
&\phi(x)=\arcsin{x}\, ,\\
&u_{\pm}=\frac{1}{s}\left(1+s A_0^2\pm \sqrt{(1+s A_0^2)^2+4 s E}\right) . 
\end{align}
\end{subequations}
Typical trajectories with negative and positive energies are depicted in Fig.~\ref{fig:Integrable_solution}. When $E<0$, the sign of the wave amplitude $A(T)$ is fixed by the initial condition $A_0$. The amplitude of the oscillations is $\sqrt{u_+}-\sqrt{u_-}$. On the other hand, when $E>0$, $A(T)$ oscillates between negative and positive values with an amplitude of $2\sqrt{u_+}$. In both cases, the maximum and minimum amplitude reached by the wave depends on the initial condition $A_0$, a feature that was previously highlighted in Refs.~\onlinecite{Pedlosky1,Pedlosky2}.

Finally, the separatrix corresponding to $E=0$ is given by Eq.~\eqref{exact_A_3}. In the limit $T\rightarrow \pm \infty$, $A(T)\rightarrow 0$, confirming that the separatrix curve corresponds to a homoclinic orbit in the phase space $(A,B)$. As we shall see, this is the first of many homoclinic orbits that emerge at nonzero values of $\gamma$, once the system is no longer Hamiltonian. 

\begin{widetext}
    \begin{align}
    & A(T)=\text{sgn}(A_0)\sqrt{u_+-(u_+-u_-)\,\text{sn}^2\Big[F\Big[\varphi\left(\frac{A_0}{\sqrt{u_+}}\right)\,\Big|\,1-\frac{u_-}{u_+}\Big]-\text{sgn}(A_0 B_0)\sqrt{\frac{s u_+}{2}}\,T\, \Big| \, 1-\frac{u_-}{u_+}\Big]}\, ,\quad E<0\, , \label{exact_A_1}\\
        &
    A(T)=\sqrt{u_+}\,\text{sn}\Big[F\Big[\phi \left(\frac{A_0}{\sqrt{u_+}}\right)\,\Big |\,\frac{u_+}{u_-}\Big]+\text{sgn}(B_0)\sqrt{-\frac{s u_-}{2}}\,T\, \Big | \, \frac{u_+}{u_-}\Big]\, , \quad E>0\, , \label{exact_A_2}\\
    & A(T)=\sqrt{2 \left(A_0^2+\frac{1}{s}\right)}\sqrt{1-\tanh^2\left(\text{arctanh}\,\left(\sqrt{1-\frac{s A_0^2 }{2( s A_0^2+1)}} \right)-\text{sgn}(B_0)\sqrt{1+s A_0^2}T \right)},\quad E=0\,.\label{exact_A_3}
    \end{align}
\end{widetext}

\section{\label{sec:stability}Linear stability analysis of the model}

Once $\gamma>0$, the Hamiltonian structure and its associated integrability are broken. Nevertheless, it is in this regime that the most interesting nonlinear dynamics arises. Our first step in studying the model \eqref{PM:model} is to perform a linear stability analysis of its fixed points, both in the limit $k_c \to \infty$ and for finite values of $k_c$.

Consequently, besides the trivial solution corresponding to $A=0$ and $\Phi=c_1 y+c_2$ ($c_1,c_2 \in \mathbb{R}$), other steady-state solutions of the model are of the form
\begin{subequations}
\begin{align}
    & A= \pm 1\, , \label{PM:PDE_fixedpoint_1} \\
    & \partial_y^2 \Phi=2 \sin(2 m \pi y) \, . \label{PM:PDE_fixedpoint_2}
\end{align}
\end{subequations}

In practice, these solutions should correspond to the fixed points of the ODE system as written in Eq.~\eqref{PM:model}, and this can be explicitly verified. Setting $\dd A / \dd T=0$, together with $\dd V_k / \dd T=0$, yields $B=\gamma A$ and $V_k=g(k) A^2/h(k)$. Substituting these constraints into $\dd B / \dd T=0$ gives the formula
\begin{align}\label{PM:formula_pi}
    &A^2 \sum_{k=1}^\infty f(k)\left(1+ \frac{g(k)}{h(k)}\right)=1 \nonumber \\ 
    &\quad\iff A^2 = \frac{\pi  m}{\pi  m \sec ^2(\pi  m)-\tan (\pi  m)} = 1, \hspace{.25cm} m \in \mathbb{Z}\, ,
\end{align}
(see Appendix~\ref{appendixB} for a proof) in agreement with the above result \eqref{PM:PDE_fixedpoint_1}. To recover the steady-state solution for the zonal-flow correction $\Phi$, the constraints for the modes $V_k$ are substituted into the second partial derivative of expression \eqref{PM:Phi_expression} with respect to $y$, yielding
\begin{align}\label{PM:Phi_steady}
    \partial_y^2 \Phi &= - 2 \sum_{k=1}^{\infty} \frac{8 m}{\left[(2 k-1)^2-4 m^2\right]\pi} \cos \left[(2k-1)\pi y\right]\nonumber \\
    &=2 \sin 2 m \pi y\, ,
\end{align}
(see formula \eqref{appA:FourierSeries}) which is in agreement with Eq.~\eqref{PM:PDE_fixedpoint_2}. In addition, and using Eq.~\eqref{PM:Phi_expression}, $\Phi$ can be evaluated at the fixed points using the same constraints on $V_k$ and the fact that $A^2=1$. One has
\begin{align}\label{formula_phi}
    \Phi(y)&=\frac{1}{\pi^3} \sum_{k=1}^{\infty}\frac{m}{(k-1/2)^2\left[(k-1/2)^2-m^2\right]} \cos \left[(2k-1)\pi y\right]\nonumber \\
    &=-\frac{\pi  m (1-2 y)+\sin 2 m \pi  y}{2 \pi ^2 m^2}\, ,
\end{align}
(see Appendix~\ref{appendixB} for a proof) whose second derivative with respect to $y$ once again agrees with Eq.~\eqref{PM:PDE_fixedpoint_2}.

In summary, the system \eqref{PM:model} admits the following fixed points,
\begin{subequations}\label{PM:EquilibriumPoints}
\begin{align}
    & \bm{X}_0 = \bm{0}\, ,  \\
    & \bm{X}_\pm=\Big(\pm 1,\, \pm \gamma,\, g(1)/h(1),\, g(2)/h(2),\, \cdots \Big)\, ,
\end{align}
\end{subequations}
and preserves the fixed points of the original PDE model~\eqref{PM:MainEquations}. Importantly, this latter feature no longer holds true in the truncated version of the system when the cut-off $k_c$ is finite. In that case, while the formulas for the fixed points of $B$ and $V_k$ remain unchanged, the equilibrium value for $A$ becomes a function of $k_c$ (omitted here for brevity, just as we do not compute the corresponding value of $\Phi$), see the top-left panel of Fig.~\ref{fig:solutions_kc}.

The linear stability of the fixed points \eqref{PM:EquilibriumPoints} can be analyzed through the eigenvalues of the Jacobian matrix $\text{D}F(\bm{X}) \equiv \partial \bm{F}/\partial \bm{Y} \big|_{\bm{Y} = \bm{X}}$, which has the form given in Eq.~\eqref{PM:JacobianMatrix} and where $\bm{D}$ is a diagonal matrix with entries $(\bm{D})_{ij}=-\gamma h(i)\delta_{ij}$. The trace of the Jacobian is coordinate-independent and given by
\begin{equation}
    \Tr \text{D}F(\bm{X})=-\gamma \left(\frac{3}{2}+\sum_{k=1}^\infty h(k) \right) .
\end{equation}
Since $h(k)>0$ for all $k$, the trace is always negative, and its absolute value is an increasing function of $k_c$ (the infinite sum over $h(k)$ diverges as $k_c\rightarrow \infty$). By Liouville's theorem, this implies exponential contraction of phase-space volumes \cite{Arnold, Strogatz}. That is, volumes in phase space contract under the flow, and any solution of the model \eqref{PM:model} will eventually converge to a zero-volume attractor in the long-time limit (fixed point, periodic orbit, strange attractor, homoclinic/heteroclinic orbits, \textit{etc.}).

The Jacobian matrix \eqref{PM:JacobianMatrix} can be evaluated at the fixed points \eqref{PM:EquilibriumPoints}. At the origin, it reduces to a block-diagonal matrix and its spectrum is the union of the spectra of the individual blocks,
\begin{align}\label{spectrum_D0}
    \text{spec}\left(\text{D}F(\bm{X}_0)\right)=&\bigg\{-\frac{3}{4}\gamma+\sqrt{1+\frac{9}{16}\gamma^2},\, -\frac{3}{4}\gamma-\sqrt{1+\frac{9}{16}\gamma^2},\nonumber \\
    &\quad-\gamma h(1),\, -\gamma h(2),\, \cdots \bigg\} ,
\end{align}
and is real. The stable directions associated with the eigenvalues $-\gamma h(k)$ are aligned with the $V_k$ axes. The first eigenvalue is always positive, and the second is always negative, so the only unstable direction lies in the $(A,B)$ plane. As a consequence, $\bm{X}_0$ is always a saddle point, regardless of the value of $\gamma$.

On the other hand, the Jacobian matrix evaluated at the two nontrivial fixed points $\bm{X}_\pm$ is no longer block-diagonal. We note that $\text{D}F(\bm{X}_+)$ and $\text{D}F(\bm{X}_-)$ are similar since $\text{D}F(\bm{X}_-)=P \, \text{D}F(\bm{X}_+)\, P^{-1}$ where $P=\text{diag}(1, 1, -1, \cdots, -1)$. Consequently, they share the same spectrum. We have verified numerically (up to $\gamma\leq 5$) that there exists a critical value $\widetilde{\gamma}$ such that, for $\gamma>\widetilde{\gamma}$, all eigenvalues are real and negative, implying that the fixed points $\bm{X}_\pm$ are stable nodes.  When $\gamma<\widetilde{\gamma}$, the spectrum contains a single pair of complex conjugate eigenvalues with a negative real part, while all other eigenvalues remain real and negative. The transition between these two regimes does not change the stability of the fixed points $\bm{X}_\pm$, but it does affect the local structure of the flow: when $\gamma<\widetilde{\gamma}$, trajectories spiral toward $\bm{X}_\pm$ in the $(A,B)$ plane. The value of $\widetilde{\gamma}$ can be determined numerically as a function of $k_c$ for fixed values of the parameters $a$ and $m$. In the case where $a=\pi \sqrt{2}$ and $m=1$, $\widetilde{\gamma}$ converges approximately to $ 3.838$ in the $k_c\rightarrow\infty$ limit.
\begin{widetext}
\begin{equation}\label{PM:JacobianMatrix}
\text{D}F(\bm{X})=\begin{pmatrix}
    -\gamma & 1 & 0 & 0 & \cdots & 0 & \cdots \\
    1 + \gamma^2/2-\sum_{k=1}^\infty f(k)\left( 3 A^2+V_k\right) & -\gamma/2 & -f(1)A & -f(2) A & \cdots & - f(l) A \delta_{kl} & \cdots \\
    2 \gamma g(1) A & 0 &  & & & &  \\
    2 \gamma g(2) A & 0 & & \ddots & & & \\
    \vdots & \vdots &  &  & \bm{D} &  &  \\
    2 \gamma g(l) A \delta_{kl} & 0 & & & & \ddots & \\
    \vdots & \vdots &  &  &  &  &  
\end{pmatrix} ,
\end{equation}
\end{widetext}

As mentioned earlier, for finite values of $k_c$, the equilibrium value of $A$ is not exactly $\pm 1$, and the Jacobian matrix in Eq.~\eqref{PM:JacobianMatrix} has size $(k_c+2)\times(k_c+2)$. For $k_c=20$, $a=\pi \sqrt{2}$, and $m=1$, the equilibrium value of $A$ is approximately given by $A=\pm 1.0000085$. This small deviation does not affect the stability analysis of the fixed points discussed above, in particular, the value of $\widetilde{\gamma}$ remains unchanged.

It is worth noting that for $k_c=1$, there exists a value of $\gamma$ below which the fixed points $\bm{X}_\pm$ become unstable. This particular cut-off of the model \eqref{PM:model} will be extensively discussed in the next section.

\section{\label{sec:toy_model}A toy model}

We now turn to the study of the wave amplitude dynamics in model \eqref{PM:model} as a function of the parameter $\gamma$, which is related to the viscosity in Pedlosky's two-layer model. For arbitrary values of the cut-off $k_c$, the model is difficult to analyze. It therefore appears natural to start with a maximally truncated version where $k_c=1$. This represents, in some sense, an extreme choice, but it will later prove to be quite fruitful for understanding the various dynamical regimes of the full model. We refer to this maximally truncated version hereafter as the \textit{toy model} and aim to gain insight from it using standard nonlinear techniques. We later demonstrate an intriguing correspondence between this toy model and the Lorenz model.

\subsection{Nonlinear analysis}

Taking $k_c = 1$ reduces the model \eqref{PM:model} to a three-dimensional system of ODEs. In what follows, we study this system through the lens of its basins of attraction, its maximal Lyapunov exponent, and ultimately attempt to build a comprehensive representation of the bifurcations encountered as $\gamma$ is varied -- a task that is considerably challenging, as we shall see. The underlying assumption is that this toy model captures the essential qualitative dynamics of the full system \eqref{PM:model} with arbitrarily large $k_c$, while being more amenable to analysis and numerical integration. The results of this section will be compared to those obtained for larger values of $k_c$ in the next section.

Explicitly, the toy model reads
\begin{subequations}\label{PM:toy_model}
\begin{align}
    &\dv{A}{T}=B-\gamma A\, , \\
    &\dv{B}{T}=-\frac{\gamma}{2}B+ \frac{\gamma^2}{2}A+A-f(1) A\left( A^2+V\right) , \\
    & \dv{V}{T}= \gamma\left(g(1) A^2-h(1) V\right) ,
\end{align}
\end{subequations}
where $V\equiv V_1$. Unless stated otherwise, the parameters are set to $a=\pi \sqrt{2}$ and $m=1$. The two nontrivial fixed points are given by $\bm{X}_\pm=(\pm A_*,\, \pm \gamma A_*,\, g(1) A_*^2/h(1))$ where $A_*^2=\pi^2 (1-4m^2)^2/(64 m^2)$. The linear stability analysis of the trivial fixed point $\bm{X}_0$ is identical to that of the full system, see Sec.~\ref{sec:stability}. By contrast, and in addition to the critical value $\widetilde{\gamma}$, there exists another critical value, denoted $\gamma_{\text{toy}}$, at which the stability of the fixed points $\bm{X}_\pm$ changes. The existence of $\gamma_{\text{toy}}$ is specific to the case $k_c=1$ and is not observed for $k_c>1$. 

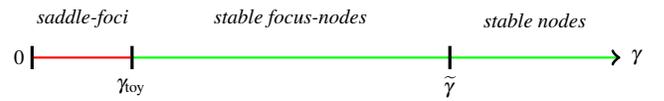
\begin{figure}[b!]
    \centering
\begin{tikzpicture}[scale=0.65, transform shape]
    \draw[very thick, |-] (-1, 0) -- (-1.02,0);
    \draw[thick, red, -] (-1, 0) -- (0.98,0);
    \draw[thick, green, -] (1.02, 0) -- (10.97,0);
    \draw[very thick, |-] (0.98, 0) -- (1.0,0);
    \draw[very thick, ->] (10.97, 0) -- (11,0);
    \draw[very thick, |-] (7.48, 0) -- (7.5,0);
    \node[] at (0.99,-0.6) {\large$\gamma_{\text{toy}}$};
    \node[] at (7.49,-0.6) {\large$\widetilde{\gamma}$};
    \node[right] at (11.1,0.0) {\large$\gamma$};
    \node[above] at (9.25,0.5) {\large\textit{stable nodes}};
    \node[above] at (4.25,0.5) {\large\textit{stable focus-nodes}};
    \node[above] at (0.0,0.5) {\large\textit{saddle-foci}};
    \node[] at (-1.3,0.0) {\large$0$};
\end{tikzpicture}
\caption{Stability of the two fixed points $\bm{X}_{\pm}$ of the toy model \eqref{PM:toy_model} as a function of the parameter $\gamma$. The transition at $\widetilde{\gamma}$ does not affect the stability of the fixed points, whereas the transition at $\gamma_{\text{toy}}$ does. The figure is not drawn to scale.}
\label{fig:toy_model_gamma} 
\end{figure}

For $\gamma>\widetilde{\gamma}\approx4.061$, the two nontrivial fixed points are stable nodes, since all eigenvalues of the Jacobian matrix of the system \eqref{PM:toy_model} evaluated at these points are real and negative. For $\gamma_{\text{toy}} \approx 0.246 < \gamma < \widetilde{\gamma}$, the fixed points become focus-nodes as the Jacobian matrix possesses a single real eigenvalue and a pair of complex-conjugate eigenvalues, all with negative real parts. Finally, for $\gamma < \gamma_{\text{toy}}$, the fixed points become saddle-foci, since the single real eigenvalue is negative and the real part of the complex-conjugate pair is positive. In Appendix~\ref{appendixD}, we provide exact expressions for $\gamma_{\text{toy}}$ and $\widetilde{\gamma}$. Figure.~\ref{fig:toy_model_gamma} summarizes the linear stability of the fixed points $\bm{X}_\pm$ as a function of $\gamma$.

\begin{figure*}[ht!]
\includegraphics[scale=0.265]{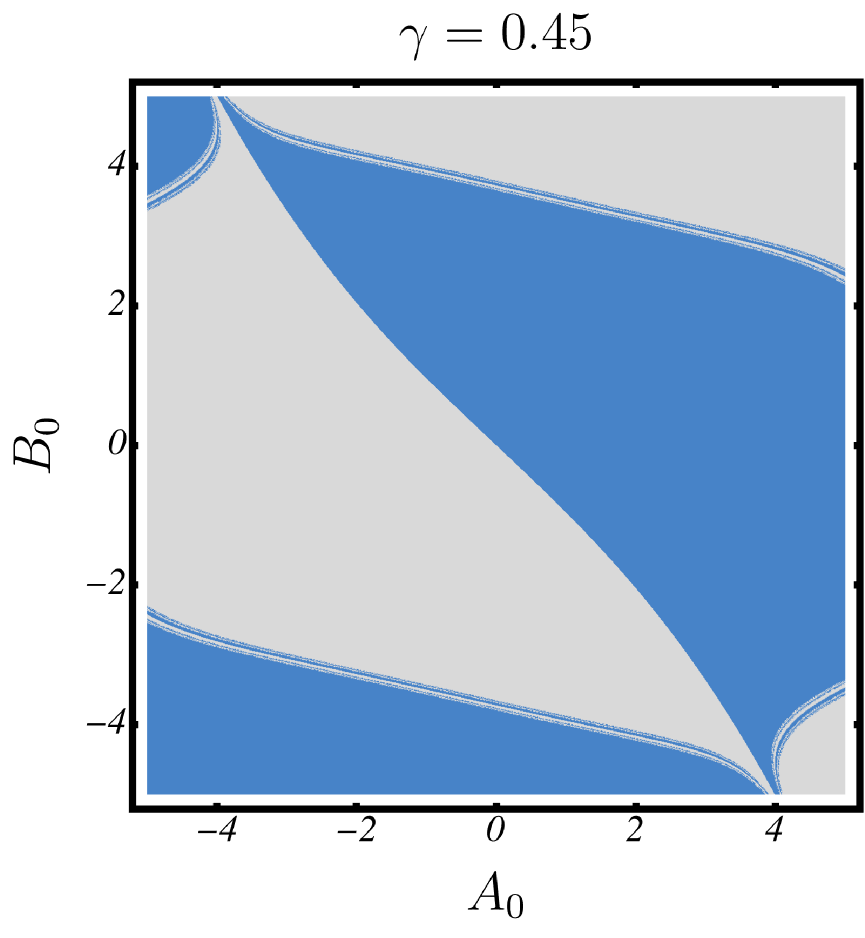}\qquad
\includegraphics[scale=0.265]{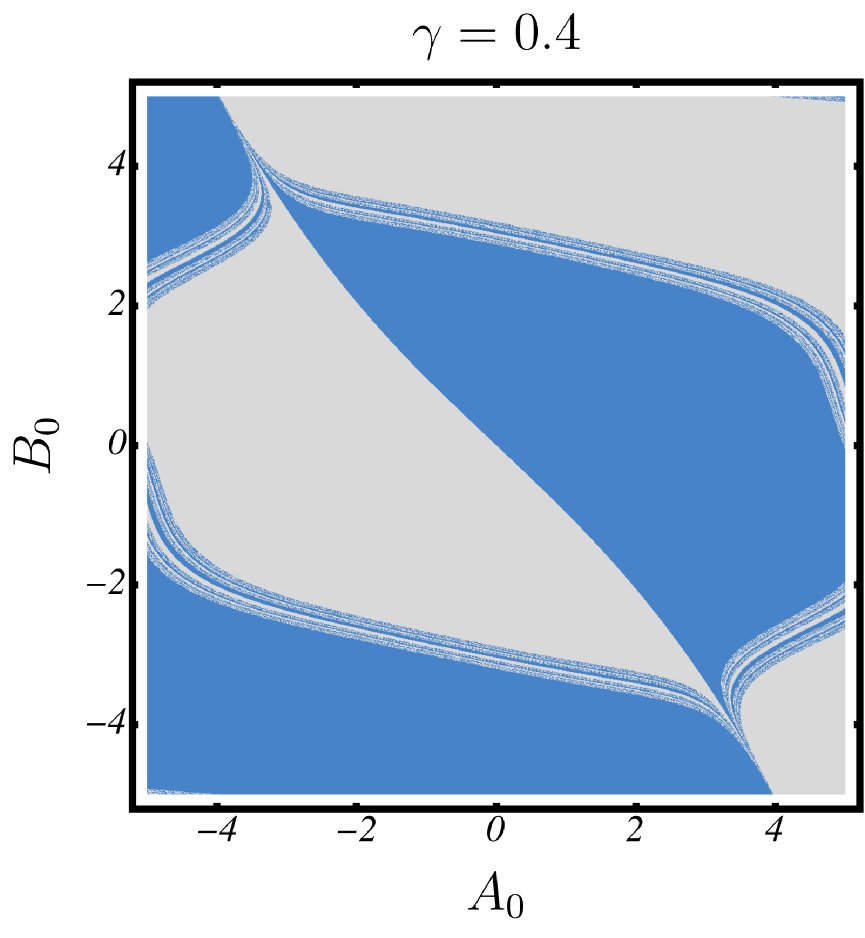}\qquad
\includegraphics[scale=0.265]{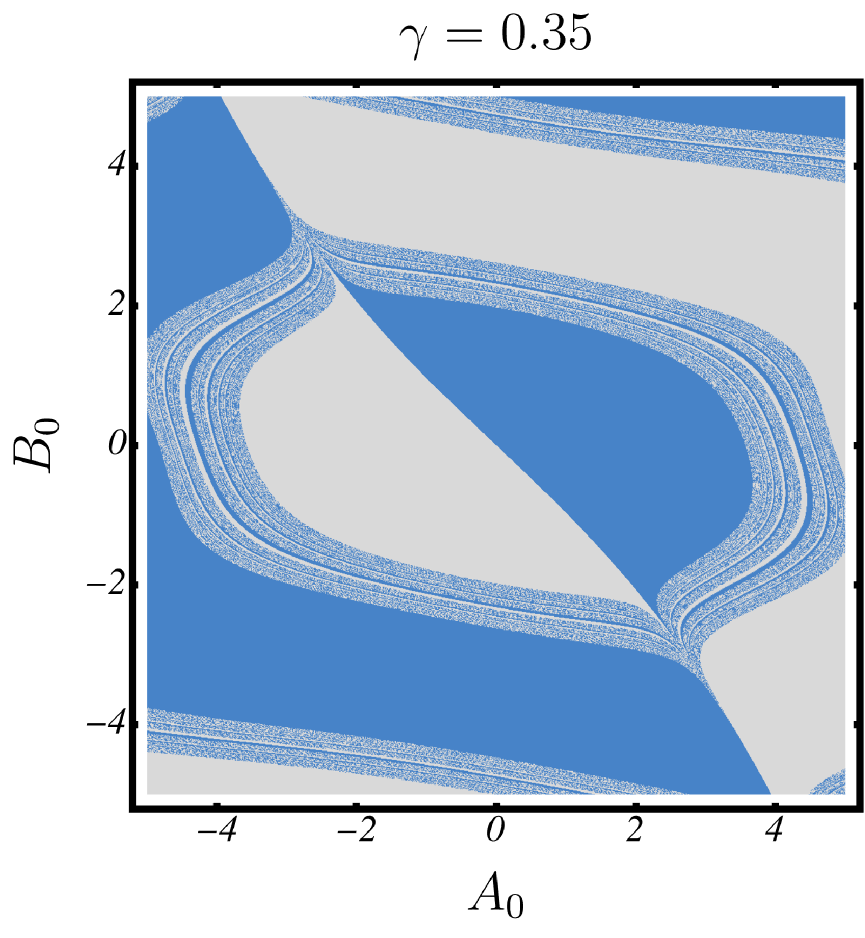}\\
\includegraphics[scale=0.265]{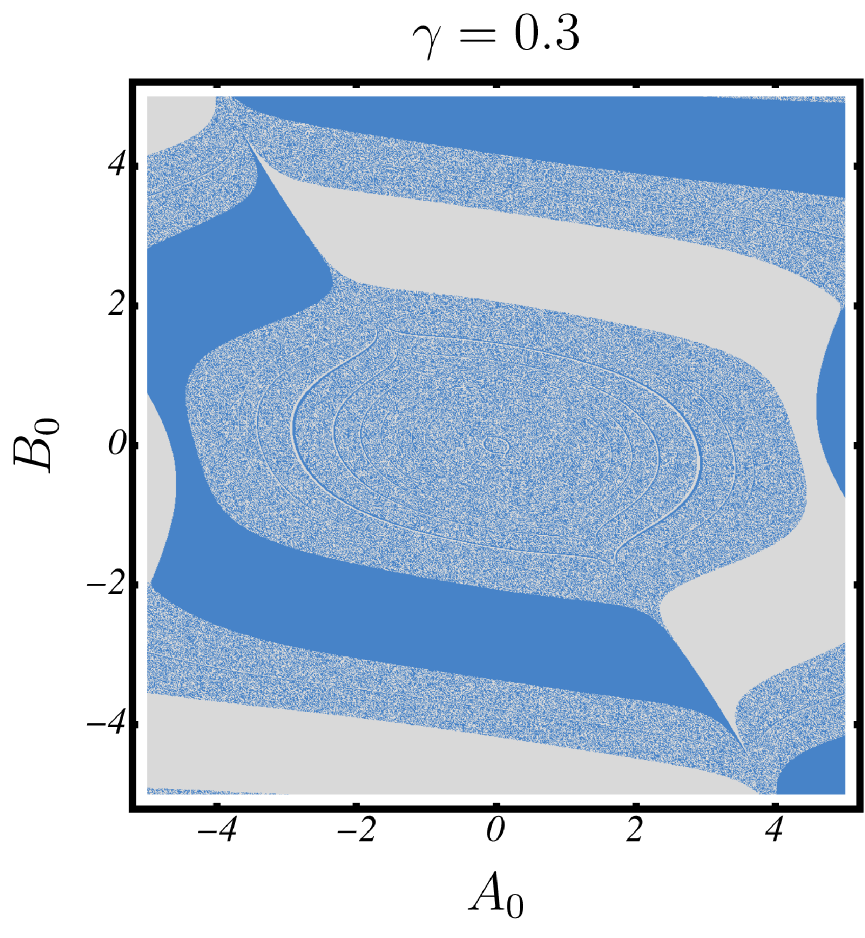}\qquad
\includegraphics[scale=0.265]{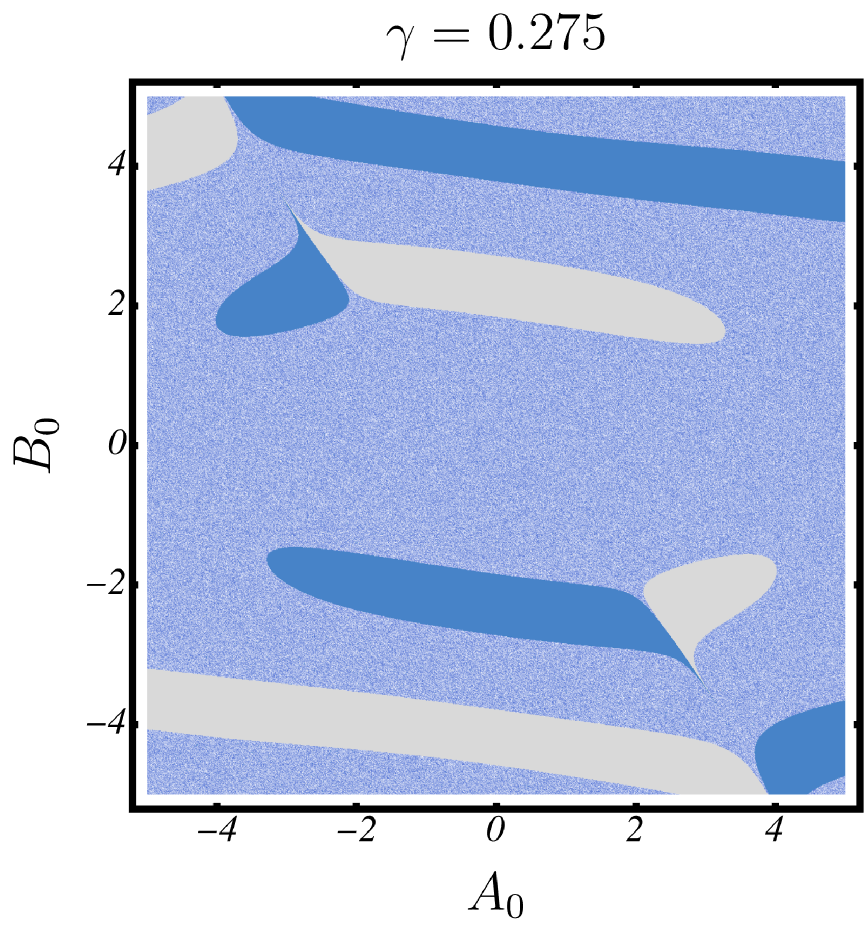}\qquad
\includegraphics[scale=0.265]{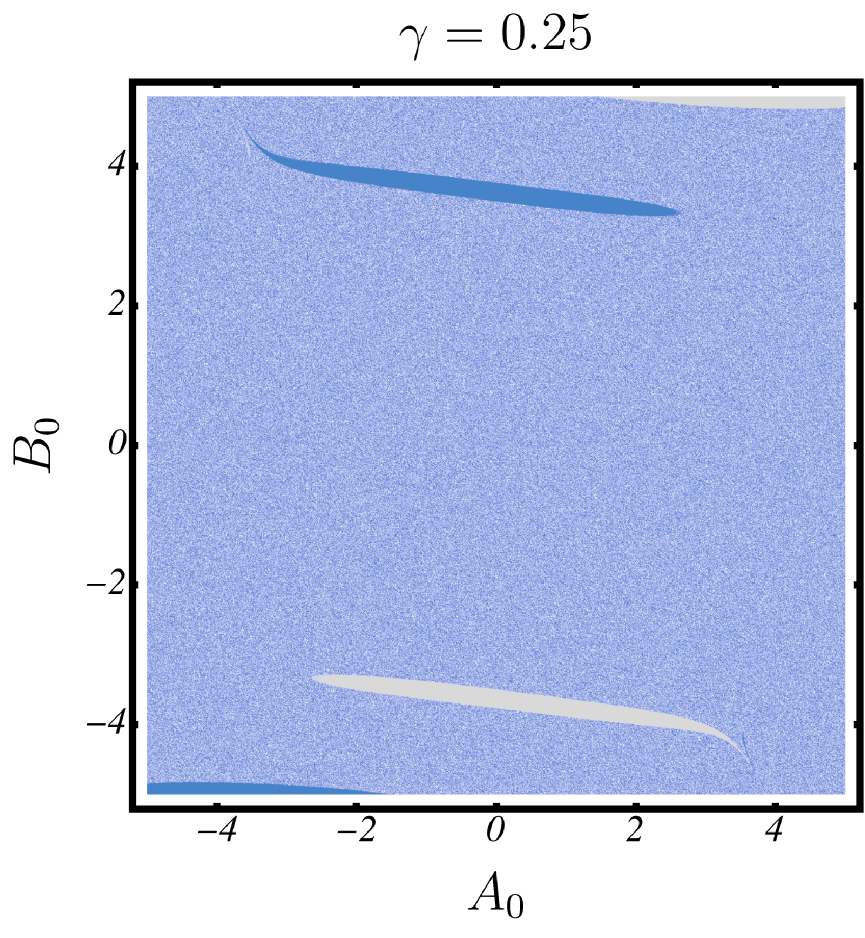}\\
\hspace{1.45cm}\includegraphics[scale=0.5]{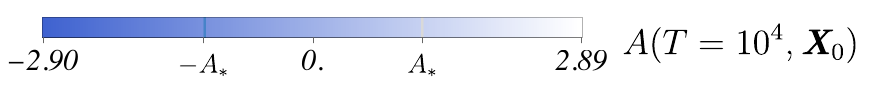}
\caption{Basins of attraction of the toy model \eqref{PM:toy_model} represented by the large-time value of the wave amplitude $A(T, \bm{X}_0)$ as a function of the initial conditions $A_0$ and $B_0$, with $A_0,B_0 \in [-5,\, 5]$. Each basin was generated by sampling $A_0$ and $B_0$ over this range with a step size of $10^{-2}$, resulting in approximately $10^6$ points per basin. The model was numerically integrated up to $T=10^4$. The basins clearly reveal which initial conditions and values of $\gamma$ lead to equilibration of the wave amplitude. However, they do not provide information about the nature of the dynamics when equilibration fails. The colorbar is intentionally non-continuous to emphasize equilibration toward one of the two fixed points.}
\label{fig:basins_attraction} 
\end{figure*}

\begin{figure*}[ht!]
\includegraphics[scale=0.375]{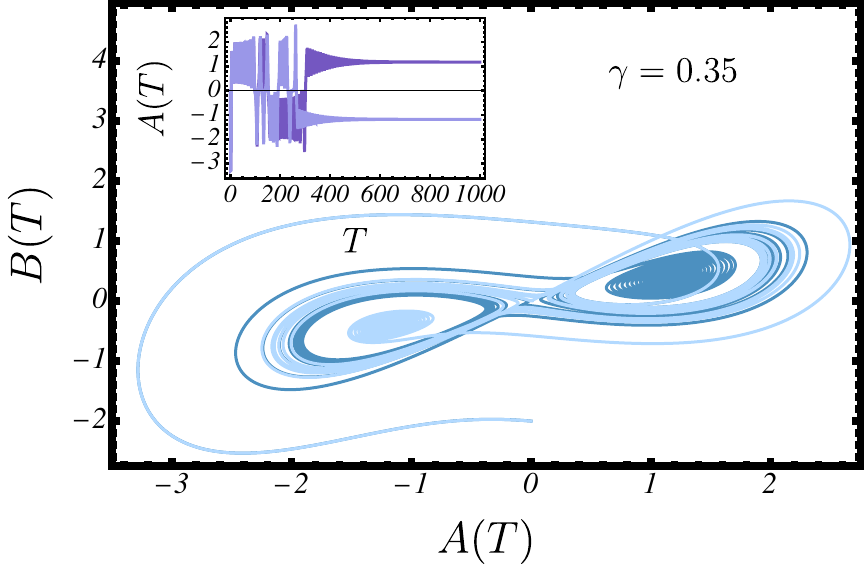}\qquad
\includegraphics[scale=0.375]{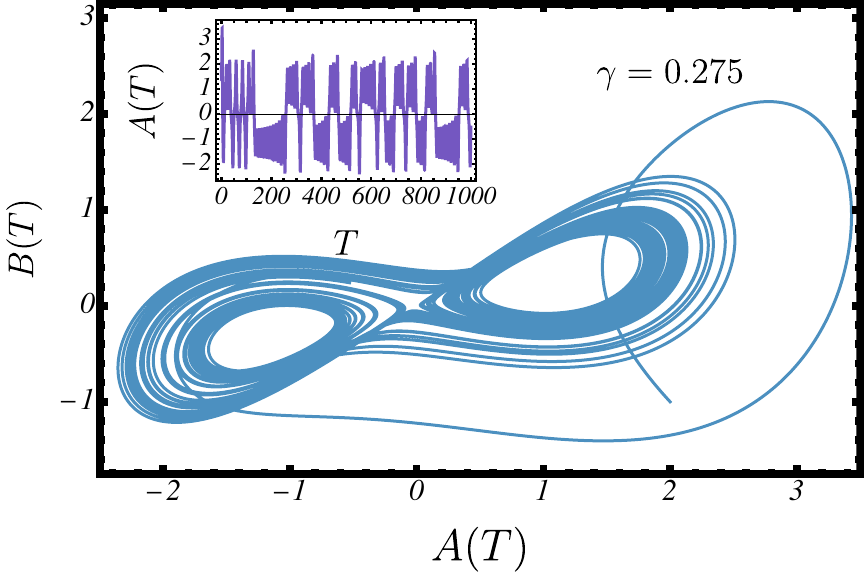}\qquad
\includegraphics[scale=0.375]{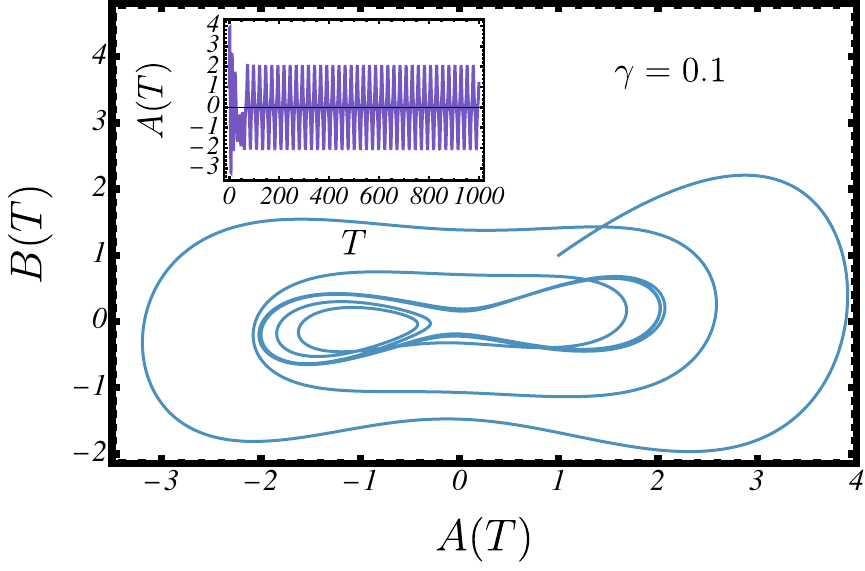}
\caption{Solutions of the toy model \eqref{PM:toy_model} shown for various initial conditions $(A_0,B_0)$ and values of $\gamma$. The solutions illustrate the different dynamics that arise in the model, including equilibration to a fixed point ($\gamma=0.35$), chaotic behavior ($\gamma=0.275$), and convergence to a stable periodic orbit ($\gamma=0.1$). In the $\gamma=0.35$ panel, sensitivity to initial conditions is illustrated. The initial condition is chosen within the intricate structures identified in the corresponding basins of attraction. The dark blue (dark purple) trajectory corresponds to the initial condition $(A_0,B_0)=(0,-2)$, while the light blue (light purple) corresponds to $(A_0,B_0)=(0,-2-10^{-3})$, leading to a different equilibration outcome.}
\label{fig:basins_attraction_solutions} 
\end{figure*}

\begin{figure*}[t]
\includegraphics[scale=0.375]{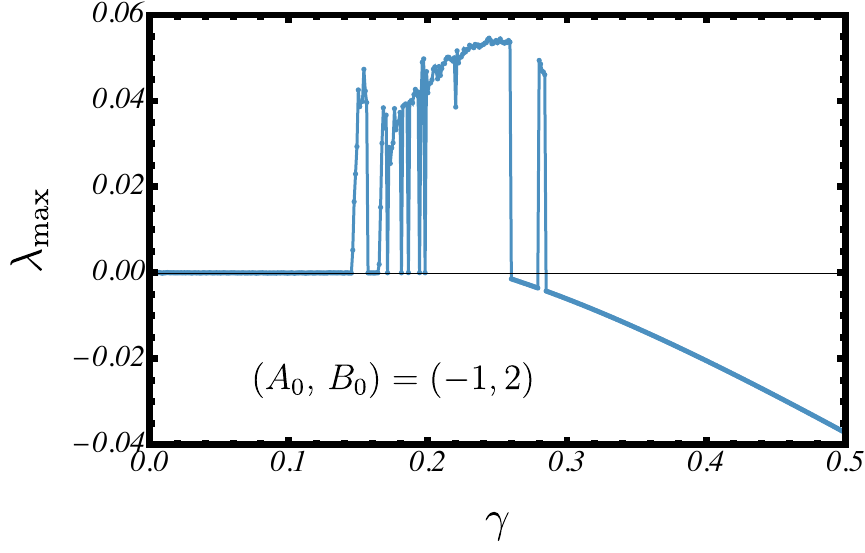}\qquad
\includegraphics[scale=0.375]{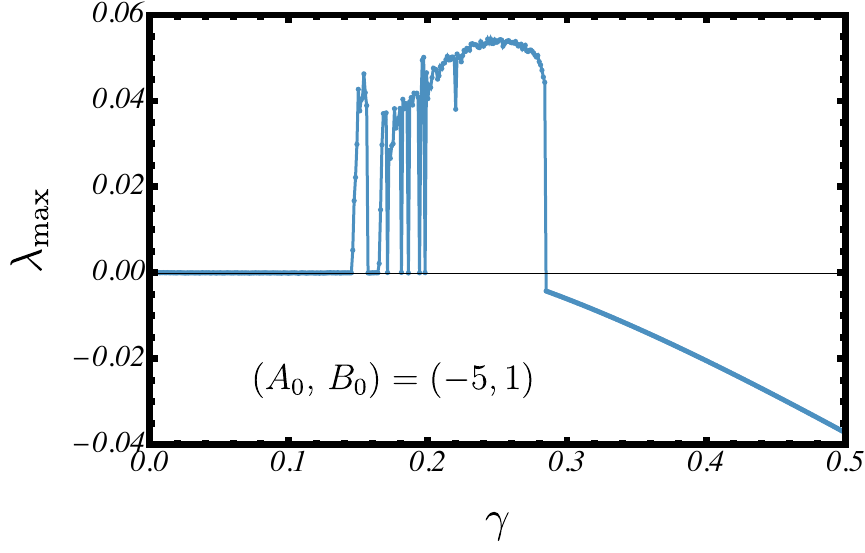}\qquad
\includegraphics[scale=0.375]{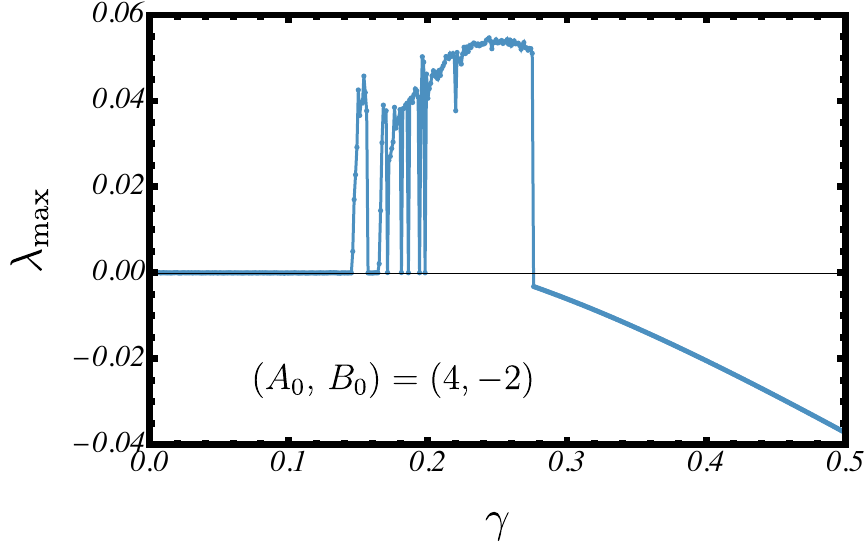}
\caption{Maximal LCE of the toy model \eqref{PM:toy_model} shown for various initial conditions. The exponents were computed using the Benettin \textit{et al.} algorithm described in Appendix~\ref{appendixC}. In our implementation, the tangent-space basis was evolved and orthonormalized at regular time intervals of $s = 0.1$, repeated for $5 \times 10^5$ iterations. The parameter $\gamma$ was sampled in the range $[0, 0.5]$ with a step size of $10^{-3}$. Positive values of $\lambda_{\text{max}}$ indicate chaotic dynamics, a zero value indicates a limit cycle (or periodic orbit), and a negative value indicates convergence to a fixed point.}
\label{fig:Lyapunov_toy_model} 
\end{figure*}

While the transition at $\widetilde{\gamma}$ modifies the local structure of the flow around each of the two fixed points, the second transition at $\gamma_{\text{toy}}$ alters their stability and corresponds to a local bifurcation of the dynamics. In particular, it is compatible with a Hopf bifurcation \cite{Strogatz} as the pair of complex-conjugate eigenvalues crosses the imaginary axis from left to right when $\gamma$ decreases through $\gamma_{\text{toy}}$. We emphasize once more that this bifurcation of the fixed points $\bm{X}_\pm$ is specific to the toy model: for $k_c>1$ these fixed points remain linearly stable for all values of $\gamma$.

\subsubsection{Basins of attraction} 

Trajectories of the system \eqref{PM:model}, and thus of the toy model \eqref{PM:toy_model}, are expected to exhibit different dynamical regimes depending on the value of the parameter $\gamma$. Here, we aim to provide a more comprehensive understanding of the toy model's dynamics by analyzing its basins of attraction. 

Recall that the initial conditions $\bm{X}_0$ of the system in Eqs.~\eqref{PM:ic} are fully determined by the values of $A_0$ and $B_0$. Specifically, for each pair of initial conditions $(A_0, B_0)$, the flow $\bm{X}(T, A_0,\, B_0,\, -A_0^2)\equiv\bm{X}(T, \bm{X}_0)$ can be computed. In particular, we focus on the first component of the flow, $A(T,\bm{X}_0)\equiv X_1(T,\bm{X}_0)$, which corresponds to the wave amplitude. The basins of attraction are defined by the two-dimensional function $A(T, \bm{X}_0)$ evaluated at a sufficiently large time $T$. We note that, owing to the symmetry relation in Eqs.~\eqref{PM:symmetry_1}, the basins are expected to be symmetric with respect to the origin.

Basins of attraction are useful diagnostic tools, but they do not capture all features of the dynamics. At best, they can indicate a loss of equilibration of the wave amplitude. The underlying idea is as follows: based on the linear stability analysis, the fixed points $\bm{X}_\pm$ are linearly stable as long as $\gamma>\gamma_{\text{toy}}$. In this parameter range (or at least for sufficiently large $\gamma$), and for a subset of initial conditions, the wave amplitude is expected to converge to $\pm A_*$. When equilibration fails, the long-time value of the wave amplitude deviates from $\pm A_*$, signaling the onset of a different dynamical regime.

These results are illustrated in Fig.~\ref{fig:basins_attraction}, where the basins are shown for some selected values of $\gamma$. By symmetry, the points $(A_0, B_0)$ and $(-A_0, -B_0)$ are assigned opposite colors, corresponding to opposite equilibration values. The basins display a rich variety of $\gamma$-dependent structures, which allow for a preliminary classification of the system’s dynamics.

Sensitivity to initial conditions increases as $\gamma$ decreases, as illustrated for $\gamma=0.45$, $0.35$ and $0.3$. For $\gamma=0.45$, the basins of attraction are relatively simple, with large regions of uniform color indicating that a wide range of initial conditions yield the same equilibration value. However, as $\gamma$ is reduced to $0.35$, the boundaries between regions associated with opposite equilibration values become increasingly intricate. In these boundary regions, small perturbations in the initial conditions can lead to opposite equilibration outcomes -- a feature that becomes even more pronounced for $\gamma=0.3$ (see also left-panel of Fig.~\ref{fig:basins_attraction_solutions}).

At lower values of $\gamma$ (see $\gamma=0.275$ and $0.25$), the basins become mixed, with only a few small regular regions of uniform color remaining. These mixed regions, where the wave amplitude fails to equilibrate, primarily emerge in regions where equilibration was sensitive to the initial condition (see the transition from $\gamma=0.3$ to $\gamma=0.275$). As $\gamma$ approaches $\gamma_{\text{toy}}$, the wave amplitude fails to equilibrate for most initial conditions. Beyond this point, the dynamics can no longer be described as simple convergence toward one of the fixed points $\bm{X}_\pm$, and more complex dynamical behavior emerges.

To gain further insight into this transition, we illustrate in Fig.~\ref{fig:basins_attraction_solutions} several trajectories of the toy model for different values of $\gamma$, all originating from the same initial conditions. While the regions of uniform color in the basins clearly correspond to equilibration toward one of the fixed points $\bm{X}_\pm$, the nature of the dynamics in the mixed regime is not immediately evident. For instance, at $\gamma=0.275$, the solution appears chaotic on the time scale shown in the figure, whereas at $\gamma=0.1$, the trajectory appears to settle into a stable periodic orbit. 

A more rigorous classification of the dynamics consequently requires additional tools. We now study the maximal Lyapunov exponent of the toy model to quantify what we identified as chaos.

\subsubsection{Lyapunov exponents} 

The basins of attraction obtained above show that variations in $\gamma$ can significantly change the nature of the toy model's asymptotic dynamics. Such changes, triggered by small parameter shifts, are known as bifurcations. In dynamical systems theory, the stability of a nonlinear system is typically assessed through its Lyapunov characteristic exponents (LCEs), which provide a quantitative measure of the system's sensitivity to initial conditions and serve as a robust method for detecting chaos. Appendix~\ref{appendixC} provides a brief overview of the theoretical framework underlying LCEs and outlines the standard numerical algorithm used to compute them.

We compute the maximal LCE as a function of $\gamma$ for several initial conditions using the well-established Benettin \textit{et al.} algorithm \cite{Benettin_1, Benettin_2, Benettin_3}. The results are shown in Fig.~\ref{fig:Lyapunov_toy_model}. Although the overall trend of $\lambda_{\text{max}}$ is similar across different initial conditions (confirmed for additional cases not shown), a careful inspection shows that local variations with $\gamma$ depend on the chosen initial condition. 

At $\gamma=0$, where the system is integrable and trajectories are periodic, $\lambda_{\text{max}}=0$. For small $\gamma$, $\lambda_{\text{max}}$ remains zero until it suddenly becomes positive, signaling the onset of chaos. This is followed by a regime in which it alternates between zero and positive values. Eventually, $\lambda_{\text{max}}$ remains positive, except for a possible brief negative window (not shown here) occurring once the nontrivial fixed points $\bm{X}_\pm$ become linearly stable above $\gamma_{\text{toy}}$, after which it turns negative and decreases monotonically. This final decay coincides with the maximal real part of the eigenvalues of the Jacobian matrix evaluated at $\bm{X}_\pm$.

This picture is consistent with the observations from the basins of attraction: below a certain value of $\gamma$ (which depends on the initial conditions), the wave amplitude no longer equilibrates. In this regime, the maximal LCE complements the basins by revealing the nature of the solutions, which alternate between chaotic dynamics and stable periodic orbits. The first value of $\gamma$ at which $\lambda_{\text{max}}$ becomes negative depends on the initial condition, mirroring the basin for $\gamma=0.275$, where regions of chaotic dynamics coexist with regions of regular equilibration. In contrast, the value of $\gamma$ below which $\lambda_{\text{max}}$ remains zero is the same for all initial conditions and is approximately given by $0.147$. We also find that chaos arises at values of $\gamma$ at which the nontrivial fixed points $\bm{X}_\pm$ are still unstable, raising questions about their possible effect.

\begin{figure}[!th]
    \centering
\begin{tikzpicture}[scale=0.65, transform shape]
    \draw[very thick, |-] (-1, 0) -- (-1.02,0);
    \draw[thick, red, -] (-1, 0) -- (0.98,0);
    \draw[thick, green, dash pattern=on 4pt off 4pt] (1.02,0) -- (7.5,0);
    \draw[thick, red, dash pattern=on 4pt off 4pt, dash phase=4pt] (1.02,0) -- (7.5,0);
    \draw[thick, green, -] (7.5, 0) -- (10.97,0);
    \draw[very thick, |-] (0.98, 0) -- (1.0,0);
    \draw[very thick, ->] (10.97, 0) -- (11,0);
    \draw[very thick, |-] (7.48, 0) -- (7.5,0);
    \node[] at (0.99,-0.6) {\large$\approx 0.147$};
    \draw[thick, <->] (7, 0.5) -- (8,0.5);
    \node[right] at (11.1,0.0) {\large$\gamma$};
    \node[above] at (9.25,0.5) {\large\textit{Fixed points}};
    \node[above] at (4.25,0.5) {\large\textit{Chaos and limit cycle}};
    \node[above] at (0.0,0.5) {\large\textit{Limit cycle}};
    \node[] at (-1.3,0.0) {\large$0$};
\end{tikzpicture}
\caption{Typical nature of the toy model's asymptotic dynamics as a function of the parameter $\gamma$, inferred from the computation of the maximal LCE. The onset of chaos occurs at a critical value of $\gamma$ that appears to be independent of the initial conditions, at $\gamma \approx 0.147$. The figure is not drawn to scale.}
\label{fig:toy_model_Lyapunov_gamma} 
\end{figure}
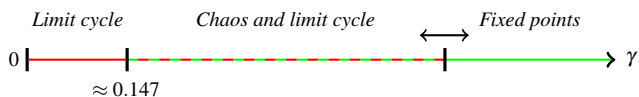

Fig.~\ref{fig:toy_model_Lyapunov_gamma} provides a summary of the overall nature of the toy model's solutions as a function of $\gamma$, based on the maximal LCE analysis presented above and as exhibited by multiple initial conditions. It should be noted that our analysis is restricted to initial conditions defined in Eq.~\eqref{PM:ic}, and therefore does not constitute a complete exploration of the toy model’s full dynamics.

These observations refine our preliminary understanding of the toy model’s dynamics while raising new questions. In particular, how can the onset of chaos be understood? Why is chaotic dynamics intertwined with windows of stable periodic orbits? In the following, we extend these investigations by employing numerical continuation methods.

\subsubsection{Continuation analysis} We now employ numerical continuation methods, which are powerful tools for tracking the evolution of solutions to nonlinear problems as parameters vary. In particular, we use the auto-AUTO Python package \cite{Demaeyer2025}, which automates the execution of the AUTO software \cite{auto07p}, monitors the continuation process, and records bifurcation points.

The procedure starts by identifying stable periodic orbits of the toy model \eqref{PM:toy_model} through numerical integration for a large set of initial conditions. If a periodic orbit attracts most (if not all) initial conditions, it is effectively considered stable. The search is initiated at $\gamma=0.1$ and terminates at the value of $\gamma$ where all trajectories originating from the sampled initial conditions settle into one of the two fixed points $\bm{X}_\pm$.

\begin{figure*}[ht!]
    \centering
    \includegraphics[scale=0.375]{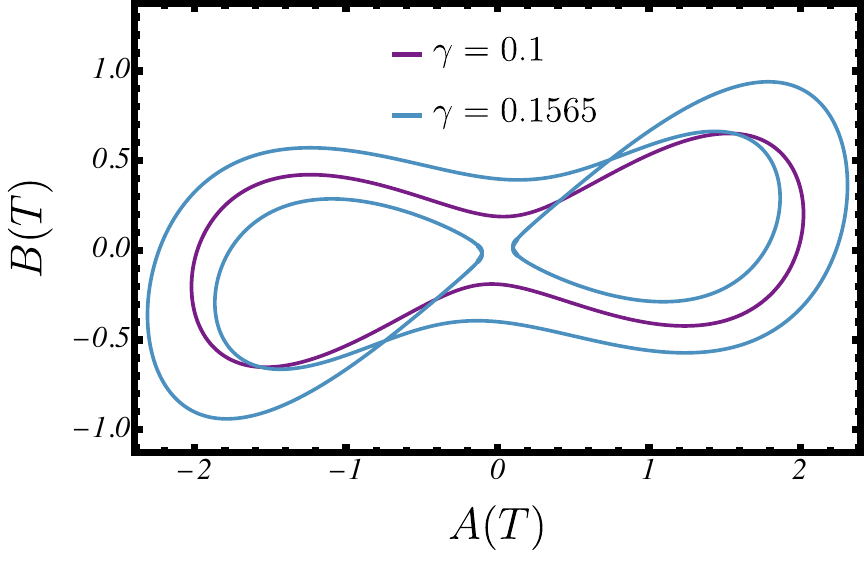}\qquad
    \includegraphics[scale=0.375]{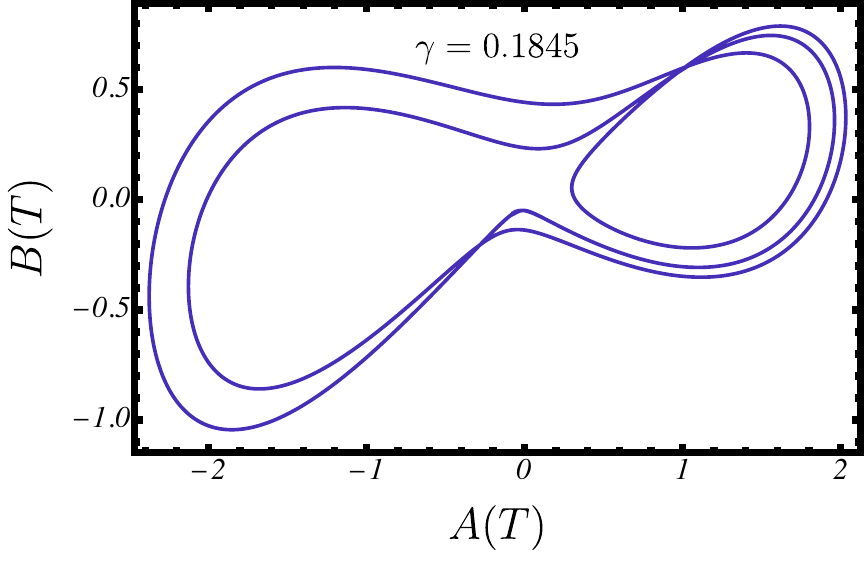}\qquad
    \includegraphics[scale=0.375]{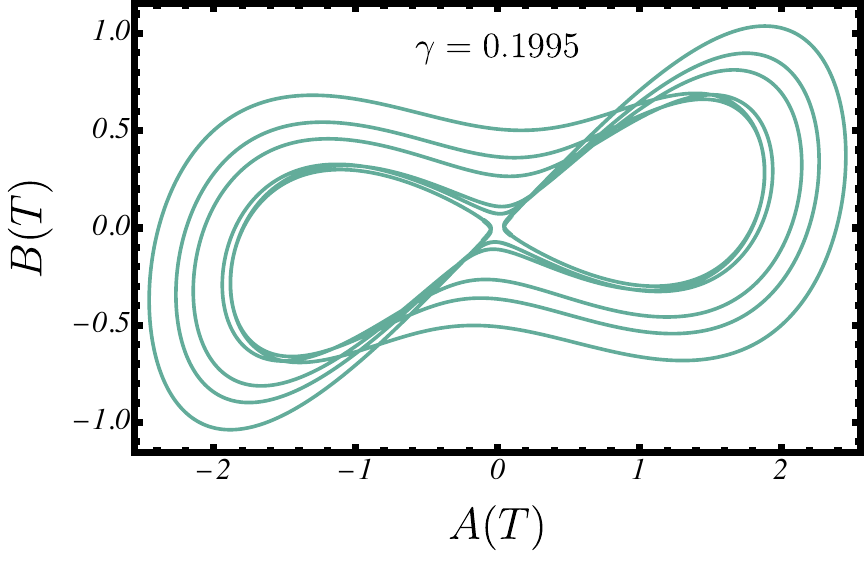}
    \caption{Some of the stable periodic orbits of the toy model listed in Tab.~\ref{tab:periodic_orbits}, shown in the $(A,B)$ plane. These orbits were obtained by numerical integration of the system, starting from a set of random initial conditions in the range $-10\leq A_0, B_0 \leq 10$.}
    \label{fig:periodic_orbit_toy_model}
\end{figure*}

As $\gamma$ approaches $0.15$ and beyond, most of the dynamics becomes chaotic, as indicated by the maximal LCE analysis, except for a few isolated windows in which the dynamics remains that of a stable periodic orbit. We identify these windows as accurately as possible and use the corresponding periodic orbits as starting points for the continuation analysis. Some of these are shown in Fig.~\ref{fig:periodic_orbit_toy_model} in the $(A,B)$ plane. In addition, the values of $\gamma$ at which periodic orbits are found, together with their corresponding periods $T$, are listed in Tab.~\ref{tab:periodic_orbits}.

\begin{table}[ht!]
\centering
\begin{tabular}{c c c c}

\begin{tabular}{c | @{\hspace{10pt}} c @{\hspace{10pt}} c}
Orbit & $\gamma$ & $T$  \\ \hline\hline
\\[-0.6em]
1 & 0.1000 & 24.46 \\
2 & 0.1513 & 152.50 \\
3 & 0.1524 & 102.24 \\
4 & 0.1565 & 50.69 \\
\end{tabular}
\qquad 

\begin{tabular}{c | @{\hspace{10pt}} c @{\hspace{10pt}} c}
Orbit & $\gamma$ & $T$   \\ \hline\hline
\\[-0.6em]
5 & 0.1692 & 201.76 \\
6 & 0.1703 & 100.38 \\
7 & 0.1752 & 75.08 \\
8 & 0.1767 & 150.07 \\
\end{tabular}
\vspace{2em}
 \\

\begin{tabular}{c | @{\hspace{10pt}} c @{\hspace{10pt}} c}
Orbit & $\gamma$ & $T$   \\ \hline\hline
\\[-0.6em]
9  & 0.1845 & 62.76 \\
10 & 0.1860 & 125.45 \\
11 & 0.1939 & 37.85 \\
12 & 0.1975 & 75.82 \\
\end{tabular}
\qquad

\begin{tabular}{c | @{\hspace{10pt}} c @{\hspace{10pt}} c}
Orbit & $\gamma$ & $T$   \\ \hline\hline
\\[-0.6em]
13 & 0.1995 & 150.55 \\
14 & 0.2079 & 125.73 \\
15 & 0.2199 & 50.17 \\
16 & 0.2211 & 100.12 \\
\end{tabular}

\end{tabular}
\caption{Values of the parameter $\gamma$ at which stable periodic orbits are found, along with their corresponding periods $T$. Each orbit can remain stable over a range of $\gamma$ values not listed here. Only the smallest value of $\gamma$ at which a stable periodic orbit is found is reported (except for $\gamma=0.1$). We do not exclude that other stable periodic orbits exist for $\gamma$ values not listed here -- we limited our search to a precision of $10^{-4}$ in $\gamma$.}
\label{tab:periodic_orbits}
\end{table}

\begin{figure*}[ht!]
    \centering
    \includegraphics[scale=0.375]{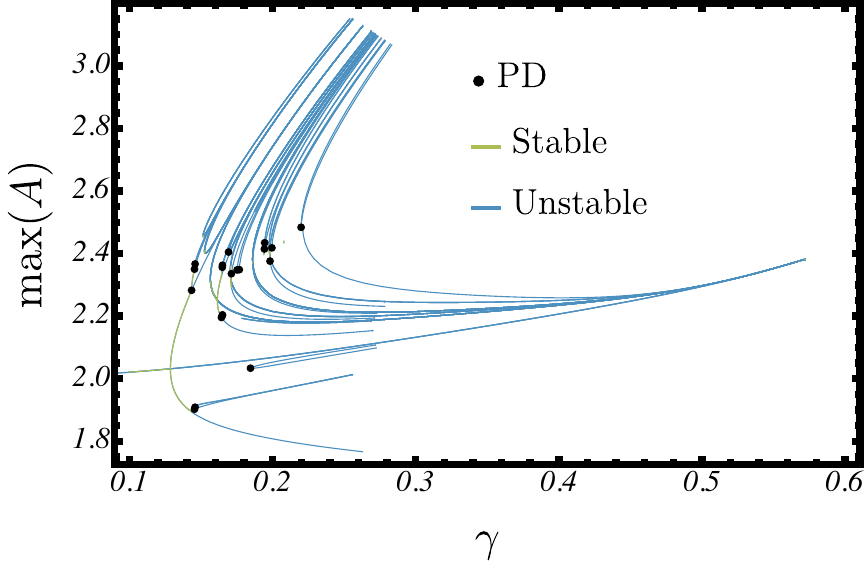}\qquad
    \includegraphics[scale=0.375]{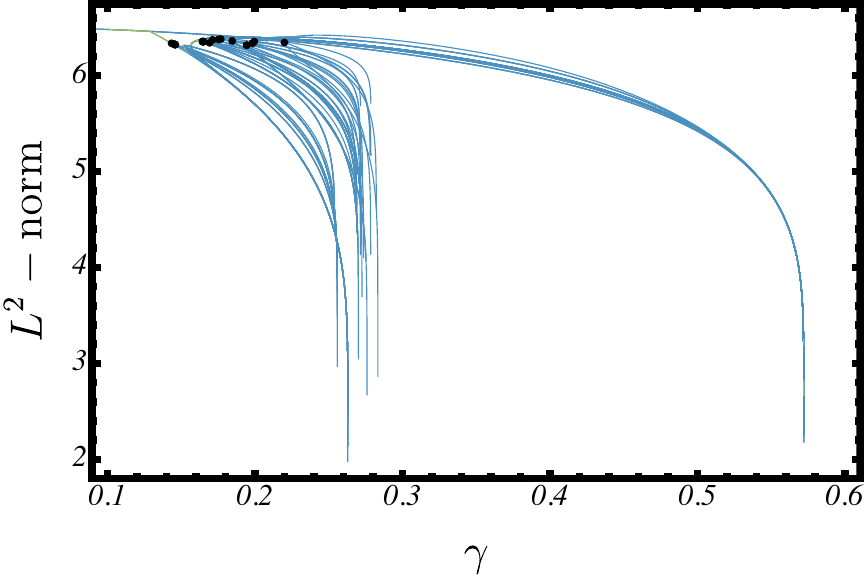}
    \caption{Bifurcation diagram of the toy model \eqref{PM:toy_model}, obtained with auto-AUTO by continuing the stable periodic orbits listed in Tab.~\ref{tab:periodic_orbits}. The left panel shows the maximal value of the wave amplitude $A$ as a function of $\gamma$, while the right panel shows the $L^2$-norm of the orbits as a function of $\gamma$. The diagram illustrates that many stable (green lines) and unstable (blue lines) periodic orbits coexist for a given value of $\gamma$, and that unstable periodic orbits all terminate at homoclinic bifurcations, where their period diverges. The black points indicate PD bifurcations identified by auto-AUTO.}
    \label{fig:bifurcation_diagram_toy_model}
\end{figure*}

\begin{figure*}[ht!]
    \centering
    \includegraphics[scale=0.375]{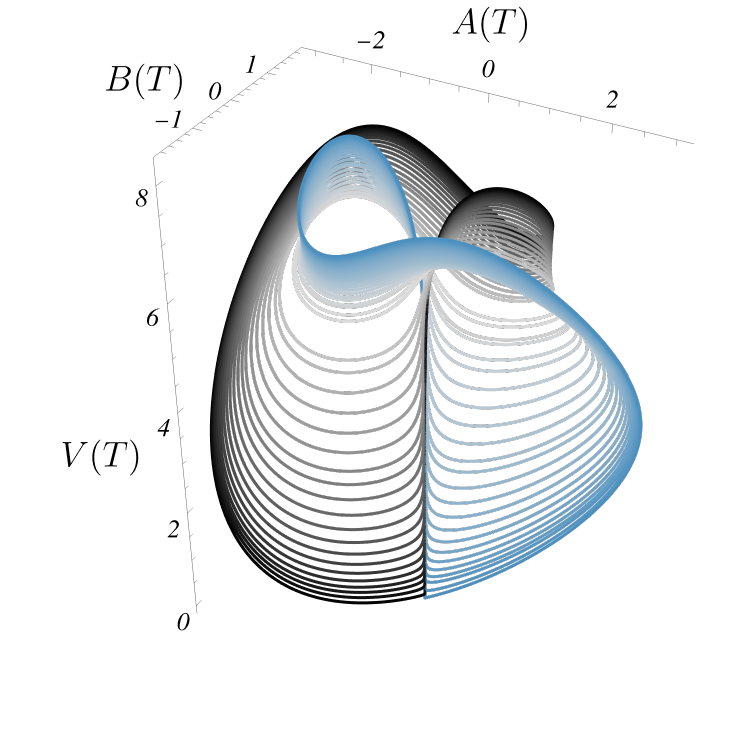}\qquad
    \includegraphics[scale=0.425]{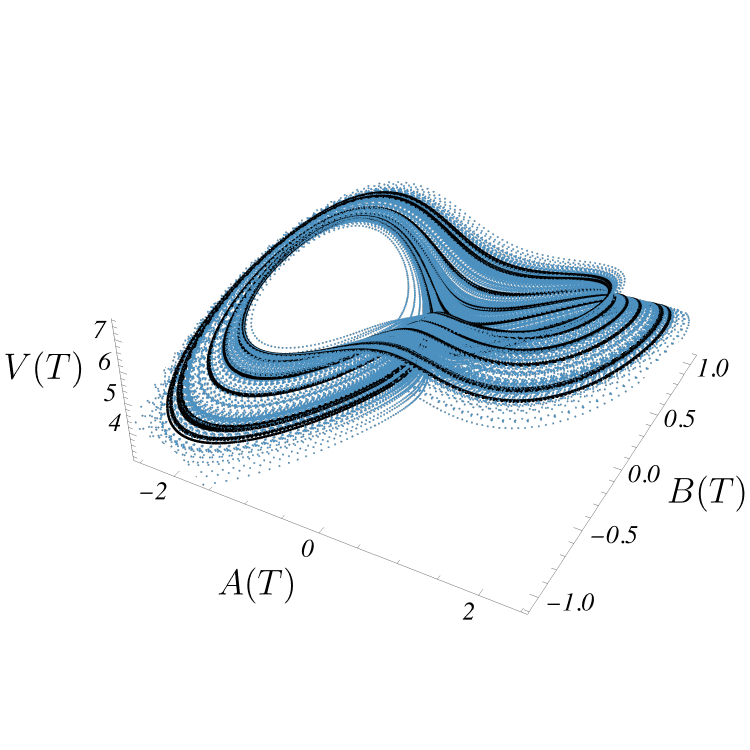}\qquad
    \includegraphics[scale=0.425]{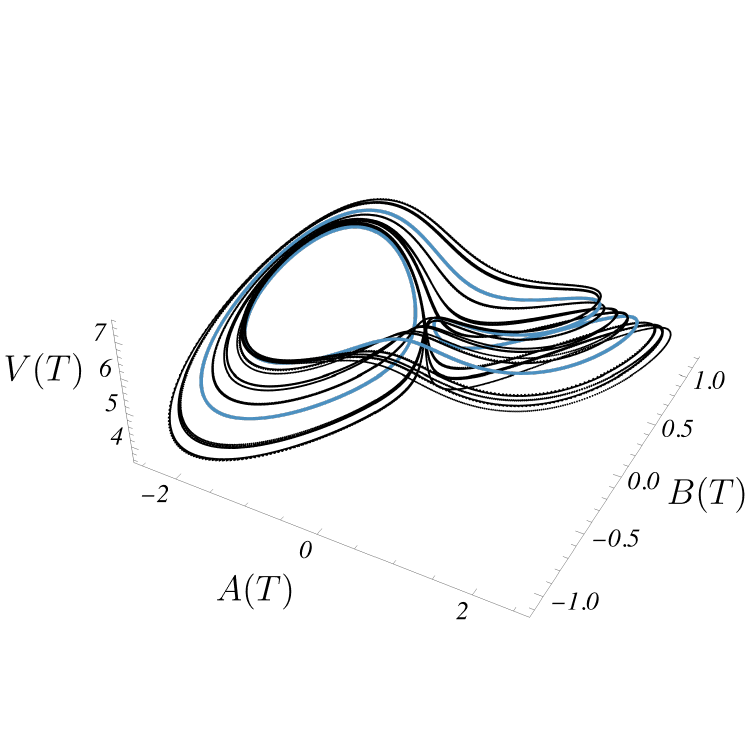}
    \caption{Continuation of the first stable periodic orbit found at $\gamma=0.1$. \textbf{(Left)} A pair of symmetric unstable periodic orbits emerges from a branching point of the first branch at $\gamma\approx 0.129$ (light gray). As $\gamma$ increases, the period of these orbits grows until it diverges at a homoclinic bifurcation at $\gamma\approx 0.263$ (black and blue homoclinic orbits), which connects to the origin. \textbf{(Middle)} Eleven unstable periodic orbits shown at $\gamma=0.155$. The blue set of points represents the trajectory obtained by integrating the initial condition $(A_0,B_0)=(-5,1)$. The trajectory is chaotic (see Fig.~\ref{fig:Lyapunov_toy_model}) and evolves on the surfaces delimited by the unstable orbits. \textbf{(Right)} The same eleven unstable periodic orbits (black) at $\gamma=0.160$, together with a trajectory obtained from the same initial condition. In this case, the trajectory is a stable periodic orbit constrained to evolve on the surface delimited by the other unstable orbits.}
    \label{fig:solution_continued}
\end{figure*}

Once identified, the stable periodic orbits are supplied to auto-AUTO, which continues them in $\gamma$ in both forward and backward directions. Using the results of the auto-AUTO continuation, we track the evolution of each periodic orbit as $\gamma$ varies and analyze their stability. This allows to identify bifurcation points at which the orbit's structure changes and, subsequently, to construct a bifurcation diagram, which is shown in Fig.~\ref{fig:bifurcation_diagram_toy_model}. In this figure, the maximal value of the wave amplitude $A$ and the $L^2$-norm of all identified periodic orbits are plotted as functions of $\gamma$. We recall that the $L^2$-norm $||\bm{X}||_2$ of a periodic orbit $\bm{X}(t)$ in $\mathbb{R}^n$ with period $T$ is defined as $||\bm{X}||_2 = \sqrt{\int_0^T \bm{X}^2(t) \, \dd t/T}$ where $\bm{X}^2(t)$ is understood as the usual scalar product $\bm{X}\cdot\bm{X}$ on $\mathbb{R}^n$. auto-AUTO identified a total of $106$ branches, some of which are barely visible in the bifurcation diagram. Stable periodic orbits are indicated by green lines, while unstable ones are shown in blue, the latter filling most of the diagram. The blue branches all terminate at homoclinic bifurcations, where the period (and thus the $L^2$-norm) of the corresponding periodic orbit diverges (clearly visible in the right panel of Fig.~\ref{fig:bifurcation_diagram_toy_model}). 

For each stable periodic orbit provided as input, auto-AUTO identified two or three period-doubling (PD) bifurcations (see the black points in the figure). This limitation is due to the small variation in $\gamma$ between successive PD bifurcations, as well as the finite precision of the continuation. Despite this, a common scenario emerges: as $\gamma$ increases, each continued periodic orbit undergoes a bifurcation at which it becomes unstable and gives rise to a stable periodic orbit with twice the period. The newly created stable periodic orbit then experiences the same bifurcation scenario as $\gamma$ is further increased, becoming unstable and generating another stable periodic orbit with twice the period, and so on. This process repeats, leading to a cascade of PD bifurcations as $\gamma$ increases, some of which are visible in Fig.~\ref{fig:bifurcation_diagram_toy_model}. We believe that some of the stable periodic orbits identified within the chaos-periodic windows, and listed in Tab.~\ref{tab:periodic_orbits}, originate from some PD bifurcation. This is supported by their periods, which for most orbits is a multiple of $T=25$ (up to some error).

The collection of periodic orbits created through the numerous PD bifurcations forms the backbone of a strange attractor where chaotic dynamics occurs, see middle panel of Fig.~\ref{fig:solution_continued}, where some of the unstable orbits are shown together with a trajectory. Simultaneously, we observe that as $\gamma$ increases, the \clearpage \noindent period of each unstable periodic orbit grows until it diverges at a homoclinic bifurcation, where a homoclinic orbit connecting to the origin is formed, see left panel of Fig.~\ref{fig:solution_continued}. Interestingly, the footprint of such an orbit was already observed in the integrable case at $\gamma=0$, see Sec.~\ref{sec:gamma_zero}. 

Notably, a first PD cascade is observed when continuing the first periodic orbit found at $\gamma=0.1$, which appears to terminate near $\gamma \approx 0.146$. This is precisely the value at which the maximal LCE becomes positive, indicating the onset of chaos regardless of the initial condition. This suggests that the first PD cascade constitutes the primary route to chaos in the toy model. Homoclinic orbits, in contrast, likely play a role in shaping the topological structure of the chaotic attractor.

Finally, we note that the toy model is symmetric under the transformation that maps $(A,B,V)\rightarrow(-A,-B,V)$, see Eqs.~\eqref{PM:symmetry_1}. Continuation of the first stable periodic orbit reaches a branching point at $\gamma\approx0.129$, where a symmetric pair of periodic orbits is created (illustrated in the left panel of Fig.~\ref{fig:solution_continued}). Any bifurcation subsequently encountered by one orbit of the pair will necessarily be mirrored by the other, which explains why multiple PD bifurcations are observed at the same value of $\gamma$ in the bifurcation diagram. We believe that this symmetry plays an important role in the structure of the toy model’s dynamics.

\subsection{Period-doubling and the onset of chaos}\label{sec:onset_chaos}

We now briefly investigate the mechanism by which chaos first arises in the toy model's dynamics. The continuation analysis performed in the previous subsection indicates a PD cascade occurring at $\gamma\approx 0.143$, followed shortly by the emergence of chaotic dynamics. Once the cascade is complete, a chaotic attractor is formed, which is underpinned by a collection of unstable periodic orbits generated during that cascade. This naturally raises the question of how the initially stable periodic orbit, identified at $\gamma=0.1$, loses stability and undergoes a sequence of PD bifurcations. 

\begin{figure}[ht!]
\includegraphics[scale=0.16]{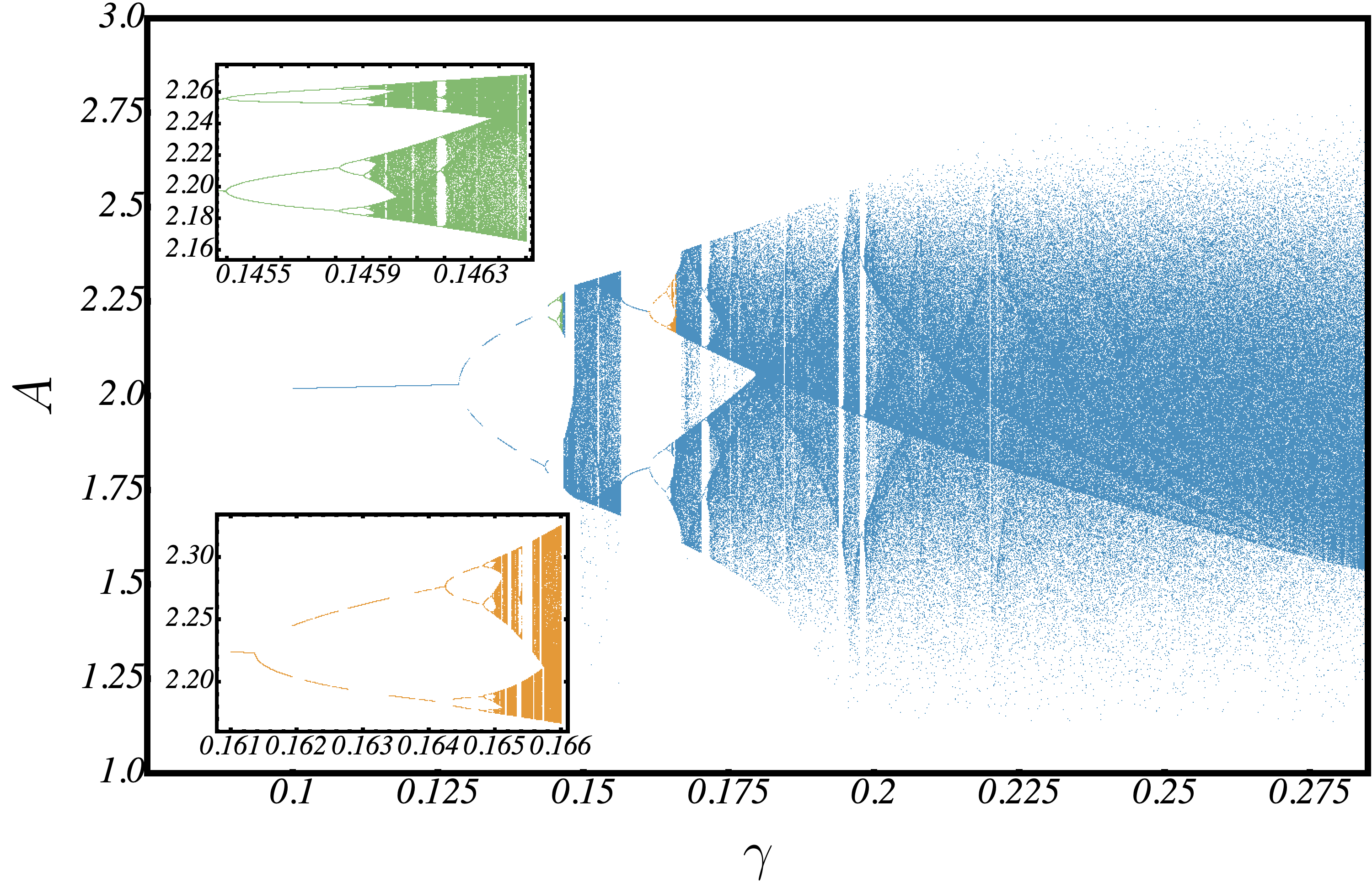}
\caption{Poincaré sections on the plane $V=6.4$ showing the wave amplitude $A$ as a function of $\gamma$ whenever the trajectory crosses the plane with $\dot{V}>0$. The initial condition is $(A_0,B_0)=(-1,1)$. Insets provide magnified views of selected PD cascades, highlighted with colors corresponding to the main plot. Chaotic regions can be distinguished from stable periodic orbits by the wide range of values attained by $A$.}
\label{fig:bifurcation_A} 
\end{figure}

In an effort to understand this first PD bifurcation, we compute the Poincaré sections on the plane $V=V_0$. Whenever the trajectory crosses this plane, the corresponding value of $A$ is recorded. Due to the symmetry of the toy model with respect to the transformation in Eqs.~\eqref{PM:symmetry_1}, whenever the trajectory crosses that plane at $A$, it will also cross it at $-A$, and we therefore restrict our analysis to positive values of $A$. Our results are shown in Fig.~\ref{fig:bifurcation_A}. The plot reveals windows of stable periodic orbits, PD cascades, and chaotic dynamics. The insets provides a zoom on the first two major PD bifurcations, confirming that these lie at the core of the onset of chaos. Subsequent cascades become sparse and difficult to identify in Fig.~\ref{fig:bifurcation_A} due to the narrow range of $\gamma$ over which they occur.

This raises a few questions that are beyond the scope of the present article but could be addressed in a subsequent work: what role do the pairs of homoclinic orbits identified in the continuation analysis play in chaotic regimes and the overall bifurcation structure of the system? In particular, do these homoclinic connections arise from Shil’nikov-type bifurcations, and if so, how do they interact with the multiple PD cascades?

\subsection{The toy model is Lorenz-like}

The toy model \eqref{PM:toy_model} shares strong similarities with the well-known Lorenz system \cite{Lorenz1963}, a three-dimensional dynamical system defined by the now-famous equations
\begin{subequations}\label{Lorenz_system}
\begin{align}
    \dot{x} &= \sigma(y-x)\, , \\
    \dot{y} &= x(\rho-z)-y\, , \\
    \dot{z} &= xy - \beta z\, ,
\end{align}
\end{subequations}
where $\sigma$, $\rho$ and $\beta$ are positive parameters. We provide a concise review of some key properties of the Lorenz system and highlight its correspondence with our toy model. We note that this correspondence was already pointed out by Pedlosky in Ref.~\onlinecite{Pedlosky3}. There, it is shown that the model \eqref{PM:MainEquations} is structurally equivalent to the Lorenz model if the zonal-flow correction has a spatial periodicity of the form $\Phi(T,y)=-Q(T)\sin 2m\pi y$ for some function $Q(T)$. Here, we show that the toy model is structurally equivalent to the Lorenz model, and in fact reduces precisely to the latter for a suitable choice of parameters.

For $\rho> 1$, in addition to the origin, the Lorenz system admits two nontrivial fixed points. These points are stable focus-nodes for $1\leq \rho \leq \sigma(\sigma+\beta+3)/(\sigma-\beta-1)$, and saddle-foci beyond this range -- a situation similar to the toy model as $\gamma$ decreases. The origin is a stable node for $\rho< 1$ and a saddle for $\rho>1$ -- consistent with the toy model in the latter case. Both systems share the same symmetry under inversion of the first two coordinates, $(x,y,z)\rightarrow(-x,-y,z)$. 

For the standard parameters $\sigma=10$ and $\beta=8/3$, at $\rho_{\text{homo}}\approx 13.926$ the Lorenz system exhibits a homoclinic bifurcation, where a symmetric pair of homoclinic orbits emerges. For $\rho$ slightly above $\rho_{\text{homo}}$, in the vicinity of these homoclinic orbits, a countable infinity of periodic orbits is created \cite{Sparrow1982, Sparrow1984, Doedel_Lorenz}. This phenomenon, known as a \textit{homoclinic explosion}, is the source of the complicated dynamics of the Lorenz system. For larger values of $\rho$, the system experiences additional homoclinic explosions, producing the orbits required for subsequent PD windows (see for example Fig.~5.12 of Ref.~\onlinecite{Sparrow1982}). The interplay between homoclinic orbits and PD bifurcations in the Lorenz system is a remarkable phenomenon that also emerges in our toy model.

The computation of the three Lyapunov exponents of the Lorenz model \cite{LorenzLyapunov} is particularly insightful, as it reveals a behavior opposite to that of our system: for $\rho$ below a certain threshold, $\lambda_{\text{max}}$ remains negative. Above this threshold, the system exhibits alternating windows of chaos ($\lambda_{\text{max}}>0$) and stable periodic orbits ($\lambda_{\text{max}}=0$). For large values of $\rho$, the system becomes almost conservative, with a stable periodic orbit that loses stability through a PD bifurcation as $\rho$ is decreased \cite{Lorenz_conservative} -- a scenario that we examined in detail for the toy model in the previous subsections. 

The seemingly opposite dynamics of the Lorenz model compared to that of the toy model can be attributed to their different dependence on dissipation: in the former, $\rho \propto 1/\nu$, whereas in the latter, $\gamma\propto \sqrt{\nu}$, where $\nu$ is the kinematic viscosity.

The structural similarities between the toy and the Lorenz model raise a natural question: can this connection be pushed further, and can the two models be mapped exactly onto one another? We show that such a correspondence exists, except that this holds true only for a specific choice of parameters $(\gamma,a,m)$. 

The Lorenz model \eqref{Lorenz_system} is equivalent to the system
\begin{subequations}\label{Lorenz_model_bis}
\begin{align}
    &\ddot{R}+(\sigma+1)\dot{R}-(D+\sigma \rho-\sigma)R+R^3=0\, ,\\
    &\dot{D}+\beta D+(2 \sigma-\beta)R^2=0\, ,
\end{align}
\end{subequations}
where $D\equiv x^2/2-\sigma z$ and $R\equiv x/\sqrt{2}$. On the other hand, the toy model \eqref{PM:toy_model} can be rewritten as
\begin{subequations}
\begin{align}
    &\ddot{A}+\frac{3\gamma}{2}\dot{A}-(1-f(1)V)A+f(1)A^3=0\, ,\\
    &\dot{V}+\gamma h(1) V-\gamma g(1)A^2=0\, .
\end{align}
\end{subequations}
Rescaling the variables as $\tilde{A}\equiv \sqrt{f(1)}A$ and $\tilde{V}\equiv -f(1) V$, and renaming $\tilde{A}\to A$ and $\tilde{V}\to V$, the system can be brought into the form
\begin{subequations}
\begin{align}
    &\ddot{A}+\frac{3\gamma}{2}\dot{A}-(V+1)A+A^3=0\, ,\\
    &\dot{V}+\gamma h(1) V+\gamma g(1)A^2=0\, .
\end{align}
\end{subequations}
By direct comparison with the Lorenz model \eqref{Lorenz_model_bis}, we observe that the two models coincide if and only if the following parameter identification holds,
\begin{equation}
    \left\{
    \begin{array}{l}
        \sigma+1=\frac{3\gamma}{2}\, ,\\
        \rho=1+\frac{1}{\sigma}\, , \\
        \beta=\gamma h(1)\,,\\
        2\sigma-\beta=\gamma g(1)\, .
    \end{array}
\right.
\end{equation}
We note that, according to the second relation above, $\rho\propto 1/\gamma$, as anticipated from the qualitative differences between the toy and the Lorenz model discussed above. These relations are satisfied if and only if the following constraint on $\gamma$ holds for all values of $a$ and $m$, 
\begin{equation}
    \gamma=\frac{2}{3-g(1)-h(1)}\, .
\end{equation}
Since $g(1)+h(1)=2$ for all values of $a$, it follows that $\gamma=2$ is required for the correspondence to be exact. Consequently, we deduce that the toy model \eqref{PM:toy_model} (expressed in the variables $\tilde{A}$ and $\tilde{V}$) is strictly equivalent to the Lorenz model with parameters
\begin{equation}
    \left\{
    \begin{array}{l}
        \sigma=2 \,,\\
        \rho=\frac{3}{2}\, , \\
        \beta=\frac{2\pi^2}{a^2+\pi^2}\,.
    \end{array}
\right.
\end{equation}
This shows that, in their respective parameter spaces, the toy model and the Lorenz model are generally distinct, except on a specific subset (\textit{i.e.} a line in the parameter space of the Lorenz model) defined by the above conditions. In particular, not all Lorenz regimes are accessible to the toy model, even though the previous analysis shows the latter to be as dynamically rich as the former. Furthermore, the mechanism for the onset of chaos is \textit{a priori} of a different nature in each model. 

\section{\label{sec:full_model}22-dimensional model and beyond}

The toy model analyzed in the previous section provided a helpful global picture of the various dynamical regimes of the model \eqref{PM:model}. We now investigate a higher-dimensional representation by setting $k_c=20$, which more faithfully approximates the \textit{true} dynamics of the original PDE model \eqref{PM:MainEquations}. Applying the same nonlinear analysis, we find that the dynamical regimes identified here do not deviate significantly from those observed in the toy model. Finally, we comment on the validity of these observations for arbitrarily large values of $k_c$.

\subsection{Nonlinear analysis}

We begin with a nonlinear analysis of a higher-dimensional representation of the model \eqref{PM:model}. The parameters $a$ and $m$ are kept at the same values, \textit{i.e.} $a=\pi \sqrt{2}$ and $m=1$. As anticipated in Sec.~\ref{sec:truncation}, a sufficiently large value of $k_c$ must be selected to accurately reproduce the dynamics of the system \eqref{PM:MainEquations}. Our analysis therein suggested that $k_c=20$ provides a reasonable compromise between precision and numerical efficiency. The next subsection further justifies this choice by showing that larger values of $k_c$ do not alter the observations reported here.

The model considered here is therefore given by the $22$ ODEs in Eqs.~\eqref{PM:model1}-\eqref{PM:model3} with $k\leq 20$. The linear stability analysis carried out in Sec.~\ref{sec:stability} remains valid. In particular, the two nontrivial fixed points are always stable, while the origin is a saddle with a $1$-dimensional unstable manifold and a $21$-dimensional stable manifold. As it follows from the spectrum of $\text{D}F(\bm{X}_0)$ in Eq.~\eqref{spectrum_D0} (and likewise from the spectrum of $\text{D}F(\bm{X}_\pm)$), increasing the number of modes $V_k$ enlarges the dimension of the corresponding stable manifolds. For this reason, the dynamics arising from higher-dimensional systems is expected to be at most as \textit{complex} (in the sense of exhibiting exotic or nonstandard dynamics) as that observed in the toy model -- this observation justifies on its own the effort devoted to studying the latter. 

We study the dynamics of the system through its basins of attraction, maximal Lyapunov exponent, and numerical continuation methods, albeit in a more condensed form. 

\subsubsection{Basins of attraction}

 The basins are constructed in the same way as for the toy model, plotting the value of the wave amplitude $A(T,\bm{X}_0)$ for $T\to\infty$ for various pairs of initial conditions $(A_0, B_0)$. The symmetry \eqref{PM:symmetry_1} applies (and more generally is independent of the specific value of $k_c$), providing the basins with a symmetry about the origin.

Two main factors are expected to modify the structure of the basins when compared to the toy model. First, the two nontrivial fixed points $\bm{X}_\pm$ remain stable for all $\gamma$ when $k_c>1$. Hence, regular regions of equilibration may persist at all values of $\gamma$, contrary to what was observed in the toy model. Second, as $k_c$ increases, the dimension of the unstable manifold becomes increasingly negligible compared to that of the stable manifolds. Under time evolution, it is expected that the $21$-dimensional stable manifold of $\bm{X}_0$ and the $22$-dimensional stable manifolds of $\bm{X}_\pm$ evolve into complicated surfaces, possibly increasing the sensitivity of equilibration to the initial condition.

\begin{figure*}[ht!]
\includegraphics[scale=0.265]{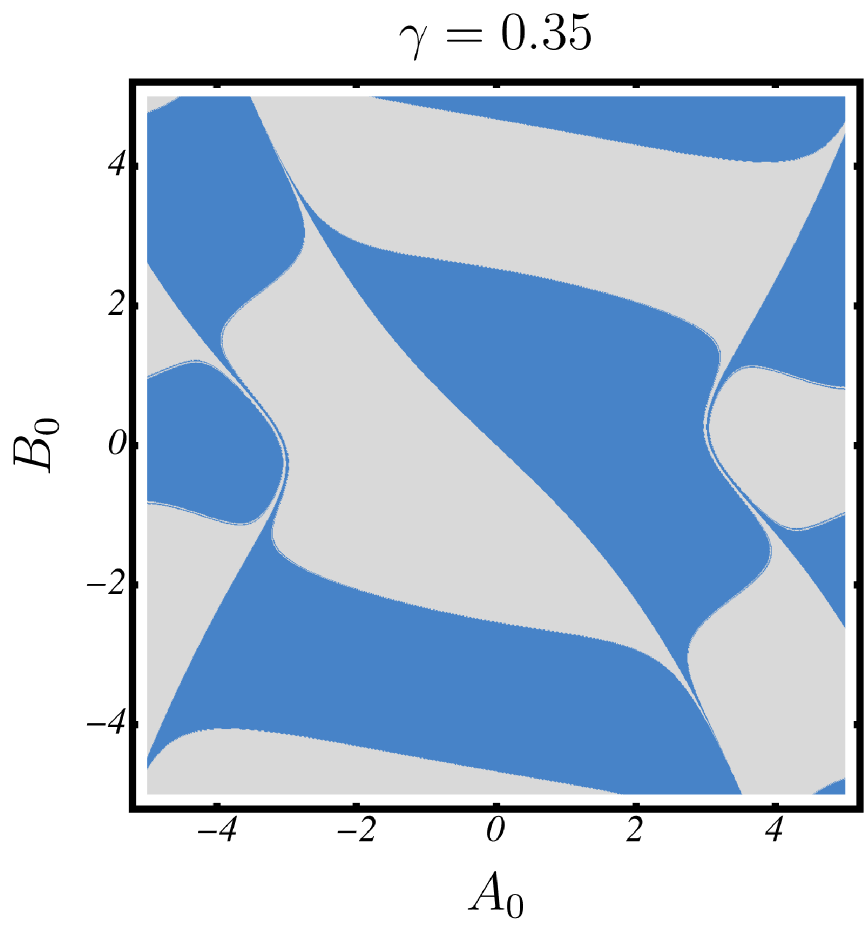}\qquad
\includegraphics[scale=0.265]{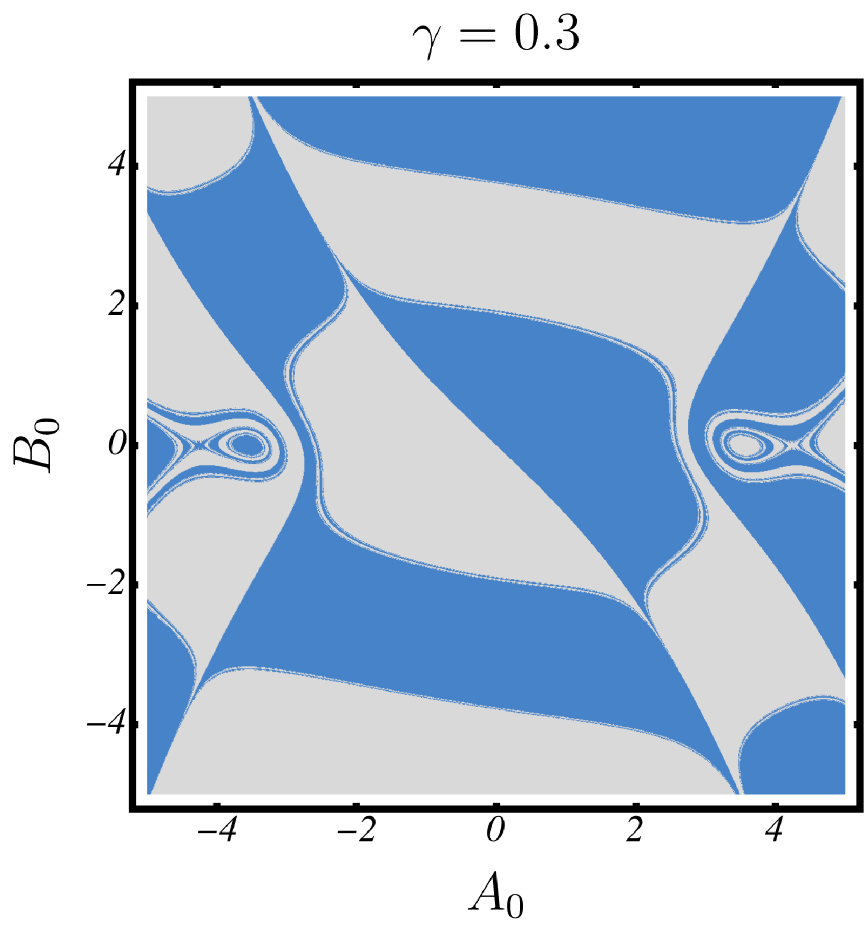}\qquad
\includegraphics[scale=0.265]{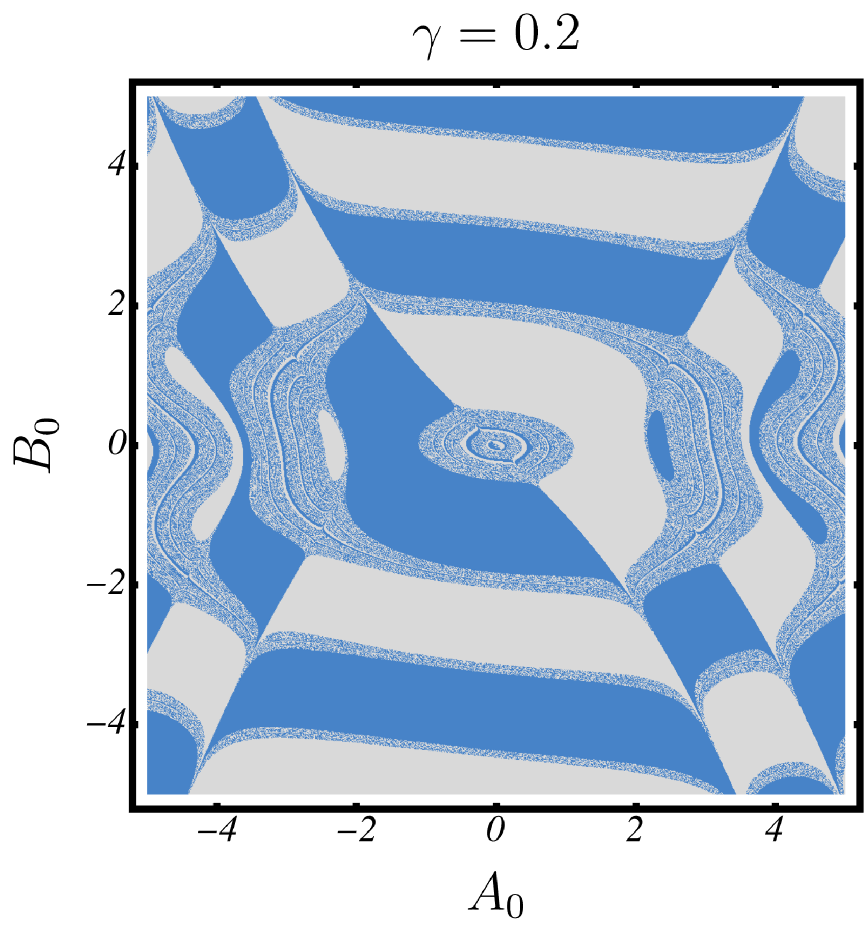}\\
\includegraphics[scale=0.265]{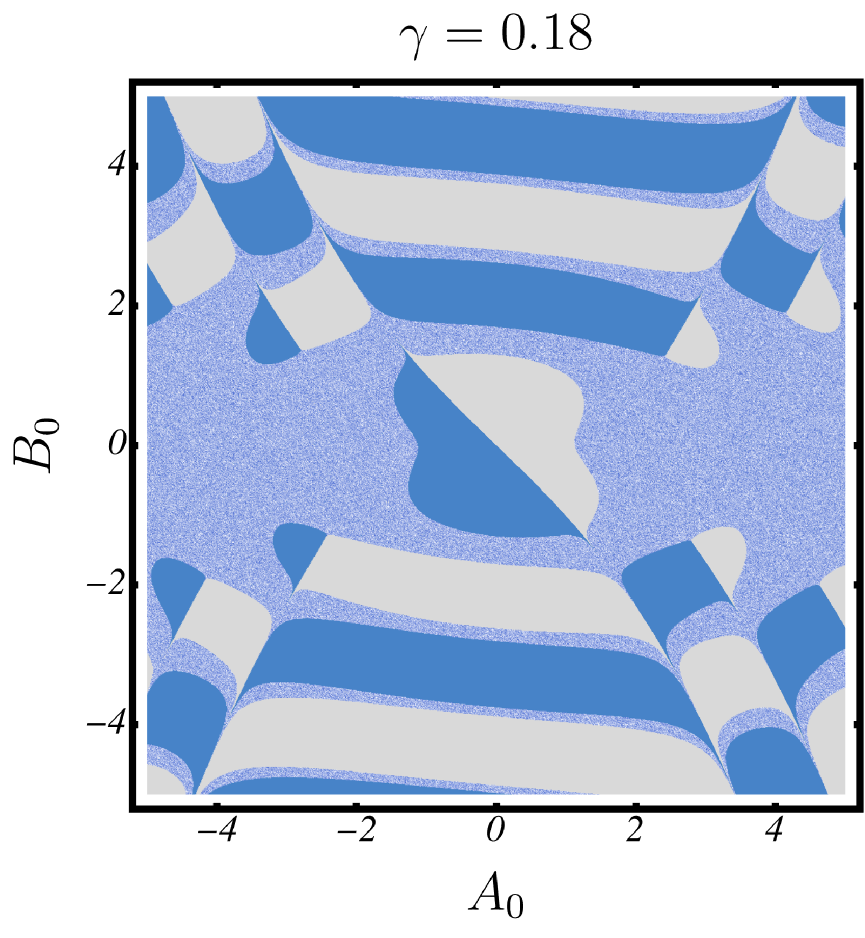}\qquad
\includegraphics[scale=0.265]{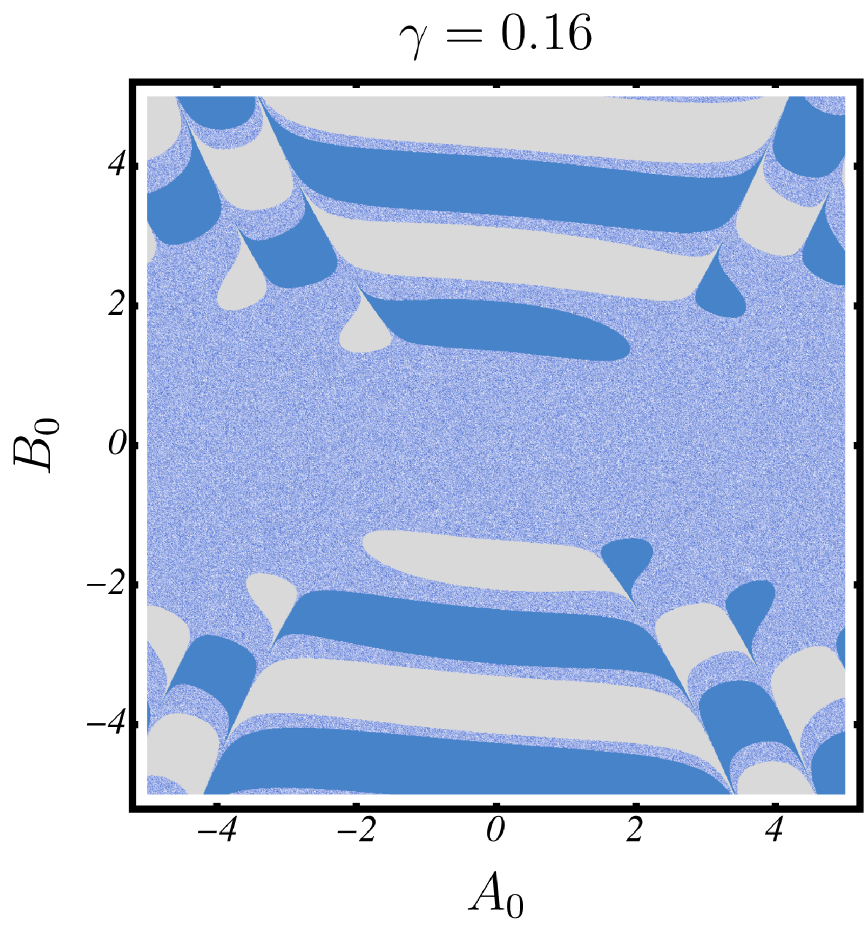}\qquad
\includegraphics[scale=0.265]{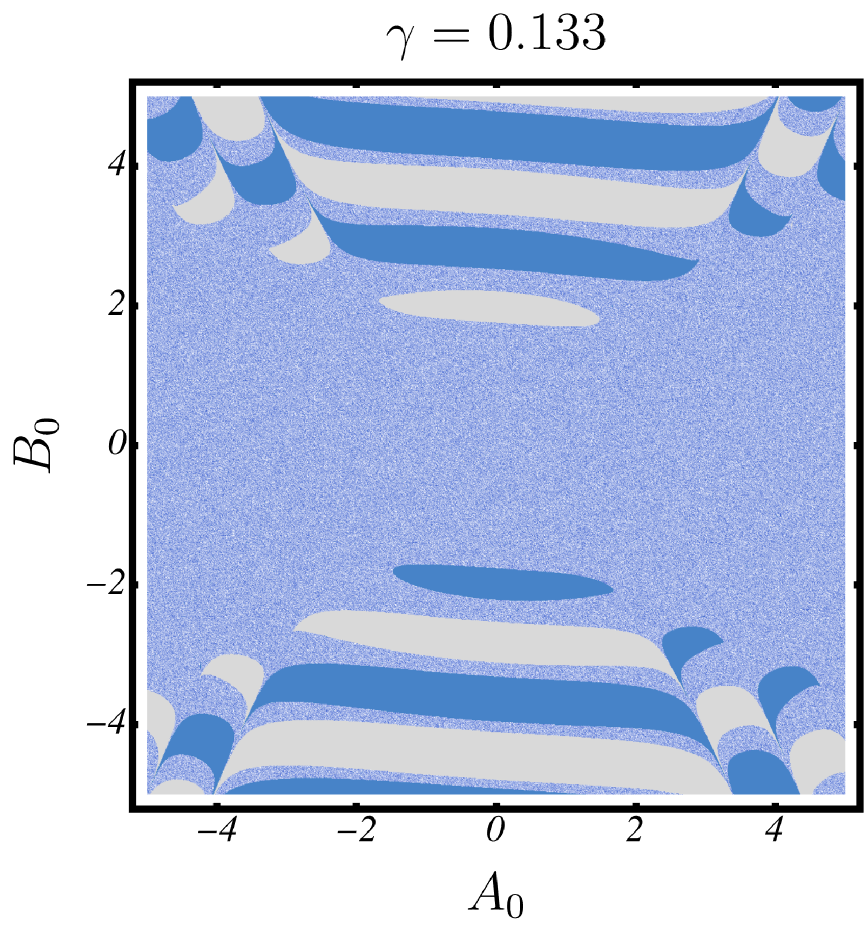}\\
\hspace{1.45cm}\includegraphics[scale=0.45]{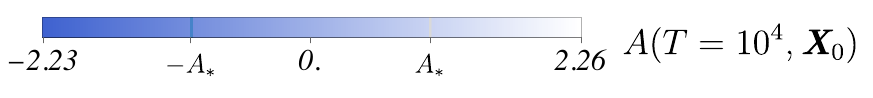}
\caption{Basins of attraction of the model \eqref{PM:model} with $k_c=20$. The basins were constructed following the same procedure as in Fig.~\ref{fig:basins_attraction} (see the caption of that figure), using adjusted values of $\gamma$ to capture the relevant dynamical regimes. In the colorbar, $A_*=1.0000085$, see the end of Sec.~\ref{sec:stability}. The basins exhibit more complex structures than those of the toy model, likely due to the increased folding of the stable manifolds. Note that the regular regions of equilibration are significantly more fragmented and persist even at low values of $\gamma$.}
\label{fig:basins_attraction_22} 
\end{figure*}

The basins of attraction of the $k_c=20$ model are shown in Fig.~\ref{fig:basins_attraction_22}. A simple comparison with those of the toy model in Fig.~\ref{fig:basins_attraction} reveals that the value of $\gamma$ at which equilibration fails is lower in the $22$-dimensional system (around $0.18-0.2$ here, compared to $0.275-0.3$ in the toy model). The basins also show that regions where equilibration is sensitive to initial condition are more intricate (see the panels for $\gamma=0.3$ and $0.2$), and that ``complexity'' increases as $\gamma$ decreases. We notice that failure of equilibration of the wave amplitude first arises in these regions, before progressively spreading to the rest of the basins as $\gamma$ continues to decrease (see the transition from $\gamma=0.2$ to $0.133$). 

Another key difference lies in the way the regular regions are fragmented. In the toy model, they consisted of large, contiguous areas of uniform colors, whereas in the $22$-dimensional system they are subdivided into much smaller cells of alternating colors, forming a \textit{mosaic}-like pattern, corresponding to equilibration toward one of the two nontrivial fixed points. We do not exclude the possibility that this behavior originates from the increased folding of the stable manifolds as $k_c$ increases.

\subsubsection{Lyapunov exponents}

\begin{figure*}[ht!]
\includegraphics[scale=0.375]{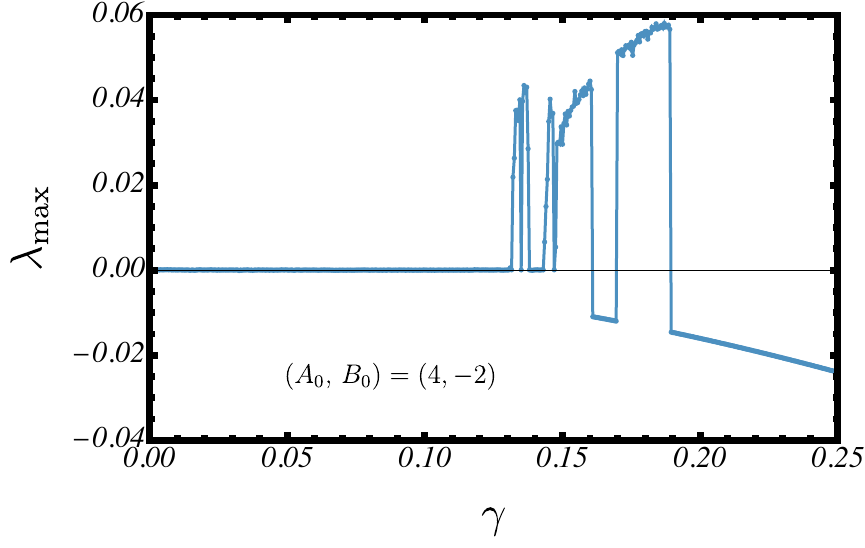}\qquad
\includegraphics[scale=0.375]{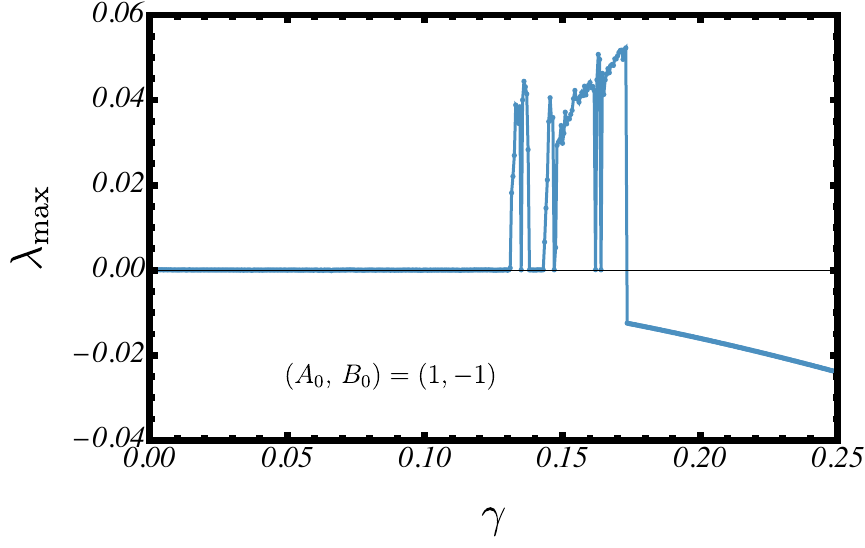}\qquad
\includegraphics[scale=0.375]{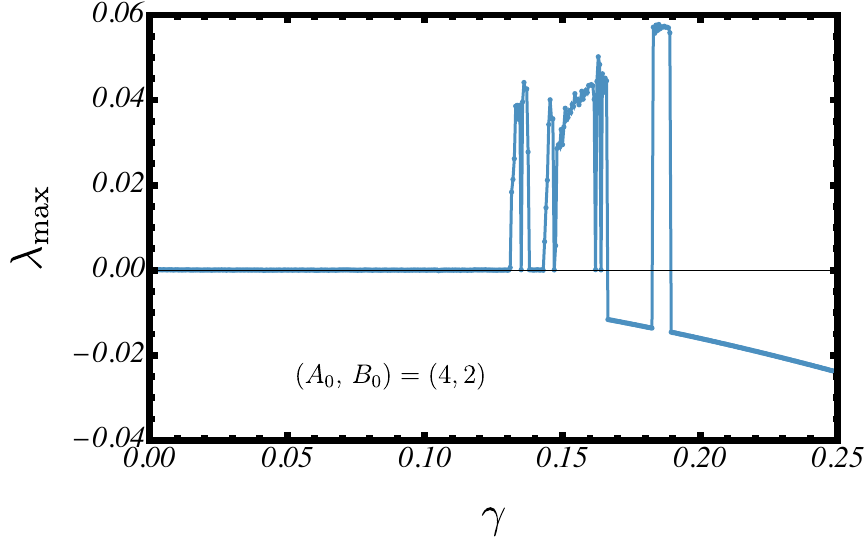}\\
\includegraphics[scale=0.375]{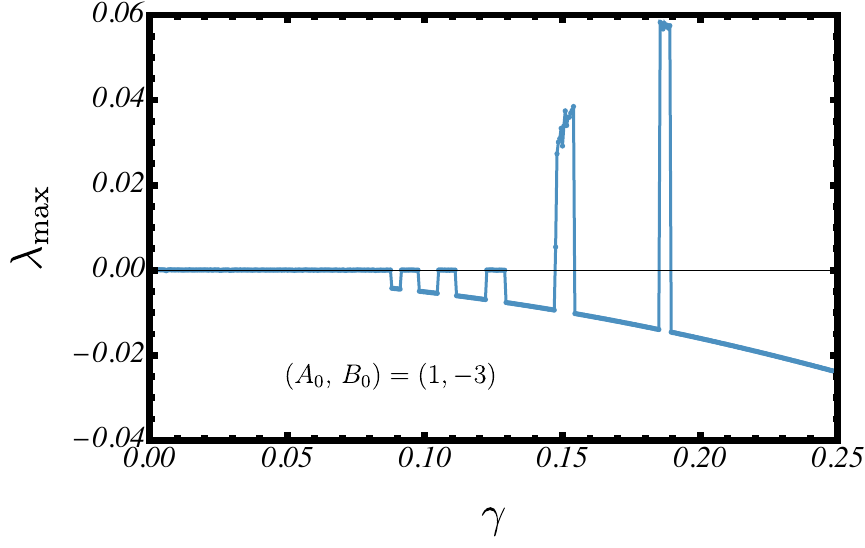}\qquad 
\includegraphics[scale=0.375]{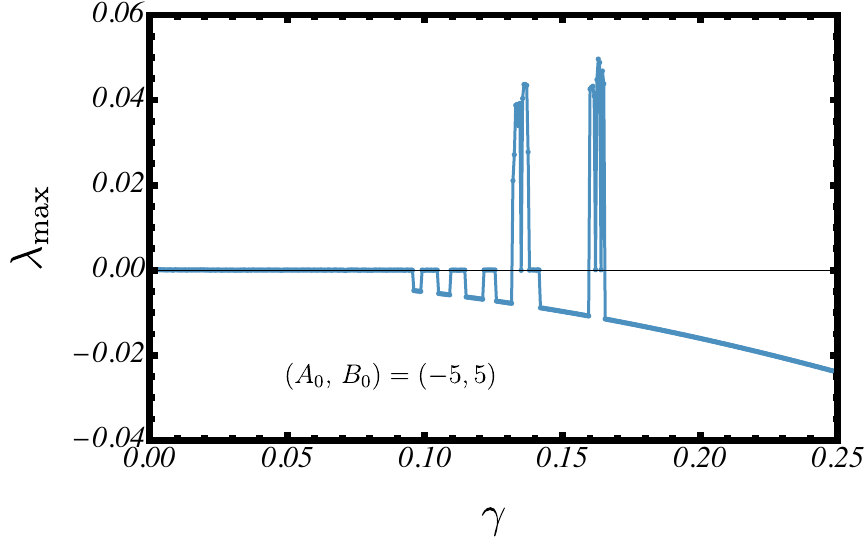}\qquad
\includegraphics[scale=0.375]{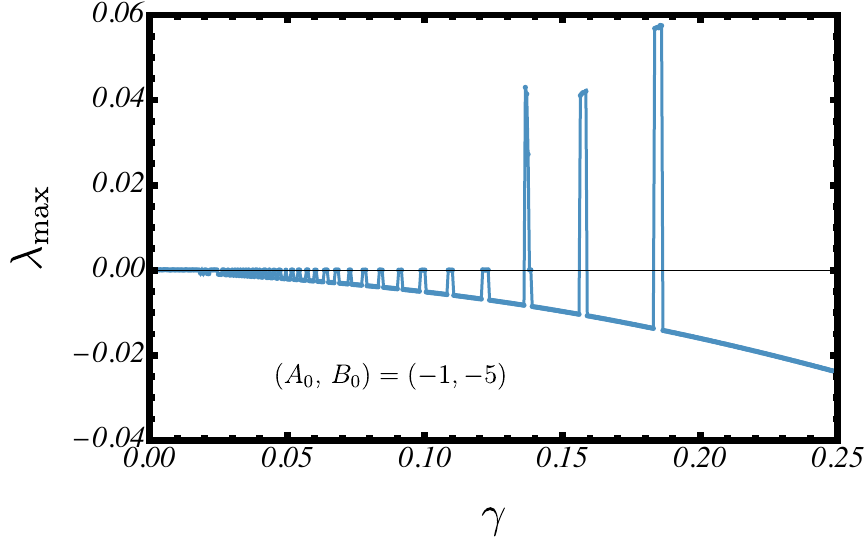}
\caption{Maximal LCE of the model \eqref{PM:model} with $k_c=20$. The maximal LCE were computed following the same procedure as in Fig.~\ref{fig:Lyapunov_toy_model} (see the caption of that figure). The vertical axis uses the same scale as in that figure, showing that the chaotic regime yields maximal LCE values similar to those of the toy model. Note that the horizontal axis range differs. The upper row displays the maximal LCE in the first regime, which closely resembles the behavior observed in the toy model, while the lower row illustrates the second regime, characterized by the persistent stability of the nontrivial fixed points across all values of $\gamma$.}
\label{fig:Lyapunov_22} 
\end{figure*}

To further characterize the asymptotic dynamics of the model, we compute the maximal LCE of the $22$-dimensional model. The resulting variations of $\lambda_{\text{max}}$ with $\gamma$ are shown in Fig.~\ref{fig:Lyapunov_22}. These differ from those of the toy model in that the dynamics falls into two distinct regimes. In the former (upper row of Fig.~\ref{fig:Lyapunov_22}), the behavior closely follows the trend identified in the toy model: prior to the onset of chaos, the dynamics is dominated by a stable periodic orbit. Once chaos is reached, the system exhibits alternating windows of chaotic and periodic behavior. The figure also reveals an early onset of chaos compared to the toy model. The value of $\gamma$ below which $\lambda_{\text{max}}$ remains zero is the same for all initial conditions, and is approximately given by $0.131$.

In the latter regime (lower row of Fig.~\ref{fig:Lyapunov_22}), the situation is markedly different: periodic orbits at small values of $\gamma$ compete with the two nontrivial fixed points, and the dynamics oscillates between them -- sometimes very rapidly, as illustrated by the case where $(A_0,B_0)=(-1,-5)$. This regime is absent in the toy model since as the two nontrivial fixed points are unstable $\gamma < \gamma_{\text{toy}}$. Regions where the maximal LCE is positive are very short, almost entirely suppressed. We believe that the ability of tuning chaos with the initial condition is a remarkable property of the model.

\begin{figure*}[ht!]
    \centering
    \includegraphics[scale=0.375]{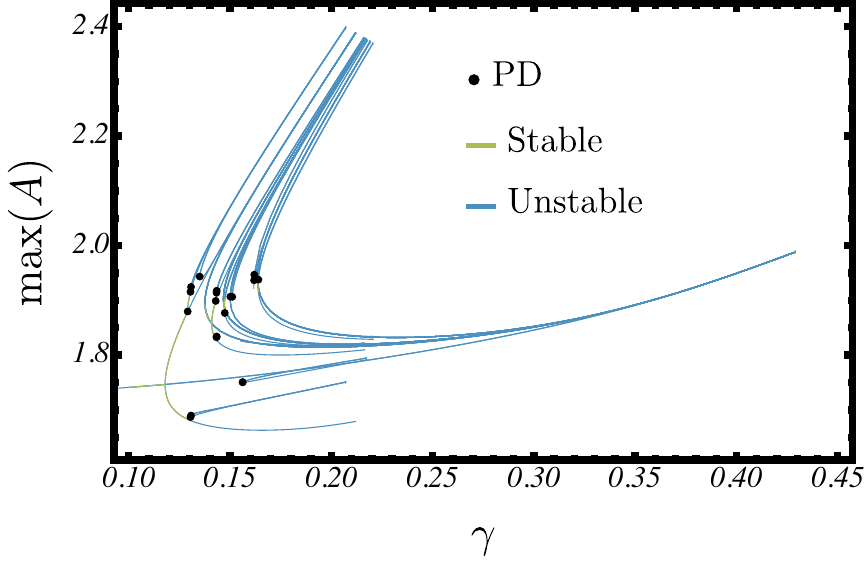}\qquad
    \includegraphics[scale=0.375]{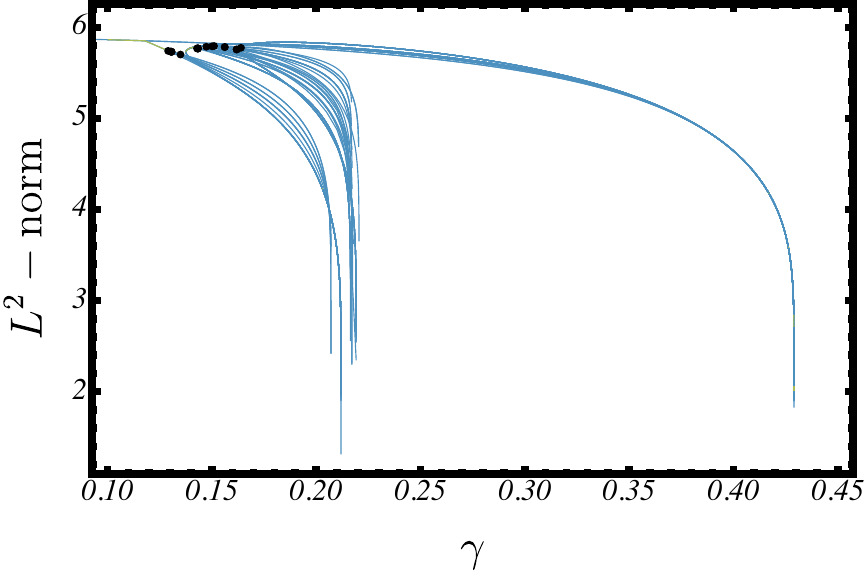}
    \caption{Bifurcation diagram of the model \eqref{PM:model} with $k_c=20$, obtained with auto-AUTO by continuing the stable periodic orbits listed in Tab.~\ref{tab:periodic_orbits_22}. The same observations noted for Fig.~\ref{fig:bifurcation_diagram_toy_model} also apply here. Overall, the bifurcation diagram closely resembles that of the toy model.}
    \label{fig:bifurcation_diagram_full_model}
\end{figure*}

\begin{figure*}[ht!]
    \centering
    \includegraphics[scale=0.375]{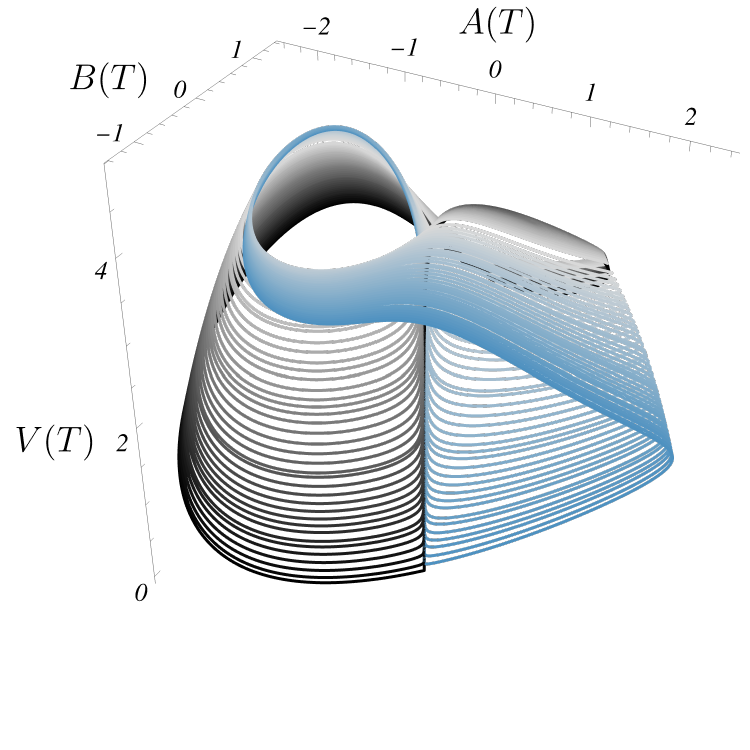}\quad
    \includegraphics[scale=0.425]{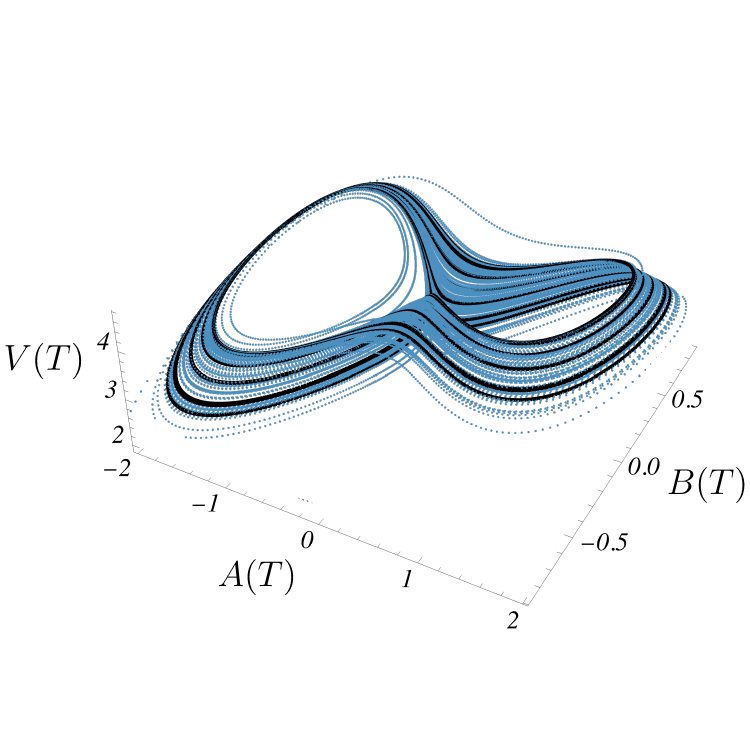}\quad
    \includegraphics[scale=0.425]{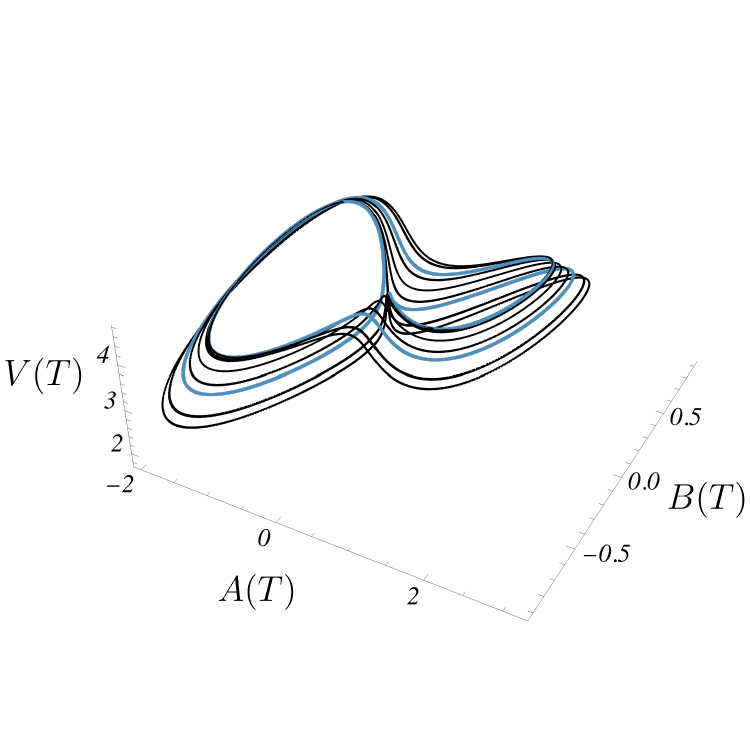}
    \caption{Continuation of the first stable periodic orbit found at $\gamma=0.1$. \textbf{(Left)} A pair of symmetric unstable periodic orbits emerges from a branching point of the first branch at $\gamma\approx 0.118$ (light gray). As $\gamma$ increases, the period of these orbits grows until it diverges at a homoclinic bifurcation at $\gamma\approx 0.212$ (black and blue homoclinic orbits), which connects to the origin. \textbf{(Middle)} Ten unstable periodic orbits are shown at $\gamma=0.135$. The blue set of points represents the trajectory obtained by integrating the initial condition $(A_0,B_0)=(1,-1)$. The trajectory is chaotic (see Fig.~\ref{fig:Lyapunov_22}) and evolves on the surfaces delimited by the unstable orbits. \textbf{(Right)} The same ten unstable periodic orbits (black) at $\gamma=0.140$, together with a trajectory obtained from the same initial condition. In this case, the trajectory is a stable periodic orbit constrained to evolve on the surface delimited by the other unstable orbits.}
    \label{fig:solution_continued_full}
\end{figure*}

\subsubsection{Continuation analysis}

To complete the full picture, we now investigate bifurcations of the $22$-dimensional model using numerical continuation methods. The approach is the same as before. We identify stable periodic orbits by fine-tuning the parameter $\gamma$, which are then supplied to auto-AUTO for continuation. The stable periodic orbits we were able to identify, along with their periods $T$, are listed in Tab.~\ref{tab:periodic_orbits_22}. Surprisingly, stable orbits are less frequent than in the toy model (or alternatively harder to localize with our precision tolerance in $\gamma$).

\begin{table}[ht!]
\centering
\begin{tabular}{c|c c}
Orbit & $\gamma$ & $T$ \\ \hline\hline
\\[-0.6em]
1 & 0.1000 & 23.25 \\
2 & 0.1349 & 98.48 \\
3 & 0.1376 & 48.82 \\
4 & 0.1468 & 96.40 \\
5 & 0.15013 & 72.11 \\
\end{tabular}
\qquad
\begin{tabular}{c|c c}
Orbit & $\gamma$ & $T$ \\ \hline\hline
\\[-0.6em]
6 & 0.15107 & 144.08 \\
7 & 0.15618 & 60.35 \\
8 & 0.15705 & 120.79 \\
9 & 0.1616 & 36.46 \\
10 & 0.1637 & 73.08 \\
\end{tabular}

\caption{Values of the parameter $\gamma$ at which stable periodic orbits are found, along with their corresponding periods $T$. Each orbit can remain stable over a range of $\gamma$ values not listed here. Only the smallest value of $\gamma$ at which a stable periodic orbit is found is reported (except for $\gamma = 0.1$). When necessary, the search precision in $\gamma$ was reduced to $10^{-5}$ to identify additional stable periodic orbits. We do not exclude that other stable periodic orbits exist for $\gamma$ values not listed here.}

\label{tab:periodic_orbits_22}
\end{table}

The limited number of stable periodic orbits proved sufficient to construct a satisfactory bifurcation diagram, shown in Fig.~\ref{fig:bifurcation_diagram_full_model}. auto-AUTO identified a total of $89$ branches, some of which are barely visible. The same color code applies with stable periodic orbits indicated in green, and unstable ones in blue. The diagram closely resembles that of the toy model, although the range of values attained by $\max(A)$ and the $L^2$-norm differ. Each stable periodic orbit undergoes a PD bifurcation, turning unstable and giving rise to a new stable periodic orbit with twice the period. The unstable periodic orbit terminates at a homoclinic bifurcation, where a homoclinic orbit connects to the origin. Repetition of this sequence leads to a cascade of PD bifurcations, as indicated by the black points in Fig.~\ref{fig:bifurcation_diagram_full_model}. In particular, the first PD bifurcation, emerging from the continuation of the stable periodic orbit at $\gamma=0.1$, occurs at $\gamma=0.129$ (earlier than in the toy model), and the cascade appears to terminate at $\gamma\approx 0.131$, fully consistent with the value of $\gamma$ at which the maximal LCE becomes positive.

Finally, Fig.~\ref{fig:solution_continued_full} illustrates some periodic orbits obtained by continuation of stable orbits, together with numerical solutions of the model. Once again, in the regions where the dynamics is chaotic, the collection of periodic orbits originating from PD bifurcations appears to form the backbone of a strange attractor.

\subsection{Higher-dimensional model}\label{sec:Higher_dimension}

We conclude by investigating the convergence of the solutions of model \eqref{PM:model} with respect to the cut-off $k_c$, an issue that was already partially addressed in Sec.~\ref{sec:truncation}. To complement that discussion, we examine how the nonlinear analysis performed for the toy model in Sec.~\ref{sec:toy_model} and for the $22$-dimensional model in Sec.~\ref{sec:full_model} extends to larger values of $k_c$.

As highlighted in the previous subsection, the basins of attraction and the maximal Lyapunov exponents of the toy and the $22$-dimensional models are structurally distinct. This raises the question of how these features are modified as the truncation $k_c$ is increased beyond $k_c=20$. Based on the results in Sec.~\ref{sec:truncation}, these differences are expected to be minimal. We verify this by computing the basins of attraction and the maximal Lyapunov exponent for $k_c=30$ and $40$. The resulting basins closely resemble those of the $22$-dimensional model, with discrepancies that are negligible. The same holds for the maximal Lyapunov exponents, although minor local deviations can be identified.

\begin{figure}[ht!]
    \centering
    \includegraphics[scale=0.55]{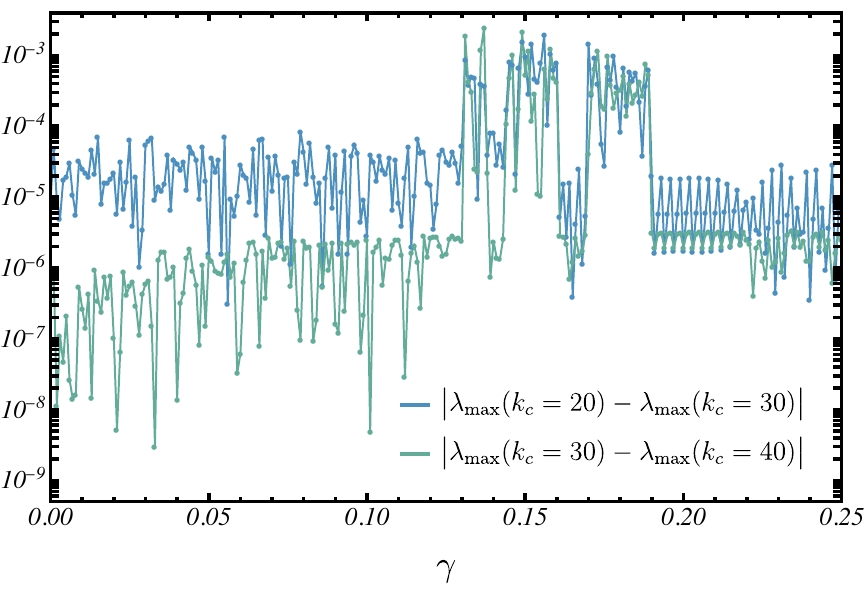}
    \caption{Absolute difference between maximal Lyapunov exponents computed for different values of the cut-off $k_c$ as a function of $\gamma$. The convergence of $\lambda_{\text{max}}$ with increasing $k_c$ is shown to be contingent upon the nature of the underlying dynamics.}
    \label{fig:lyapunov_difference}
\end{figure}

To quantify these differences, we analyze the absolute variation between the maximal Lyapunov exponents for $k_c=20$, $30$ and $40$. The results, shown in Fig.~\ref{fig:lyapunov_difference}, indicate that the differences decrease as $k_c$ increases, except within the $\gamma$ intervals where chaotic behavior occurs. In those chaotic regions, the differences appear to remain approximately constant and saturated regardless of the increase in $k_c$.

These results are significant as they confirm the persistence of the various dynamical regimes of the model \eqref{PM:model} observed throughout this work. In particular, they allow us to conclude that the model in its original formulation \eqref{PM:MainEquations} possesses all the structural and dynamical features identified in the $22$-dimensional model. 

\section{\label{sec:conclusion_outlooks}Conclusion \& Outlook}

In this work, we conducted a comprehensive investigation of the nonlinear dynamics of baroclinic instability in Pedlosky’s two-layer model, extending the original studies by Pedlosky. Despite the model’s simplified formulation and assumptions, we demonstrated that the amplitude of the baroclinic wave exhibits a wide range of dynamical regimes. Our analysis focused on the nature of the solutions of the model \eqref{PM:MainEquations}, in which the geophysically relevant parameter $\gamma$ -- controlling the level of dissipation -- was varied. This model corresponds to an infinite-dimensional system of ordinary differential equations given by \eqref{PM:model}, which we analyzed using standard nonlinear techniques to understand how the system bifurcates as $\gamma$ varies.

Before drawing any definitive conclusions about the model, it was essential to understand how the dynamics of \eqref{PM:model} depends on the truncation $k_c$. We showed that the rate of convergence of both trivial and periodic solutions as $k_c$ increases typically follows an inverse power law in $k_c$. This result provides confidence that a moderately large truncation value (above $k_c = 20$ in our case) is sufficient to capture the dynamics of the original model. This was later confirmed by higher-dimensional analysis, where further increases in $k_c$ did not alter the validity of our results.

The analyses of the dynamical regimes of both the toy and the $22$-dimensional model can be synthesized as follows. In the absence of dissipation, the system is Hamiltonian and integrable. Once $\gamma>0$, a rich variety of nonlinear effects emerge, featuring intricate basins of attraction and a complicated bifurcation structure revealed by the Lyapunov exponents and by continuation methods . The dynamics of the toy model already provides a good representation of the various regimes arising in higher-dimensional systems. A main findings is that both models could potentially exhibit strong sensitivity to initial conditions, including in regimes where solutions are stable and converge to one of the fixed points. Consequently, even in the absence of deterministic chaos, nearby initial conditions can lead to qualitatively different equilibration states, ultimately resulting in contrasting geophysical outcomes.

In regimes where $\gamma$ is small, this sensitivity is lost, and a single stable periodic orbit attracts all trajectories. This orbit plays a central role in the bifurcation structure of the model, as it undergoes a period-doubling cascade leading to chaos. The cumulative effect of these cascades is the creation of many unstable periodic orbits that form the backbone of a chaotic attractor, interspersed with parameter windows in which the dynamics again settles onto a stable periodic orbit. As $\gamma$ increases, these unstable orbits generate homoclinic connections. We conjecture that an infinite sequence of period-doubling bifurcations exists in the system, although a rigorous proof lies beyond the scope of our analysis. Overall, the chaotic regime occupies only a small portion of the full parameter space, which becomes even more restricted for certain initial conditions.

Finally, inspired by results from Pedlosky, we further investigated the structure of the toy model and showed that, in specific parameter regions, it reduces exactly to the Lorenz model. Away from these regions, the two systems remain structurally equivalent but dynamically distinct. The bifurcation structure of the Lorenz model is intriguingly similar yet reversed compared to the toy model: large values of $\gamma$ correspond to small values of $\rho$, and \textit{vice versa}. It illustrates the relevance of the Lorenz model in atmospheric dynamics, but more importantly suggests that Pedlosky’s model represents a more general framework covering a wider range of possible dynamics for the baroclinic waves.

This work also opens the door to interesting questions. In particular, what is the role of the homoclinic orbits generated by successive period-doubling bifurcations of the many stable periodic orbits, or on the origin of the stable periodic orbit for small values of $\gamma$. Some of these questions could potentially be addressed, at least for the minimal toy model. Tackling them could further deepen our understanding of the mechanisms governing the transition to chaos in reduced-order atmospheric models.

\clearpage

\begin{acknowledgments}

Nicolas De Ro thanks Joseph Pedlosky for fruitful discussions during the inception of this work.
This research has been partly supported by the Belgian Federal Science Policy Office (BELSPO) under contract number B2/233/P2/PRECIP-PREDICT.
Some of the numerical computations for this work were performed using the Baobab HPC service at the University of Geneva.

\end{acknowledgments}

\section*{Data Availability Statement}

The data and codes that support the findings of
this study are openly available on Github (\url{https://github.com/Climdyn/Pydlosky}) and archived on Zenodo~\cite{de_ro_2026}.

\appendix

\section{\label{appendixA}System of PDEs into a system of ODEs}

This appendix shows how the system of PDEs given by Eqs.~\eqref{PM:MainEquations} can be transformed into the system of ODEs given by Eqs.~\eqref{PM:model}, the latter being used throughout this work to study the wave dynamics. 

On the one hand, Eq.~\eqref{PM:MainEquations2} provides differential equations for $A$ and $B$. To see this, the function $\partial_y \Phi$ is expanded in a Fourier sine series, 
\begin{equation}\label{appA:FourierSeries}
    \partial_y\Phi(T,y)=\sum_{j \text{ odd}}^\infty U_j(T) \sin j \pi y\, ,
\end{equation}
where $U_j(T)\equiv8 j m(\pi^2 j^2 +a^2)\left(A^2(T)+V_j(T) \right)/((j^2-4m^2)(\pi^2 j^2 +a^2))$. Recalling that the dimensionless variable $y$ is defined on the interval $[0,1]$, it can then be shown that 
\begin{equation}
    \int_0^1\sin\left(2 m\pi y\right)\partial_y^2\Phi\, \dd y = \sum_{j \text{ odd}}^\infty \frac{32 j^2 m^2 \left( A^2+V_j\right)}{(j^2-4m^2)^2(j^2 \pi^2+a^2)}\, .
\end{equation}
Defining the function $B(T)\equiv \dd A/\dd T+\gamma A(T)$, and inserting the result of the previous integral into Eq.~\eqref{PM:MainEquations2}, we obtain
\begin{align}
        \dv{B}{T} = &-\frac{\gamma}{2}B+ \frac{\gamma^2}{2}A+A\nonumber \\
        &\quad-\frac{2 m^2}{\pi^2}A\sum_{k=1}^\infty\frac{ (k-1/2)^2 \left( A^2+V_k\right)}{((k-1/2)^2-m^2)^2\left((k-1/2)^2+\frac{a^2}{4\pi^2}\right)}\, ,
\end{align}

where the sum over odd values of $j$ is converted into a sum over all integers $k$.

On the other hand, Eq.~\eqref{PM:MainEquations1} provides a differential equation for the modes $V_k$. Using the expansion \eqref{appA:FourierSeries}, the left-hand side of Eq.~\eqref{PM:MainEquations1} takes the form
\begin{equation}
    \sum_{p\text{ odd}}^\infty\left( p\pi \dv{U_p}{T}+\frac{a^2}{p \pi} \dv{U_p}{T} +\gamma p \pi U_p\right)\cos p \pi y\, .
\end{equation}
Using the Fourier series of the function $\sin 2 m \pi y$, \textit{i.e.}
\begin{equation}\label{appA:FourierSineSeries}
    \sin 2 m \pi y= -\sum_{n\text{ odd}}^\infty \frac{8m}{(n^2-4m^2)\pi} \cos n \pi y\, ,
\end{equation}
Eq.~\eqref{PM:MainEquations1} then reads
\begin{align}\label{appA:transit}
    &\sum_{p\text{ odd}}^\infty\left( p\pi \dv{U_p}{T}+\frac{a^2}{p \pi} \dv{U_p}{T} +\gamma p \pi U_p\right)\cos p \pi y \nonumber \\
    &\quad= \left(\dv{A^2}{T}+2\gamma A^2\right)\sum_{n\text{ odd}}^\infty \frac{8m}{(n^2-4m^2)\pi} \cos n \pi y\,.
\end{align}
We note that $\left\{\cos n \pi y\right\}_{n=1}^\infty$ forms a family of orthogonal functions with respect to the inner product $(f,g)=2\int_{0}^1 f(y) g(y)\, \dd y$, \textit{i.e.} $(\cos k \pi y, \cos l \pi y)=\delta_{kl}$. Consequently, Eq.~\eqref{appA:transit} can be projected onto a specific mode, say $(\cdot, \cos l \pi y)$, yielding
\begin{equation}
    \left(l \pi +\frac{a^2}{l \pi}\right)\dv{U_l}{T}+\gamma l \pi U_l = \frac{8m}{(l^2-4m^2)\pi} \left(\dv{A^2}{T}+2\gamma A^2\right) .
\end{equation}
Straightforward algebraic manipulation yields an evolution equation for the mode $V_k$ (with $l=2k-1$)
\begin{align}
    \dv{V_k}{T}&=\gamma\left(\frac{\left((k-1/2)^2+a^2/2\pi^2\right)}{(k-1/2)^2+a^2/4\pi^2}\,A^2\right.\nonumber \\
    &\quad \left.- \frac{(k-1/2)^2}{(k-1/2)^2+a^2/4\pi^2}\,V_k\right).
\end{align}

\section{Exact expression of the wave amplitude in the inviscid limit}\label{appendixGamma0}

In this appendix, we derive the analytical solution of Eq.~\eqref{equation_gamma_0} for the wave amplitude $A(T)$ as given by Eqs.~\eqref{exact_A_1} or \eqref{exact_A_2} depending on the sign of the energy $E$. Finally, we address the special case $E=0$.

Using conservation of energy along trajectories, \textit{i.e.} $H(A,B)=E$, one finds
\begin{equation}\label{app:starting_equation}
    \dv{A}{T}=\text{sgn}(B_0) \sqrt{2E+\alpha_1 A^2-\alpha_2 A^4}\, ,
\end{equation}
where, for convenience, we have introduced the parameters $\alpha_1\equiv 1+s A_0^2$ and $\alpha_2\equiv s/2$. We note that the sign of $\dot{A}$ is determined by that of $B_0$. The expression under the square root can be factorized as
\begin{align}
    2E+\alpha_1 A^2-\alpha_2 A^4&=\alpha_2 u_+ u_-\left(1-\tilde{A}^2\right) \left(k \tilde{A}^2-1 \right)\, \label{app:factorization_1}  \\
    &=-\alpha_2 u_+ u_-\left(1-\tilde{A}^2\right) \left(1-k \tilde{A}^2 \right)  ,\label{app:factorization_2}
\end{align}
where $k\equiv u_+/u_-$, $\tilde{A}\equiv A/\sqrt{u_+}$, and where $u_\pm$ are the roots of $2E+\alpha_1 u-\alpha_2 u^2=0$, namely $u_\pm= (\alpha_1\pm \sqrt{\alpha_1^2+8 \alpha_2 E})/(2\alpha_2)$. While $u_+$ is positive for all values of $A_0$ and $B_0$, $u_-$ is positive provided that $B_0\in [-A_0\sqrt{1+A_0^2 s/2},\, A_0\sqrt{1+A_0^2 s/2}  ]$. In this case, the square root of the expression in Eq.~\eqref{app:factorization_1} is well defined. Otherwise, $u_-<0$, and the square root of the expression in Eq.~\eqref{app:factorization_2} is instead well defined. We note that $E$ and $u_-$ always have the same sign. 

By separation of variables, Eq.~\eqref{app:starting_equation} takes the form,
\begin{equation}\label{app:starting_equation_bis}
    \text{sgn}(B_0)(T-T_0)=\int_{A_0}^{A(T)}\frac{\dd A}{\sqrt{2E+\alpha_1 A^2-\alpha_2 A^4}}\, .
\end{equation}
The two scenarios, depending on the sign of $u_-$, must be treated separately as the integration proceeds differently in each case. For later convenience, we recall the following identities.

For $k> 1$ and $x\in\, ]-1, \, -1/\sqrt{k}[$ or $x\in\, ]1/\sqrt{k}, \, 1[$ (such that $\text{sgn}(x_0)=\text{sgn}(x_1)$), one has
\begin{align}
    &\int_{x_0}^{x_1} \frac{\dd x}{\sqrt{(1-x^2)(kx^2-1)}}\nonumber \\
    &=-\frac{\text{sgn}(x_0)}{\sqrt{k}}\int_{\varphi(x_0)}^{\varphi(x_1)} \frac{\dd \theta}{\sqrt{1-(1-1/k)\sin^2\theta}}\nonumber \\
    &=\frac{\text{sgn}(x_0)}{\sqrt{k}}\Big(F[\varphi(x_0)\,|\,(1-1/k)]-F[\varphi(x_1)\,|\,(1-1/k)] \Big)\, ,\label{app:integral_1}
\end{align}
where $\varphi(x)\in\,]0,\pi/2[$ is defined by the equality $\sin^2 \varphi=(1-x^2)/(1-1/k)$. The function $F$ denotes the incomplete elliptic integral of the first kind, defined as 
\begin{equation}
    F[\varphi\,|\, k]=\int_0^\varphi \frac{\dd \theta}{\sqrt{1-k \sin^2\theta }}\, .
\end{equation}

For $k\leq 0$ and $x\in\,]-1,1[\,$,
\begin{align}
    \int_{x_0}^{x_1} \frac{\dd x}{\sqrt{(1-x^2)(1-k x^2)}}&=\int_{\phi(x_0)}^{\phi(x_1)} \frac{\dd \theta}{\sqrt{1-k\sin^2\theta}}\nonumber \\
    &=F[\phi(x_1)\,|\, k]-F[\phi(x_0)\,|\, k]\, ,\label{app:integral_2}
\end{align}
where $\phi(x)\in [-1,1]$ is defined by $\phi(x)=\arcsin{x}$.
We now treat the two cases separately.

\noindent \textit{Case 1}: $u_->0$ ($\Leftrightarrow k>1$). Using Eq.~\eqref{app:starting_equation_bis} (with $T_0=0$) and the integral in \eqref{app:integral_1}, we obtain
\begin{align}
    \varphi(\tilde{A}(T))&=\text{am}\Big[F\Big[\varphi(\tilde{A}_0)\,\Big|\,1-\frac{u_-}{u_+}\Big]\nonumber\\
    &\quad-\text{sgn}(A_0 B_0)\sqrt{\alpha_2 u_+}\,T\, \Big| \,1-\frac{u_-}{u_+}\Big]\, ,
\end{align}
where $\text{am}$ is the Jacobi amplitude function and the inverse of the incomplete elliptic integral of the first kind, \textit{i.e.} $\text{am}[F[\varphi\,|\, k]\,|\, k]=\varphi$. The final solution for the wave-amplitude is therefore given by
\begin{align}
    A(T)^2&=u_+-(u_+-u_-)\text{sn}^2\Big[F\Big[\varphi(\tilde{A}_0)\,\Big|\,1-\frac{u_-}{u_+}\Big]\nonumber \\
    &\quad-\text{sgn}(A_0 B_0)\sqrt{\alpha_2 u_+}\,T\, \Big| \, 1-\frac{u_-}{u_+}\Big]\, ,
\end{align}
where $\text{sn}$ is the Jacobi elliptic sine function. The sign of $A(T)$ is fixed by its initial value $A_0$.

\noindent \textit{Case 2}: $u_-<0$ ($\Leftrightarrow k<0$). Using Eq.~\eqref{app:starting_equation_bis} and the integral in \eqref{app:integral_2}, we obtain 
\begin{align}
    \phi(\tilde{A}(T))&=\text{am}\Big[F\Big[\phi(\tilde{A}_0)\,\Big |\,\frac{u_+}{u_-}\Big]\nonumber \\
    &\quad+\text{sgn}(B_0)\sqrt{-\alpha_2 u_-}\,T\, \Big | \, \frac{u_+}{u_-}\Big]\, .
\end{align}
The final solution for the wave-amplitude is therefore given by
\begin{equation}
    A(T)=\sqrt{u_+}\,\text{sn}\Big[F\Big[\phi(\tilde{A}_0)\,\Big |\,\frac{u_+}{u_-}\Big]+\text{sgn}(B_0)\sqrt{-\alpha_2 u_-}\,T\, \Big | \, \frac{u_+}{u_-}\Big]\, .
\end{equation}

We conclude this appendix with the simpler case $E=0$. Here, the integral in \eqref{app:starting_equation_bis} reduces to
\begin{align}
    &\text{sgn}(B_0) T\nonumber \\
    &=\frac{1}{\sqrt{\alpha_1}}\int_{\tilde{A}_0}^{\tilde{A}(T)}\frac{\dd \tilde{A}}{\tilde{A}\sqrt{1-\tilde{A}^2}}\nonumber \\
    &=\frac{1}{\sqrt{\alpha_1}}\left[\text{arctanh} \left(\sqrt{1-\tilde{A}_0^2}\right)-\text{arctanh}\left( \sqrt{1-\tilde{A}(T)^2}\right)\right] ,
\end{align}

where now $\tilde{A}\equiv \sqrt{\alpha_2/\alpha_1} A$. Solving for $A(T)$ then yields Eq.~\eqref{exact_A_3} for the separatrix. 

\section{Fixed points of the PDE system}\label{appendixB}

In this appendix, we provide a proof of formulas \eqref{PM:formula_pi} and \eqref{formula_phi}, which correspond to the steady-state solutions of the system \eqref{PM:MainEquations}, derived from the fixed points of the model \eqref{PM:model} in the limit $k_c\rightarrow \infty$. \\

\noindent \textit{Part 1}: Proof of the formula for the wave amplitude $A$, see Eq.~\eqref{PM:formula_pi}. This formula amounts to showing that the following relation holds
\begin{equation}
   \sum_{k=1}^\infty f(k)\left(1+ \frac{g(k)}{h(k)}\right)=\frac{1}{\cos^2 \pi m}-\frac{\tan \pi m}{\pi m}\, ,
\end{equation}
where the functions $f(k)$, $g(k)$, and $h(k)$ are defined in Eqs.~\eqref{PM:function}. First, we note that the left-hand side can be rewritten as
\begin{equation}\label{appB:rewriting}
    \sum_{k=1}^\infty f(k)\left(1+ \frac{g(k)}{h(k)}\right)=\sum_{k=1}^\infty \frac{64 m^2}{\pi^2}\frac{1}{((2k-1)^2-4 m^2)^2}\, .
\end{equation}
Consider now the infinite product representation of the cosine function \footnote{See for example \url{https://dlmf.nist.gov/4.22}},
\begin{equation}
    \cos x = \prod_{k=1}^{\infty}\left(1-\frac{4 x^2}{\pi^2 (2k-1)^2}\right) .
\end{equation}
Taking the logarithm on both sides of the equality (assuming $x$ lies within a domain where $\cos x > 0$) and differentiating with respect to $x$ gives
\begin{equation}\label{appB:tan}
    \tan x = -\frac{\dd}{\dd x}\ln \cos x = \sum_{k=1}^\infty \frac{8 x}{(2k-1)^2 \pi^2 -4 x^2}\, .
\end{equation}
A further derivative with respect to $x$ yields
\begin{equation}\label{appB:sec}
    \frac{1}{\cos^2 x} = \frac{\dd}{\dd x}\tan x = \sum_{k=1}^\infty 8 \frac{ (2k-1)^2 \pi^2 +4 x^2}{\left((2k-1)^2 \pi^2 -4 x^2\right)^2}\, .
\end{equation}
Combining Eqs.~\eqref{appB:tan} and \eqref{appB:sec} leads to
\begin{equation}
    \frac{1}{\cos^2 x}-\frac{\tan x}{x} = \sum_{k=1}^\infty \frac{64 x^2}{\left((2k-1)^2 \pi^2 -4 x^2\right)^2}\, .
\end{equation}
In particular, setting $x=\pi m$ reproduces the expected result, see Eq.~\eqref{appB:rewriting}. \\

\noindent \textit{Part 2}: Proof of the formula for the zonal-flow correction $\Phi$, see Eq.~\eqref{formula_phi}. Consider the piecewise function $f(y)$ defined on the interval $[-1,1]$ by
\begin{equation}
    f(y)=\left\{
    \begin{array}{ll}
        1+2y\, , & -1\leq y \leq 0\, , \\
        1-2y\, , & 0 \leq y \leq 1\, .
    \end{array}
\right.
\end{equation}
It is straightforward to show that its Fourier cosine series is given by
\begin{equation}\label{appC:Fourier_triangle}
    f(y)=\sum_{k=1}^\infty \frac{8}{(2k-1)^2 \pi^2}\cos\left[(2k-1)\pi y\right] .
\end{equation}
The zonal-flow correction evaluated on the fixed points of the system \eqref{PM:model} is given by
\begin{align}
    \Phi(y)&=\frac{1}{\pi^3} \sum_{k=1}^{\infty}\frac{m}{(k-1/2)^2\left[(k-1/2)^2-m^2\right]} \cos \left[(2k-1)\pi y\right]\nonumber \\
    &=\frac{1}{\pi^3} \sum_{k=1}^{\infty}\left(\frac{1}{m\left[(k-1/2)^2-m^2 \right]}-\frac{1}{m(k-1/2)^2} \right)\nonumber \\
    &\quad\times\cos \left[(2k-1)\pi y\right] .
\end{align}
The first sum can be evaluated using the formula \eqref{appA:FourierSineSeries},
\begin{align}
    &\frac{1}{\pi^3} \sum_{k=1}^{\infty}\frac{1}{m\left[(k-1/2)^2-m^2 \right]}\cos \left[(2k-1)\pi y\right]\nonumber \\
    &=\frac{1}{2\pi^2 m^2} \sum_{n\text{ odd}}^\infty \frac{8m}{(n^2-4m^2)\pi} \cos{n \pi y}=-\frac{\sin{2m\pi y}}{2\pi^2m^2}\, ,
\end{align}
while the second sum can be evaluated using the formula \eqref{appC:Fourier_triangle} derived above,
\begin{align}
    &\frac{1}{\pi^3} \sum_{k=1}^{\infty}\frac{1}{m(k-1/2)^2} \cos \left[(2k-1)\pi y\right]\nonumber \\
    &=\frac{1}{2\pi m} \sum_{k=1}^{\infty}\frac{8}{(2k-1)^2 \pi^2} \cos \left[(2k-1)\pi y\right]=\frac{1-2y}{2\pi m}\, ,
\end{align}
where we have used that $y\in[0,1]$. Combining these two contributions yields Eq.~\eqref{formula_phi}.

\section{Exact values of \texorpdfstring{$\gamma_{\rm toy}$ and $\widetilde{\gamma}$}{gamma toy and tilde gamma}}\label{appendixD}

In this appendix, we provide the exact expressions for $\gamma_{\text{toy}}$ and $\widetilde{\gamma}$ introduced in Sec.~\ref{sec:toy_model}. Recall that $\gamma_{\text{toy}}$ is defined as the value of $\gamma$ at which the stability of the two nontrivial fixed points $\bm{X}_\pm$ of the toy model changes. Specifically, this corresponds to the value of $\gamma$ at which the real part of the pair of complex eigenvalues of the Jacobian matrix at $\bm{X}_\pm$ changes sign. $\widetilde{\gamma}$ is defined as the value of $\gamma$ at which the two nontrivial fixed points $\bm{X}_\pm$ transition from stable focus-nodes to stable nodes. That is, it corresponds to the value of $\gamma$ at which the pair of complex eigenvalues of the Jacobian matrix at $\bm{X}_\pm$ becomes real.

We present the calculations for the specific case $a=\pi \sqrt{2}$ and $m=1$ (which are the values used throughout this work). These results may be generalized to other values of $a$ and $m$ if desired. The Jacobian matrix of the toy model \eqref{PM:toy_model} at the fixed points $\bm{X}_\pm$ is given by
\begin{equation}
    \left(
\begin{array}{ccc}
    -\gamma  & 1 & 0 \\
    \frac{\gamma ^2}{2}-\frac{1}{3} & -\frac{\gamma }{2} & -\frac{4}{9 \pi } \\
    \frac{5 \pi }{4} \gamma & 0 & -\frac{\gamma }{3} \\
\end{array}
    \right) ,
\end{equation}
and its characteristic polynomial is given by
\begin{equation}
    p(\lambda) =-\lambda ^3-\frac{11 \gamma}{6} \lambda ^2-\left(\frac{\gamma ^2 }{2}+\frac{1 }{3}\right)\lambda-\frac{2 \gamma }{3}\, .
\end{equation}
More generally, the general form of the characteristic polynomial $p(\lambda)$ of a matrix $M$ of size $3\times 3$ is 
\begin{align}\label{appC:general_form}
    p(\lambda) = -\lambda^3 + \Tr(M) \lambda^2 - \frac{1}{2}\left(\Tr(M)^2 - \Tr(M^2)\right)\lambda + \det(M)\, .
\end{align}

\noindent \textit{Calculation of $\gamma_{\text{toy}}$.}\hspace{0.2cm} By definition, at $\gamma=\gamma_{\text{toy}}$, the characteristic polynomial $p(\lambda)$ has a pair of purely imaginary roots, say $\pm i\, \omega$, and a third real root, say $r$. Consequently, $p(\lambda)$ can be factorized as
\begin{equation}\label{appC:factorization}
    -(\lambda - r)(\lambda-i\, \omega)(\lambda+i\, \omega) = -\lambda^3 + r \lambda^2 - \omega^2 \lambda + r \omega^2\, .
\end{equation}
Equating the coefficients of $p(\lambda)$ with those in Eq.~\eqref{appC:factorization} yields the following system
\begin{subequations}
\begin{align}
    &r = \Tr(M) = -\frac{11 \gamma}{6}\, ,\\
    &-\omega^2 = -\frac{1}{2}\left(\Tr(M)^2 - \Tr(M^2)\right) = - \left(\frac{\gamma ^2}{2}+\frac{1}{3}\right)  ,\\
    &r \omega^2 = \det(M) = -\frac{2 \gamma}{3}\, . 
\end{align}
\end{subequations}
Solving this system yields a unique positive solution for $\gamma$, which is given by $\gamma_{\text{toy}} =\sqrt{2/33}\approx 0.246$.\\

\noindent \textit{Calculation of $\widetilde{\gamma}$.}\hspace{0.2cm} By definition, at $\gamma=\widetilde{\gamma}$, the characteristic polynomial $p(\lambda)$ has only real roots. This condition is equivalent to requiring that the discriminant $\Delta$ of $p(\lambda)$ satisfies $\Delta\geq 0$. Recall that the discriminant of a cubic polynomial $p(\lambda)=a \lambda^3 + b \lambda^2 + c \lambda + d$ is $\Delta=18abcd - 4b^3 d + b^2 c^2 - 4 a c^3 - 27 a^2 d^2$, which in our case is equal to
\begin{equation}
    \Delta =\frac{1}{1296}\left(441 \gamma ^6-6884 \gamma ^4-6428 \gamma ^2-192 \right) .
\end{equation}
The critical case $\Delta=0$ corresponds to solving a cubic equation in $\gamma^2$. Using Cardano's method, the only positive root is given by
\begin{equation}
    \widetilde{\gamma}^2 = -\frac{1}{1323}\left(-6884 + e^{-i \frac{4\pi}{3}} C + \frac{55893700}{e^{-i\, \frac{4\pi}{3}} C} \right),\quad \widetilde{\gamma}^2\in \mathbb{R} ,
\end{equation}
where $C=2\  \sqrt[3]{25} \sqrt[3]{-2072739667+i\,3124926 \sqrt{7086}}\in \mathbb{C}$. Numerical evaluation yields $\widetilde{\gamma}\approx 4.061$.

\section{\label{appendixC}Benettin \textit{et al.} algorithm for computing the Lyapunov characteristic exponents}

This appendix provides a brief review of LCEs and on the Benettin \textit{et al.} algorithm for computing the LCE spectrum. For further details, we refer the reader to the abundant
literature on this topic, see for example the standard references~\onlinecite{ LCE_numerical_approach,LCE_time_series,LCE_computation,LCE_book}\\

\noindent \textit{Part 1: General considerations.} Consider a generic $n$-dimensional dynamical system of the form 
\begin{equation}\label{app:dynamical_system}
    \dot{\bm{X}}=\bm{F}(\bm{X})\, ,
\end{equation}
where $\bm{X}=\left(X_1,\cdots,X_n \right)$ is a vector of the dynamical variables, and the functions $F_i$ ($i=1,\cdots,n$) describe the dynamical processes governing the evolution of $\bm{X}$. The latter may also depend on a set of parameters $\mu_1, \mu_2, \cdots$ that we don't write explicitly. While we assume the system to be autonomous, the following considerations also apply to non-autonomous systems.

The stability of trajectories of such a system is a central question in dynamical systems theory. In essence, a trajectory is considered stable if all other trajectories starting in its vicinity remain close over time. Conversely, it is unstable if an infinitesimal perturbation of the trajectory grows exponentially with time. The concept of LCEs arises naturally when assessing the stability of generic trajectories. LCEs quantify the exponential rate of separation of infinitesimally close trajectories and are a fundamental tool for determining the stability of a given dynamical regime.

In the context of deterministic chaos, a primary signature is sensitivity to initial conditions, which limits the ability to predict the system’s evolution beyond a certain time. Traditionally, a system is considered chaotic if it has at least one positive LCE, with the magnitude of the exponent reflecting the time scale over which the dynamics becomes unpredictable. A positive LCE corresponds to an exponential rate of separation of infinitesimally close trajectories. Since nearby trajectories represent nearly identical states, even tiny differences -- often too small to resolve in practice -- can rapidly lead to divergent outcomes, resulting in a loss of predictive power.

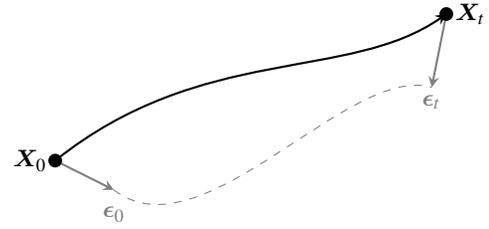
\begin{figure}
    \centering
\begin{tikzpicture}[scale=1.3, >=stealth]
\draw[gray, thick, ->] (0,0) -- (0.6,-0.3);
\node[text=gray] at (0.6,-0.55) {$\bm{\epsilon}_0$};

\draw[gray, thick, ->] (4,1.5) -- (3.85,0.75);
\node[text=gray] at (3.85,0.6) {$\bm{\epsilon}_t$};

\draw[thick, black, ->]
  (0,0) .. controls (1.5,1.2) and (3,0.8) .. (4,1.5)
  node[right] {$\bm{X}_t$};

\fill (0,0) circle (2pt) node[left] {$\bm{X}_0$};
\fill (4,1.5) circle (2pt);

\draw[gray, dashed]
  (0.6,-0.3) .. controls (1.5,-1) and (3,1) .. (3.85,0.75);
\end{tikzpicture}
\caption{Time evolution of a trajectory from initial condition $\bm{X}_0$ to $\bm{X}_t$, and the corresponding growth of an infinitesimal perturbation from $\bm{\epsilon}_0$ to $\bm{\epsilon}_t$.}
\label{fig:perturbation}
\end{figure}

Let $\bm{X}_0$ be an initial condition of the system~\eqref{app:dynamical_system}, and let $\bm{\epsilon}_0$ be an infinitesimal perturbation from $\bm{X}_0$, as illustrated in Fig.~\ref{fig:perturbation}. The time evolution of this infinitesimal perturbation, $\bm{\epsilon}(t)$, is obtained by linearizing the equations of motion,
\begin{equation}
    \dot{\bm{\epsilon}}=\text{G}\left[\bm{X}(t)\right]\,\bm{\epsilon}\, ,
\end{equation}
where $\text{G}\left[\bm{X}\right]_{ij}\equiv \left(\partial F_i/\partial X_j\right)\big|_{\bm{X}}$ is usually referred to as the linear stability (Jacobian) matrix. The solution of this linearized system can be written as
\begin{align}\label{appE:fundamental_matrix}
    & \bm{\epsilon}(t)=\text{M}\left[\bm{X}_0,t\right] \bm{\epsilon}_{0}\, ,
\end{align}
where $\text{M}\left[\bm{X}_0,t\right]$ is the fundamental matrix, which can formally be expressed as
\begin{equation}
    \text{M}\left[\bm{X}_0,t\right]=\mathcal{T}\exp{\int_0^t \dd \tau\,  G\left[\bm{X}(\tau) \right]}\, ,
\end{equation}
where $\mathcal{T}$ denotes the time-ordering operator. This matrix plays an important role in the dynamics of error growth, as the squared amplitude of the perturbation $\bm{\epsilon}(t)$ can be expressed as
\begin{equation}
    \parallel \hspace{-0.075cm}\bm{\epsilon}(t)\hspace{-0.075cm}\parallel^2=\parallel \hspace{-0.075cm}\text{M}\left[\bm{X}_0,t\right]\bm{\epsilon}_0\hspace{-0.075cm}\parallel^2=\bm{\epsilon}_{0}^T \text{M}^T\left[\bm{X}_0,t\right]\text{M}\left[\bm{X}_0,t\right]\bm{\epsilon}_{0}\, ,
\end{equation}
where we assume that $\parallel \hspace{-0.075cm} \cdot \hspace{-0.075cm} \parallel$ denotes the usual Euclidean norm on $\mathbb{R}^n$. This shows that the growth of the perturbation amplitude is governed by the (positive) eigenvalues $\sigma_n(t)\leq \cdots\leq \sigma_2(t)\leq \sigma_1(t)$ of the symmetric matrix $\text{M}^T\left[\bm{X}_0,t\right]\text{M}\left[\bm{X}_0,t\right]$. By the multiplicative ergodic theorem of Oseledets \cite{Oseledets}, for almost every initial condition $\bm{X}_0$, the following limit exists
\begin{equation}
    \lim_{t\rightarrow\infty}\left(\text{M}^T\left[\bm{X}_0,t\right]\text{M}\left[\bm{X}_0,t\right] \right)^{\frac{1}{2t}}\equiv \Lambda\left[\bm{X}_0\right] .
\end{equation}
The asymptotic LCE spectrum is then defined as
\begin{equation}
    \lambda_k=\lim_{t\rightarrow \infty}\frac{1}{t}\ln\sqrt{\sigma_k(t)}\, .
\end{equation}
There exist as many LCE exponents as there are linearly independent initial perturbations $\bm{\epsilon}_0$. We highlight that the LCEs are defined in the long-time limit. 

Given the LCE spectrum, the nature of the trajectory can be classified as follows,
\begin{align*}
    & \lambda_k<0 \quad\forall\, k=1,\cdots,n\hspace{0.7em} \rightarrow \hspace{0.7em} \text{fixed point,}\\
    & \lambda_1=0\text{ and } \lambda_k<0\quad \forall\, k=2,\cdots,n\hspace{0.7em} \rightarrow \hspace{0.7em} \text{periodic motion,}\\
    & \lambda_{1}, \lambda_{2}=0\text{ and } \lambda_k<0\quad \forall\,  k=3,\cdots,n\hspace{0.7em} \rightarrow \hspace{0.7em} \text{toroidal motion,}\\
    & \exists\, \text{ at least one }  \lambda_k>0\hspace{0.7em} \rightarrow \hspace{0.7em}\text{chaotic motion.}\\
\end{align*}

\noindent \textit{Part 2: Benettin et al. algorithm.} Important theoretical and numerical results on LCEs can be found in the seminal works of Benettin \textit{et al.} \cite{Benettin_1, Benettin_2, Benettin_3}, where an explicit method for computing the full LCE spectrum was developed. The central idea is to follow the evolution of an orthonormal basis of vectors $\bm{\epsilon}_k$ ($k=1,\cdots,K$, with $K\leq n$) initially chosen at random in the tangent space of the trajectory $\bm{X}(t)$. A direct consequence of Oseledets' theorem is that the exponential growth or contraction of the volume of the $K$-dimensional parallelepiped spanned by the vectors $\bm{\epsilon}_k$, denoted $\text{vol}_K\left(\bm{\epsilon}_1,  \cdots, \bm{\epsilon}_K\right)$, is given by the sum of the first $K$ LCEs, 
\begin{equation}
    \sum_{k=1}^K \lambda_k = \lim_{t\rightarrow \infty}\frac{1}{t}\ln\left( \frac{\text{vol}_K\left(\bm{\epsilon}_1(t),  \cdots, \bm{\epsilon}_K(t)\right)}{\text{vol}_K\left(\bm{\epsilon}_1(0),  \cdots, \bm{\epsilon}_K(0)\right)}\right).
\end{equation}
However, this formula cannot be applied directly in numerical computations. Without intervention, the vectors $\bm{\epsilon}_k$ tend to align along the most unstable direction associated with the largest LCE, causing a loss of linear independence. Furthermore, in chaotic systems, the norm of the vectors typically grows exponentially, which can lead to numerical overflow. To overcome these issues, the set of vectors $\bm{\epsilon}_k$ is periodically reorthonormalized using the standard Gram-Schmidt (GS) procedure.

The algorithm proceeds as follows. An initial condition $\bm{X}(0)$ and an orthonormal basis of vectors $\bm{\epsilon}_k(0)$ ($k=1,\cdots,K$) chosen at random are fixed. The total number of iterations $N$ and the time interval $s$ between successive GS reorthogonalizations are also specified.\\

\begin{algorithmic}
\For{$i = 1$ \textbf{to} $N$}
    \State Evolve the vectors $\bm{\epsilon}_k((i-1)s)$ using Eq.~\eqref{appE:fundamental_matrix} with \\\hspace{.55cm}initial conditions $\bm{X}((i-1)s)$
    \State 
    \State Define $\alpha_1(i)=\parallel \hspace{-0.075cm}\bm{\epsilon}_1(is)\hspace{-0.075cm}\parallel $ and $\tilde{\bm{\epsilon}}_1(is)=\bm{\epsilon}_1(is)/\alpha_1(i)$ 
    \State
    \For{$k = 2$ \textbf{to} $K$}
    \State $\alpha_k(i)=\parallel \hspace{-0.075cm} \bm{\epsilon}_k(is)-\sum_{l=1}^{k-1}\left(\tilde{\bm{\epsilon}}_l(is)\cdot \bm{\epsilon}_k(is)  \right)\tilde{\bm{\epsilon}}_l(is) \hspace{-0.075cm}\parallel $
    \State $\tilde{\bm{\epsilon}}_k(is)=\left(\bm{\epsilon}_k(is)-\sum_{l=1}^{k-1}\left(\tilde{\bm{\epsilon}}_l(is)\cdot \bm{\epsilon}_k(is)  \right)\tilde{\bm{\epsilon}}_l(is) \right)/\alpha_k(i)$
    \EndFor
    \State 
    \For{$k = 1$ \textbf{to} $K$}
    \State Reinitialize $\bm{\epsilon}_k(is)=\tilde{\bm{\epsilon}}_k(is)$
    \EndFor
\EndFor
\State 
\For{$k = 1$ \textbf{to} $K$}
\State $\lambda_k=\sum_{i=1}^N \ln \alpha_k(i)/(Ns)$ 
\EndFor
\State
\end{algorithmic}
The algorithm yields the LCE evaluated at time $t=Ns$, and convergence as $N\rightarrow\infty$ gives the proper asymptotic LCEs. We also note that the choice of norm is irrelevant, as LCEs are independent of it.

\section*{References}
\nocite{*}
\bibliography{bibli_project}

\end{document}